\documentclass[3p,11pt]{elsarticle}
\date{September 11, 2020}

\usepackage{amssymb}
\usepackage{latexsym}

\usepackage{amsmath}
\usepackage{xfrac}
\usepackage{trfsigns}
\usepackage{tikz} 

\usepackage{url}
\usepackage{xcolor}
\definecolor{newcolor}{rgb}{.8,.349,.1}

\newcommand{\dd}[2]{\frac{D #1}{D #2}}
\newcommand{\bfm}[1]{\mbox{\boldmath $#1$}}
\newcommand{\bm}[1]{\bfm{#1}}
\newcommand{\nop}{\bm \nabla}
\newcommand{\pp}[2]{\frac{\partial{#1}}{\partial{#2}}}
\newcommand{\ddb}[2]{\frac{\bar{D} #1}{D #2}}
\newcommand{\ddt}[2]{\frac{\widetilde{D} #1}{D #2}}

\newcommand{\ten}[1]{\bfm{\sf{#1}}}

\newcommand{\bstar}[1]{{\bfm{#1}}^*}

\newcommand{\sstar}[1]{{#1}^*}
\newcommand{\mm}{\mathrm{\,mm}}
\newcommand{\mys}{\mathrm{\,\mu s}}
\newcommand{\Hz}{\mathrm{\,Hz}}
\newcommand{\mps}{\mathrm{\,m/s}}
\newcommand{\dB}{\mathrm{\,dB}}
\newcommand{\Pa}{\mathrm{\,Pa}}
\newcommand{\m}{\mathrm{\,m}}
\newcommand{\s}{\mathrm{\,s}}

\newcommand{\rec}[1]{{#1}^{-1}}
\newcommand{\qfrac}[2]{{#1}/{#2}}

\newcommand{\bcal}[1]{\bfm{\mathcal{#1}}}

\journal{Journal of Computational Physics}

\begin{document}

\begin{frontmatter}

\title{Hydrodynamic/acoustic splitting approach with flow-acoustic feedback for universal subsonic noise computation}

\author[1]{Roland Ewert\corref{cor1}}
\cortext[cor1]{Corresponding author}
\ead{Roland.Ewert@dlr.de}

\author[2]{Johannes Kreuzinger}
\ead{J.Kreuzinger@km-turbulenz.de}

\address[1]{Inst. of Aerodynamics and Flow Technology, German Aerospace Center DLR, Braunschweig, Germany}

\address[2]{KM-Turbulenz GmbH, Munich, Germany}

\begin{abstract}
A generalized approach to decompose the compressible Navier-Stokes equations into an equivalent set of coupled equations for flow and acoustics is introduced. As a significant extension to standard hydrodynamic/acoustic splitting methods, the approach provides the essential coupling terms, which account for the feedback from the acoustics to the flow. A unique simplified version of the split equation system with feedback is derived that conforms to the compressible Navier-Stokes equations in the subsonic flow regime, where the feedback reduces to one additional term in the flow momentum equation. Subsonic simulations are conducted for flow-acoustic feedback cases using a scale-resolving run-time coupled hierarchical Cartesian mesh solver, which operates with different explicit time step sizes for incompressible flow and acoustics. The first simulation case focuses on the tone of a generic flute. With the major flow-acoustic feedback term included, the simulation yields the tone characteristics in agreement with reference results from K\"uhnelt~\cite{kuehnelt2016} based on Lattice-Boltzmann simulation. On the contrary, the standard hybrid hydrodynamic/acoustic method with the feedback-term switched off lacks the proper tone. As the second simulation case, a thick plate in a duct is studied at various low Mach numbers around the Parker-$\beta$-mode resonance. The simulations reveal the flow-acoustic feedback mechanism in very good agreement with experimental data of Welsh et al.~\cite{welsh84}. Simulations and theoretical considerations reveal that the feedback term does not reduce the stable convective flow based time step size of  the flow equations.
\end{abstract}

\end{frontmatter}



\section{Introduction}\label{sec:introduction}

Many aeroacoustics problems in engineering are related to vortex-sound problems at low Mach numbers in the range $M<0.3$. Examples among others of problems that fall into this category are automotive noise, airframe noise of commercial aircraft at landing approach, and wind turbine noise generation and radiation problems.  

Hydrodynamic/acoustic splitting (HAS) methods frame an efficient approach for scale resolved noise computation in the low Mach number range. They are based on a decomposition of the continuity and momentum equation of the compressible Navier-Stokes equations into an hydrodynamic part governed by the incompressible Navier-Stokes equations plus equations for the residual compressible corrections that include the acoustic components. As long as the feedback from acoustics to the flow is of little relevance, this hybrid approach yields a low Mach number alternative to a direct noise computation (DNC) using the compressible Navier-Stokes equations to resolve both, flow and acoustics~\cite{bailly2010}. 

In the first proposed HAS method of Hardin \& Pope~\cite{hardin94,hardin95}, viscous terms in the acoustic equations are neglected. A simplified caloric equation of state~\cite{hardin94} is applied following some reasoning of Batchelor~\cite{batchelor67} that the fast varying pressure fluctuations are effectively isentropic in origin. Using a low Mach number expansion, Slimon et al. derived an alternative approach to HAS (Expansion over Incompressible Flow, EIF)~\cite{slimon99,slimon2000}. Shen \& S{\o}rensen~\cite{shen99a,shen99b} proposed a revised version of the Hardin \& Pope~\cite{hardin94,hardin95} splitting, where the partial time derivative of incompressible pressure occurs as the major sound source term in the acoustic equations. Results for the sound radiated from NACA airfoils in laminar and turbulent flow were reported~\cite{shen2001,shen2009,zhu2011}.    

\begin{figure}[!h]
  \centering
    \includegraphics[width=0.49\textwidth]{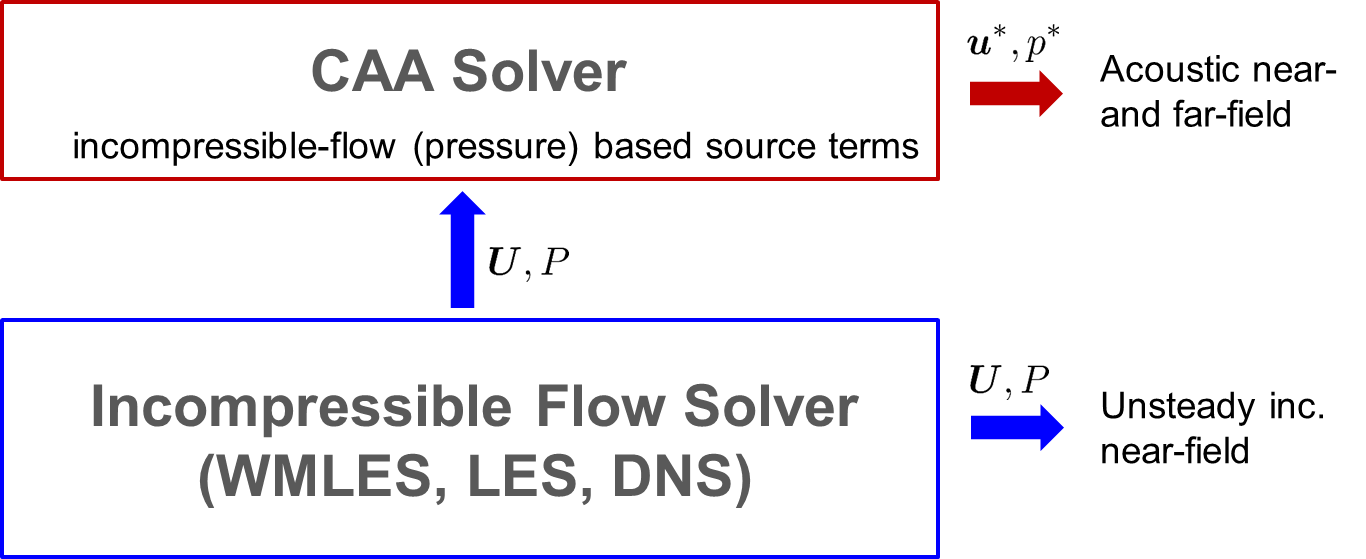}  
    \caption{Open loop hydrodynamic-acoustic splitting method. The incompressible flow solver provides the unsteady flow $\bfm U$, which serves as the background medium of a computational aeroacoustics (CAA) simulation. The flow is simulated applying either wall-modeled large-eddy simulation (WMLES), large-eddy simulation (LES), or direct numerical simulation (DNS); the incompressible pressure $P$ defines the essential sound source.}
    \label{fig:open_loop_HAS}  
\end{figure}

For each of the HAS equations optimized numerical methods can be applied, i.e., dedicated Computational Fluid Dynamics (CFD) methods to resolve the unsteady incompressible flow part, and Computational AeroAcoustics (CAA) methods for the sound propagation task. Furthermore, different meshes for flow and acoustics can be applied in both simulations. This way, the multi-scale resolution problem present for low Mach flow~\cite{tam95} is mitigated, refer to the more detailed discussion in Section~\ref{sec:HADvsDNC}. Figure~\ref{fig:open_loop_HAS} gives a schematic of the open loop hybrid HAS approach which indicates the incompressible pressure $P$ as the main element of the sound source term in the acoustic equations. Since the hydrodynamic pressure follows from the solution of a Poisson equation, the anisotropy of its turbulent source term is effectively leveled out. The resulting incompressible pressure field therefore represents a well suited fluctuating vortex sound source to be transferred by regular time-step updates to the typically more isotropic acoustic mesh. 
  
The acoustic momentum equation of the HAS versions discussed so far comprise besides the acoustic also the vorticity mode and include non-linear terms, which can give rise to flow instabilities and residual turbulence production in the residual compressible equations. To fully exploit the potential of the HAS approach in overcoming the low Mach number multi-scale stiffness, it would be desirable to have a clear separation between acoustic and non-acoustic modes and associated scales.  

Furthermore, acoustic fluctuations are small so that linear acoustic equations are well suited for the acoustic simulation step. However, linearization of the residual compressible equations yields linearized Euler type of equations (LEE) which include growing convective Kelvin-Helmholtz instabilities~\cite{delfs96,bogey2002} in free shear layers and global instabilities behind bluff bodies~\cite{boppana2011}.\footnote{The convective instabilities are limited by the missing non-linear and viscous terms; when present they cause the development of residual turbulence.} 

Ewert \& Schr\"oder~\cite{ewert2003} proposed linear acoustic perturbation equations (APE) that result from the LEE by filtering out the vorticity and entropy modes, i.e. only support the acoustic mode and suppress the development of all types of flow instabilities. Their APE-1 and APE-2 systems represent HAS variants based on a triple decomposition of the flow into a time averaged mean plus fluctuations further decomposed into a hydrodynamic and an acoustic part. APE-1 involve the gradient and APE-2 the substantial time derivative of fluctuating incompressible pressure as the major source term. 

Schwertfirm \& Kreuzinger~\cite{schwertfirm2012,kreuzinger2013,schwertfirm2014,kreuzinger2016} have demonstrated an impressive agreement with experimental data applying APE-1 run-time coupled with incompressible Large Eddy Simulation for complex geometries such as realistic automotive heating, ventilation, and air conditioning (HVAC) exhaust nozzle units.

Seo \& Moon~\cite{seo2006} improved the general HAS approach by avoiding the usage of an simplified caloric equation of state through deriving their split acoustic equations from the complete compressible Navier-Stokes equations. Based on a term-by-term analysis to simplify their non-linear perturbed compressible equations (PCE)~\cite{seo2005}, they arrive at a system of linearized equations (LPCE) that combine the APE momentum equation over unsteady incompressible base flow with a modified linearized equation for pressure. To arrive at this result, they conclude that the perturbed Lamb vector\footnote{The Lamb vector defined by flow velocity $\bfm u$ in terms of $\bcal L := \bfm \omega \times \bfm u$, $\bfm \omega = \nop \times \bfm u$, will be introduced and discussed in more detail in Section~\ref{sec:split_eqns} and in the context of the identity Eq.~(\ref{eq:identity_lamb}). Its perturbed form is given by $\bcal L' = \bcal L - \overline{\bcal L}$, where $\overline{\bcal L}$ denotes the time average of $\bcal L$.} can be neglected as an order $M^4$ term for low Mach number problems. As part of this work, the substantial time derivative of incompressible pressure was rigorously derived as the major sound source term from the exact governing equations. Many successful direct noise computations based on incompressible LES with LPCE have been reported since by Moon and co-workers~\cite{moon2010,seo2011,bae2011,moon2013}. 

Derivatives of the HAS approach based on this three key elements, viz, (i) incompressible scale resolving simulation, (ii) incompressible pressure based vortex sound source term, and (iii) linear convective wave propagation and their successful application to various low Mach number flow problems have been reported in the mean time by several groups, e.g., refer to H\"uppe and Kaltenbacher~\cite{hueppe2014,kaltenbacher2017,schoder2019}.   

\begin{figure}[!h]
  \centering
    \includegraphics[width=0.49\textwidth]{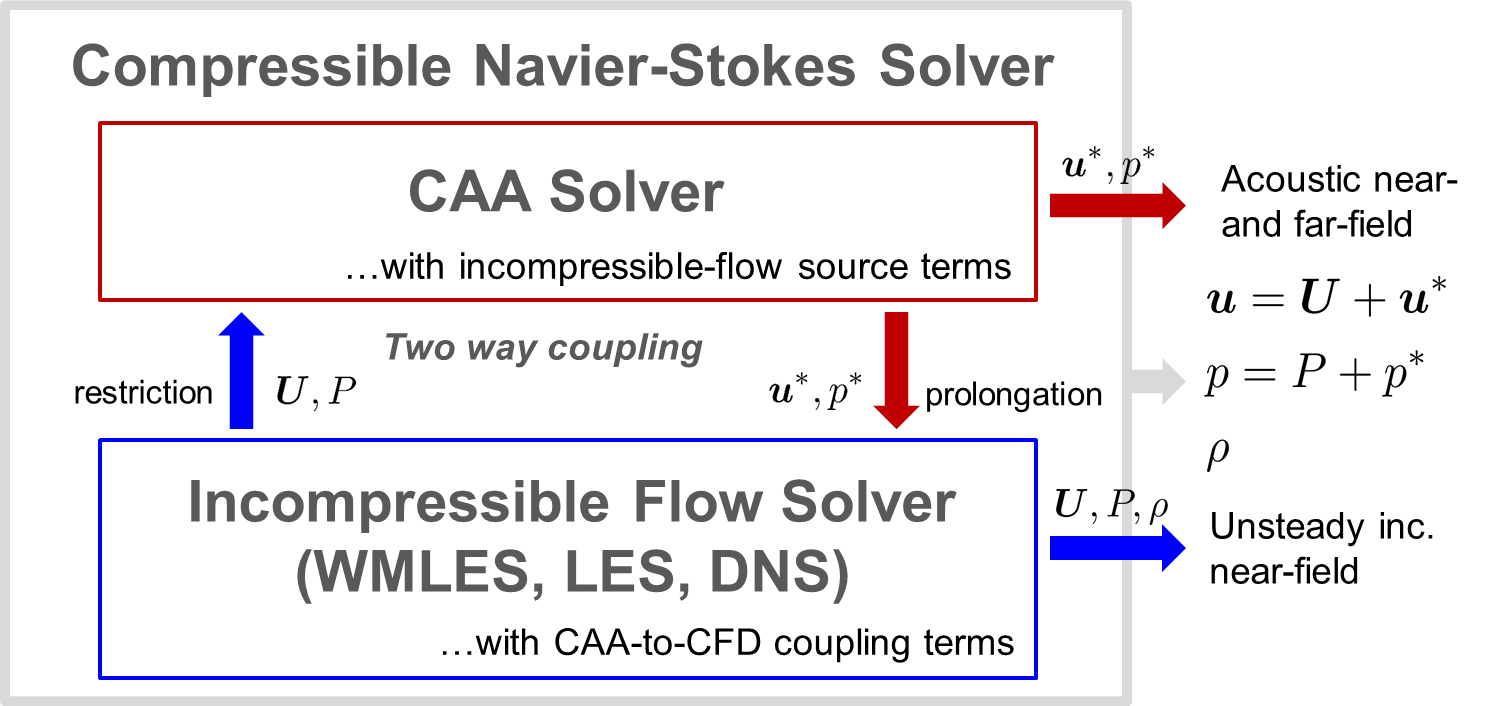}  
    \caption{A run-time coupled hydrodynamic-acoustic splitting method with feedback provides an alternative way to solve the compressible Navier-Stokes equations.}
    \label{fig:installed_directivity}   
\end{figure}

However, the splitting of the governing equations into two equation systems for flow and acoustics and their causal solution within a hybrid two-step approach is based on the assumption that the back action of the acoustic field onto the sound generating flow is negligible. This prevents to capture any feedback related tonal phenomena. In engineering, unexpected tones are the most annoying of all aeroacoustics noise phenomena. The understanding of the underlying source mechanism and their suppression by means of measures derived from simulation therefore represent an important tasks for CAA. 

In this work an alternative derivation of the hydrodynamic/acoustic splitting is introduced that provides a class of split equation systems that are equivalent to the compressible Navier-Stokes equations. It is demonstrated that the acoustic equations can be well approximated in terms of a generalized non-linear APE momentum equation and vortex sound source term based on the substantial time derivative of incompressible pressure. In the linearized low-Mach number limit the acoustic system reduces to the LPCE equations. 

In addition, the major coupling terms are identified that provide the feedback from acoustics to the flow. It is shown that even for low Mach number flows the perturbed Lamb vector cannot be neglected but rather provides the essential flow-acoustic feedback term in the hydrodynamic momentum equation.

This paper is organized as follows: Section~\ref{sec:HADvsDNC} provides some additional reasoning about the efficiency of the HAS approach relative to DNC if applied to low Mach number generic trailing edge noise. Section~\ref{sec:split_eqns} introduces the generalized splitting approach of the compressible flow regime that enables the identification of the essential flow-acoustic feedback terms in the hydrodynamic momentum equation. Section~\ref{sec:low_subsonic_vortex_sound} introduces additional simplifications in the acoustic equations that hold for low Mach number flow and uses the result to show theoretically that the flow-acoustic feedback terms do not affect the numerical stability of the combined system. Section~\ref{sec:validation} presents the validation of the proposed HAS method with feedback terms for two test problems, viz, the sound generated by a generic flute and for the resonant flow-acoustic interaction of a thick plate in a duct. Final conclusions are drawn in Section~\ref{sec:conclusions}.

   
\section{Low subsonic performance characteristics of HAS and DNC}\label{sec:HADvsDNC}

The notion Direct Noise Computation (DNC) was introduced by Bailly et al.~\cite{bailly2010} to indicate the simultaneous solution of the aerodynamic and the acoustic field in one computational step by means of the compressible Navier-Stokes equations. These method requires low dispersive and dissipative numerical methods to properly support wave propagation over extended distances, e.g. refer to the review of Tam~\cite{tam95}. Standard numerical methods applied to solve the compressible Navier-Stokes equations resolve wave propagation only over short distances and require hybrid methods that employ an extra wave propagation step to propagate the sound waves from the acoustic near- to the far-field~\cite{ewert2014,akkermans2018}. Hybrid methods utilize acoustic analogies as first introduced in the seminal papers of Lighthill~\cite{lighthill52,lighthill54}. The recent state-of-the-art is highlighted by the work of Schlottke-Lakemper et al.~\cite{schlottkelakemper2017}, who proposed a run-time coupled massive parallel hybrid APE-4/compressible LES solver on hierarchical meshes~\cite{niemoeller2020}.

HAS methods represent an alternative to acoustic analogies~\cite{sun2018} in the low subsonic regime. In comparison to DNC, hybrid and HAS approaches mitigate potential accuracy problems related to the simultaneous resolution of fluctuations of turbulent flow together with acoustics of much smaller magnitude, refer to the discussion of Tam~\cite{tam95}.   

Unlike the DNC and the acoustic analogy approach, the HAS approach provides a decomposition into acoustic and hydrodynamic near field components that might be useful for the further analysis of the aeroacoustics problem at hand.   
   
A DNC at low Mach number, realized in a scale resolving simulation framework such as direct numerical simulation (DNS), large eddy simulation (LES), or wall modeled LES (WMLES), inevitably introduces the numerical stiffness problem related to the simultaneous resolution of acoustic and convective wave-modes on meshes adapted to anisotropic turbulent source dynamics. 

The resolution constraints of DNC and HAS are discussed in the next two subsections for the example of noise generated at the trailing edge of an airfoil. The indicated DNC requirements to properly resolve the acoustic near field also hold for general compressible Navier-Stokes approaches. 

\subsection{Wall resolved LES}

\subsubsection{Trailing Edge Noise Example}

\begin{figure}[!h]
	\centering
	\includegraphics[width=0.40\textwidth]{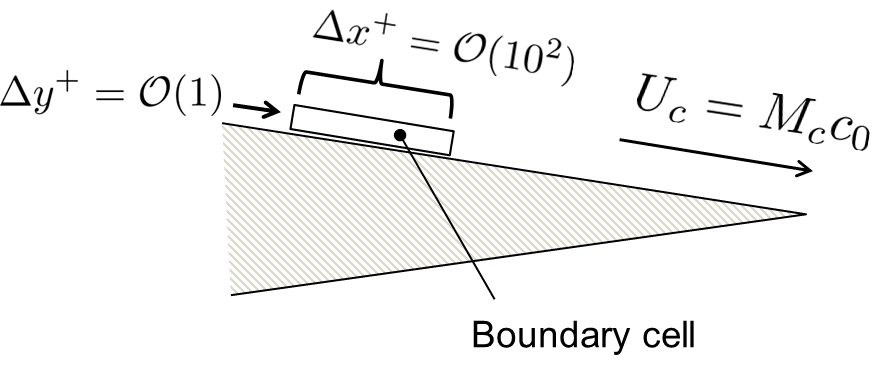}  
	\caption{Sketch of trailing edge with typical wall-normal and streamwise resolution requirements.}
	\label{fig:TE_Sketch}   
\end{figure}

To simulate trailing edge noise with a wall resolved LES, stretched near-wall cells with typical wall normal spacing of the order of one wall unit are required, $\Delta y^+= \mathcal{O}(1)$, and the streamwise resolution is up to two orders of magnitude higher, $\Delta x^+=\mathcal{O}(10^2)$~\cite{wagner2007}, refer to Figure~\ref{fig:TE_Sketch}. 

Trailing edge noise is generated by the convection of turbulent eddies past the trailing edge. The associated characteristic time-scale is given by the characteristic streamwise turbulent length-scale and boundary layer convection velocity $U_c\approx 0.6\ldots 0.7 U_\infty$. Note, the turbulent decay time-scale is typically larger than the convective time-scale. For example, trailing edge noise models provide good predictions based on the assumption of frozen turbulence (decay time-scale $\to \infty$)~\cite{herr2015}. A resolved convection time-scale is of order $\Delta t \sim \Delta x^+/U_c$. The time-scale of the sound waves generated at the trailing edge must comply to the convective time-scale. Hence, the acoustic length-scale is given by $\Delta_\lambda^+ = \Delta x^+ /M_c$, where $M_c=U_c/c_\infty$ is the convection Mach number defined by the speed of sound $c_\infty$.       

For a DNC, the Courant-Friedrichs-Lewy (CFL) condition that restricts the numerical time step size is defined by the speed of sound and the smallest spacing, $\Delta t_{DNC}\sim \Delta y^+/c_\infty$.  

The HAS approach aims to alleviate the stiffness problem by means of a complementary solution of the different propagation modes with each type of split equation applied. For example, utilizing a fractional step method for the incompressible flow part, the time step size for the intermediate velocity update is governed by a  convective velocity scale. Usage of explicit Runge-Kutta time integration for the first step ensures good dispersive properties in the resolution of turbulent dynamics. This first step is followed by a projection step by means of solving an elliptic Poisson problem that yields pressure and velocity at the next time level. 

For the trailing edge noise example, the CFL-condition of the first step is defined by the convection velocity $U_c$ and the streamwise mesh resolution $\Delta x^+=\mathcal{O}(10^2)$, i.e. $\Delta t_{HAS} \sim \Delta x^+/U_c$. A typical Mach number of interest is in the range $M_\infty \approx 0.15$~\cite{herr2015}. 

To conclude, for the present case the difference in explicit time step size between the HAS flow equations and DNC reaches three orders of magnitude, $\Delta t_{HAS}/\Delta t_{DNC} = M_c^{-1} \Delta x^+/\Delta y^+=  \mathcal{O}\left ( 10^3 \right )$.  

The wall resolved acoustic stiffness problem is circumvented in the HAS approach by using a dedicated isotropic acoustic mesh more suitable to resolve the acoustic scale $\Delta_\lambda^+$, which yields comparable explicit time step sizes for incompressible flow and acoustics. For the projection step in the HAS flow treatment, an efficient Poisson solver for pressure is of paramount importance~\cite{friedrich2001}. With this prerequisite considered, the indicated order of magnitude of the time step ratios between HAS and DNC can be preserved for the effective computational time step ratios as well. Good overall performance characteristics for scale resolved incompressible simulation of subsonic flow problems have been reported e.g. in~\cite{schwamborn2014}. 

\subsubsection{Implicit time integration for DNC}

In general, the stiffness problem of the compressible Navier-Stokes equations could be mitigated using implicit time integration. However, the implicit solution procedure has to be applied to five coupled flow variables, compared to one implicit procedure for the incompressible pressure in the HAS flow equation and to catch up in efficiency with the latter poses an ambitious goal. Furthermore, the second order backward difference Euler method often applied for implicit time integration due to their favorable robustness properties exhibits only poor dispersive and dissipative properties compared with explicit Runge-Kutta time integration and therefore the latter appears better suited for the resolution of both turbulent flow dynamics and acoustic waves. On the contrary, usage of an implicit Runge-Kutta method for improved dispersive and dissipative properties poses challenging numerical robustness issues that are avoided from the outset in the HAS framework.  

\subsubsection{Simplified low-subsonic compressible Navier-Stokes equations}

It has been proposed to simplify the compressible Navier-Stokes equations in the low Mach number limit to reduce the numerical effort. Tam and Kurbatski~\cite{tam2000,tam2001,tam2011} have used the compressible Navier-Stokes equations in suitably simplified form for a direct numerical simulation (DNS) of flow induced sound generation at subsonic Mach number $M_\infty \approx 0.1$. In their simulations the dissipation terms in the pressure equation are omitted. Furthermore, the viscous right-hand side term of the momentum equation is computed based on a constant viscosity ($\mu = const.$). The dissipation of mechanical energy is still governed by the momentum equation. However, the simplification implies that the mechanical energy dissipation is negligible relative to the thermodynamic energy content. The modification reduces to an isentropic relationship between pressure and density.   

Pont et al.~\cite{pont2018} propose to use a pressure equation similar to artificial compressibility methods with the true compressibility derived from the isentropic relationship between pressure and density derivatives. Due to the reduced computational effort by skipping the density equation, they report a higher efficiency of their implicit compressible solver relative to their incompressible flow solver implementation. Furthermore, they point out that the compressible Navier-Stokes equations are able to resolve the two-way coupling between flow and acoustics that is not captured by the standard segregated HAS approach as shown in Figure~\ref{fig:open_loop_HAS}. Contrary to these results, Kiris \& Kwak~\cite{kiris2002} find the pressure projection method to be computationally more efficient than an implicit method based on artificial compressibility for an unsteady flow with small physical time step size. 

\subsection{Wall modeled LES}

For wall modeled LES (WMLES), the wall normal resolution requirements at the wall are less restrictive and therefore soften the stiffness problem. A suitable resolution is prescribed by the outer boundary layer scale given by $\Delta y \approx 0.02\ldots 0.05\delta$ and $\Delta x \approx 0.08 \delta$, refer to Larsson et al.~\cite{larsson2016}, where $\delta$ indicates the boundary layer thickness. With the shear stress Reynolds number $Re_\tau = u_\tau \delta/\nu$ based on shear stress velocity $u_\tau = \sqrt{\rho \tau_w}$ from wall shear stress $\tau_w$ and fluid density $\rho$, this translates into a smallest resolution in terms of wall units of $\Delta y^+\approx 0.02 Re_\tau$ and $\Delta x^+\approx 0.08 Re_\tau$. Hence, for WMLES $\Delta x^+/\Delta y^+ = \mathcal{O}\left ( 1 \right )$ and the time step ratio reduces to $\Delta t_{HAS}/\Delta t_{DNC} \approx  M_c^{-1}= \mathcal{O}(10^1)$.  

Specifically, if WMLES is applied for the HAS flow part on a hierarchical Cartesian mesh, it becomes beneficial to use the same mesh for the acoustic computation with a reduced explicit time step size so that a couple of intermediate acoustic steps are conducted per one flow update cycle. Due to the smaller numerical effort necessary to conduct one acoustic step, a good overall performance can be maintained for typical WMLES resolution in low subsonic flow. Approximately, the simulation time doubles relative to the stand-alone flow simulation. Furthermore, interpolation errors are avoided using the same grid for flow and acoustics.


\section{Modified Split Equation System}\label{sec:split_eqns}
\subsection{Compressible Navier-Stokes Equations}\label{ssec:compNS}

Starting point of the split equation derivation are the compressible Navier-Stokes-Equations for ambient air (assumed as an thermally and caloric ideal gas) without further body accelerations. In terms of primitive notation for density $\rho$, velocity vector $\bfm u$, and pressure $p$, they read   
\begin{align}
  \label{eq:ns_mass}
  &\pp{\rho}{t} +  \nop \cdot \left ( \rho \bfm u \right ) = 0 \\
  \label{eq:ns_momentum}
  &\pp{\bfm u}{t} + \left ( \bfm u \cdot \nop \right ) \bfm u + \frac{\nop p}{\rho}  - \frac{\nop \cdot \ten{\tau}}{\rho}  = \bfm 0 \\
  \label{eq:ns_energy}
  &\pp{p}{t} + \bfm u \cdot \nop p + \gamma p \nop \cdot \bfm u = \theta.   
\end{align}
Here, $\gamma=C_p/C_v$ is the isentropic exponent from the ratio of the specific heats at constant pressure and volume, respectively, which for ambient air takes on a constant value of $1.40$. The shear-stress tensor is denoted $\ten{\tau}$, which for an assumed Newtonian fluid without second viscosity is given by 
\begin{equation}\label{eq:visc_stress_compr}
	\ten{\tau}= 2\mu \ten{S} - \frac{2}{3} \mu \left ( \nop \cdot \bfm u \right ) \ten{I}  ,
\end{equation}
where $\ten{I}$ indicates the unity matrix, 
\begin{equation}
	\ten{S} = 1/2 \left ( \nop \bfm{u} +  \left ( \nop \bfm{u} \right )^T \right )
\end{equation}
denotes the rate-of-strain tensor, and $\mu$ is the dynamic viscosity. 
The term on the right-hand side (r.h.s.) of the pressure equation~(\ref{eq:ns_energy}) describes dissipation and thermal conductivity, 
\begin{equation}
	\theta = \left ( \gamma-1 \right ) \Phi + \left ( \gamma -1 \right ) \nop \cdot \left ( \kappa \nop T \right ) .
\end{equation}
The dissipation is defined by 
\begin{equation}
	\Phi = \ten{\tau} : \nop \bfm u, 
\end{equation}
where '$:$' indicates a double tensor contraction. The thermal conductivity term stems from the divergence of the heat flux 
\begin{equation}
	\bfm q = - \kappa \nop {T},
\end{equation}
with thermal conductivity parameter $\kappa = \mu C_p/Pr$ ($Pr\approx 0.72$ for ambient air) and temperature $T=p/(\rho R)$. For ambient air, the gas constant is 
$R=287.03$~$\mathrm{m^2/(s^2K)}$ and the specific heat at constant pressure is given by $C_p = \gamma/(\gamma-1)R$. The dynamic viscosity is a function of temperature alone, e.g., defined by Sutherland's law~\cite{wagner2007}. 
Based on these definition, and neglecting gradients of dynamic viscosity, the divergence of the stress tensor in Eq.~(\ref{eq:ns_momentum}) becomes  
\begin{equation}\label{eq:div_tau} 
	\nop \cdot \ten{\tau} \simeq  \mu \Delta \bfm u + \frac{1}{3} \mu \nop \left ( \nop \cdot \bfm u \right ), 
\end{equation}
where $\Delta \equiv \nop^2$ denotes the Laplacian operator. 

\subsection{Split Equation Derivation}\label{ssec:derivation}

To split-up the equation system~(\ref{eq:ns_mass}-\ref{eq:ns_energy}), a velocity decomposition 
\begin{equation}\label{eq:split_u}
	\bfm u = \bfm U + \bstar u
\end{equation}
is introduced. At this point no further constraints are imposed on the decomposition. In particular, both velocity contributions can carry vorticity, i.e.,
\begin{equation}\label{eq:split_omega}
	\bfm \omega = \bfm \Omega + \bstar \omega
\end{equation}
holds for the total and the sum of the split vorticity as defined by their respective flow velocities, 
$\bfm \omega := \nop \times \bfm u$, $\bfm \Omega := \nop \times \bfm U$, and $\bstar \omega := \nop \times \bstar u$.

In a next step, the momentum equation~(\ref{eq:ns_momentum}) is split-up into two separated equation systems, one for each velocity time-derivative. As the only constraint in place, we demand that the sum of the split system must reproduce the momentum equation~(\ref{eq:ns_momentum}). Therefore, the splitting is not unique, different options are possible. The approach taken here reads   
\begin{align}
  \label{eq:ns_momentum_U}
  &\pp{\bfm U}{t} + \underbrace{\left ( \bfm U \cdot \nop \right ) \bfm U - \frac{\nop \cdot \ten{\tau}}{\rho} 
     + \bfm \Omega \times \bstar u + \bstar \omega  \times \bfm u}_{\bfm l}   
  = -\bfm{X}  \\
  \label{eq:ns_momentum_up}
  &\pp{\bstar u}{t}  + \nop \left ( \bfm U \cdot \bstar u   + \frac{{\bstar u}^2}{2} \right ) + \frac{\nop p}{\rho} = \quad \quad \quad  \quad \quad \;  +\bfm{X}  .
\end{align}
The spatial derivative terms result from the decomposition Eq.~(\ref{eq:split_u}) and using the identities 
\begin{equation}\label{eq:identity_lamb}
 \left ( \bstar u \cdot \nop \right )\bfm U  + \left ( \bfm U \cdot \nop \right ) \bstar u  \equiv  \nop \left ( \bfm U \cdot \bstar u \right )  
	 + \underbrace{\bfm \Omega \times \bstar u + \bstar \omega  \times \bfm U}_{=:\bstar L}  
\end{equation}
and
\begin{equation*}
	\left ( \bstar u \cdot \nop \right )\bstar u \equiv \nop \frac{ {\bstar u}^2}{2} + \underbrace{\bstar \omega  \times \bstar u}_{=:\bfm{L}^{**}}.  
\end{equation*}
In the first and second identities, $\bstar L$ and $\bfm{L}^{**}$ indicate parts of the Lamb vector defined by $\bcal L := \bfm \omega \times \bfm u$ that follow from the decomposition introduced by Eqs.~(\ref{eq:split_u}) and (\ref{eq:split_omega}). Specifically, $\bcal L = \bfm L + \bstar L + \bfm{L}^{**}$, where furthermore $\bfm L := \bfm \Omega \times \bfm U$. It is easy to check that the sum of Eq.~(\ref{eq:ns_momentum_U}) and Eq.~(\ref{eq:ns_momentum_up}) reproduces Eq.~(\ref{eq:ns_momentum}). 

As indicated, an arbitrary unsteady gauge field $\bfm X(\bfm x,t)$ can be complementarily added and subtracted to both the velocity equations, without affecting the reconstruction of the momentum equation by means of summing up the split-equations. The additional gauge field can be used to impose an additional constraint on one of the split velocities. Specifically, we chose $\bfm X$ so that a projection on divergence-free velocity is enforced in Eq.~(\ref{eq:ns_momentum_U}). Hence, based on this specific definition, the split velocity therein satisfies
\begin{equation}\label{eq:div_free}
	\nop \cdot \bfm U = 0.
\end{equation} 
{\it Corollary:} A divergence-free velocity field can be achieved through a gauge field $\bfm X(\bfm x,t)$ defined in terms of a properly chosen real-valued scalar field $P(\bfm x,t)$, real-valued positive field $g(\bfm x,t)>0$, and given arbitrary real-valued vector field $\bfm h(\bfm x,t)$ (where $\bfm h$ is assumed to be not a function of $P$), if combined in the form    
\begin{equation}
\bfm X = \frac{\nop P}{g} + \bfm h.  
\end{equation}
{\it Proof:} With the such defined gauge function used together with the left-hand side (l.h.s.) part indicated by $\bfm l$ in Eq.~(\ref{eq:ns_momentum_U}), we can deduce a partial differential equation (PDE) for $P$ that will achieve the proper projection.       

To be precise, conducting the following two steps to both sides of Eq.~(\ref{eq:ns_momentum_U}), i.e., (i) subtraction of $\bfm l$ and (ii) taking the divergence, we arrive at  
\begin{equation}
	\pp{}{t} \left ( \nop \cdot \bfm U \right ) = -\nop \cdot \left (\frac{\nop P}{g} \right ) - \nop \cdot \left ( \bfm l + \bfm h \right ).
\end{equation}
If the velocity field satisfies Eq.~(\ref{eq:div_free}) at initial time level $t=0$, with the help of the previous equation this also holds for all later time level $t>0$, if $P$ satisfies the PDE 
$\nop \cdot \left ({\nop P}/{g} \right ) = - \nop \cdot \left ( \bfm l + \bfm h \right )$ for given $g$, $\bfm l$, and $\bfm h$. The PDE specifies $P$ up to an arbitrary constant. For the special choice $g\equiv \rho_0$, with constant ambient density $\rho_0$, the PDE reduces to a Poisson equation. Here, we make the choice $g \equiv \rho$. For both choices, $P$ has the unit of pressure. For the latter, the PDE corresponds to the PDE solved for incompressible pressure in the framework of the variable density incompressible Navier-Stokes equations,~\cite{guermond2001}. 

$P$ can be calibrated by proper far-field boundary conditions to yield the undisturbed far-field pressure. This newly introduced pressure $P$ together with compressible pressure $p$ implies a further pressure decomposition,  
\begin{equation}\label{eq:p_decomp}
	p = P+\sstar p,
\end{equation}
which introduces the residual pressure $\sstar p$. 

The fields $\bfm h$ is still free for definition, besides the constraint that it does not depend on $P$. Note, its choice will not affect the projection property Eq.~(\ref{eq:div_free}). However, the resulting field $P$ eventually will depend on the explicit choice made for the additional term. Consequently, the residual pressure function as defined by  Eq.~(\ref{eq:p_decomp}) would be modified as well. 

Our specific choice is 
\begin{equation}\label{eq:hchoice}
	\bfm h := \frac{\sstar  p\nop \rho}{\rho^2} . 
\end{equation}

Based on this extra term added, we finally arrive at two coupled equation systems. The first system that yields the compressible corrections based on variables $(\bstar u, \sstar p)$  reads    
\begin{align}
\label{eq:APE-u}
& \pp{\bstar u}{t}  +
  \nop \left ( \bfm U \cdot \bstar u  + \frac{{\bstar u}^2}{2} \right ) + \underbrace{\nop \left ( \frac{\sstar p}{\rho} \right )}_{(I)} =  \bfm 0   \\
\label{eq:APE-mod_p}
&\dd{\sstar p}{t} + \gamma p \nop \cdot \bstar u   = -\dd{P}{t}  + \theta, 
\end{align}
where ${D}/Dt :=  \partial/\partial{t} + \bfm u \cdot \nop $ denotes the substantial time derivative operator based on $\bfm u$.

The second system for variables $(\rho, \bfm U)$ reads  

\begin{align}
\label{eq:inc_rho}
& \ddb{\rho}{t}  + \underbrace{\nop \cdot \left ( \rho \bstar u \right )}_{(II)}  = 0\\
&\ddb{\bfm U}{t} + \frac{\nop P}{\rho}- \frac{\nop \cdot \bfm{\underline{\tau}}}{\rho} +
 \underbrace{\bfm \Omega \times \bstar u}_{(III)} + \underbrace{\frac{\sstar p\nop \rho}{\rho^2}}_{(IV)} = \bfm 0 \label{eq:inc_U}\\
 \label{eq:inc_divU}
 &\nop \cdot \bfm U = 0.
\end{align}
where ${\bar D}/Dt :=  \partial/\partial{t} + \bfm U \cdot \nop $ denotes the substantial time derivative operator based on $\bfm U$.

In agreement with other hydrodynamic/acoustic splitting methods, the substantial time derivative of the incompressible pressure $P$ appears as the major vortex-sound source term in Eq.~(\ref{eq:APE-mod_p}). The substantial time derivative results from (i) absorbing the initial gradient of incompressible pressure term on the r.h.s. of the momentum equation into the l.h.s. pressure term that changes this term into an expression of the compressible pressure correction $\sstar p$, and (ii) by substituting the pressure in the pressure equation~(\ref{eq:ns_energy}) through the decomposition Eq.~(\ref{eq:p_decomp}) and shifting all resulting incompressible Pressure terms to the r.h.s..

As an additional source term, the dissipation function $\theta$ appears on the r.h.s. of Eq.~(\ref{eq:APE-mod_p})  

Because of the specific choice Eq.~(\ref{eq:hchoice}), the pressure in combination with variable density eventually shows up in Eq.~(\ref{eq:APE-u}) in the form of a curl-free gradient term~$(I)$. To this specific notation of the pressure term we will hereafter refer to as gradient form. As a result, taking the curl of this equation we arrive at $\partial \bstar \omega /\partial t = \bfm 0$. Without further restriction, assuming an initial solution $\bstar \omega (t=0) = 0$, this implies the velocity component $\bstar u$ to be curl-free, i.e. $\bstar \omega  = \nop \times \bstar u =0$ for $t>0$. This result has already been incorporated in Eq.~(\ref{eq:inc_U}) dropping the otherwise present term $\bstar \omega  \times \bfm u$.   

Therefore, the presented split-equation system is based on a Helmholtz decomposition, i.e. the split velocities defined by Eq.~(\ref{eq:split_u}) satisfy 
\begin{equation}\label{eq:helmholtz-1}
	\nop \cdot \bfm U=0, \quad \nop \times \bstar u = 0,
\end{equation}
which allows to define the velocity in terms of scalar and vector potentials,  
\begin{equation}\label{eq:helmholtz-2}
	\bstar u = \nop \varphi, \quad \bfm U = \nop \times \bfm A.
\end{equation}
The previous result would also have been obtained by directly starting with a Helmholtz decomposition. However, 
the previous discussion includes the more general derivation of all other possible split equation variants.

A brief discussion of the specific features of a Helmholtz based derived system and other variants will be given in the subsequent sections.

\subsection{Discussion of Split Equations}\label{ssec:discussion1}

\subsubsection{Acoustic System}

Equation~(\ref{eq:APE-u}) with gradient form pressure term represents the characteristic velocity equation of the Acoustic Perturbation Equation (APE) system~\cite{ewert2003} as being used in Computational Aeroacoustics (CAA) for the simulation of acoustic mode propagation over non-uniform base flow. From Eq.~(\ref{eq:APE-u}) $(\bfm U + \bstar u/2)$ can be identified as the effectively used unsteady base-flow. 

The system Eqs.~(\ref{eq:APE-u},\ref{eq:APE-mod_p}) describes wave-propagation over the in general unsteady base flow. To demonstrate this essential feature, we can recast the system into an equivalent generalized convective wave-equation for acoustic potential $\varphi$ by taking into account that $\bstar u$ is curl-free, so that it can be expressed by  
\begin{equation}\label{eq:acoustic_phi}
\bstar u=:\nop \varphi. 
\end{equation}
Introduction of the acoustic potential into Eq.~(\ref{eq:APE-u}), effectively this equation can be reformulated to express pressure $\sstar p$ in terms of the potential, viz.
\begin{equation}\label{eq:p_phi}
	\sstar p = -\rho \ddt{\varphi}{t}, 
\end{equation}  
where the substantial time derivative operator ${\widetilde{D}}/Dt=\partial/\partial t + \widetilde{ \bfm u} \cdot \nop$ based on flow convection velocity $\widetilde{\bfm u}:=\bfm U + \bstar u/2$ is introduced. Upon substituting the pressure $\sstar p$ in Eq.~(\ref{eq:APE-mod_p}) using Eq.~(\ref{eq:p_phi}) and velocity $\bstar u$ by means of Eq.~(\ref{eq:acoustic_phi}), we arrive at an inhomogeneous generalized convective wave equation for the acoustic potential and wave speed $c^2:=\gamma p/\rho$,  
\begin{equation}\label{eq:gen_ac_wave_eqn}
	\dd{}{t} \left ( \rho \ddt{\varphi}{t} \right ) - \rho c^2 \Delta \varphi =  \dd{P}{t} - \theta .
\end{equation}
The sources are still be given by the substantial time derivative of incompressible pressure $P$ and an entropy source based on dissipation function $\theta$. 

 Generalized convective wave equations could come in many forms~\cite{campos2007}. The notion 'generalized convective wave equation' is used here to indicate any partial differential equation of one scalar variable, which reduce in the weakly coupled limit and for negligible base-flow gradients into the well-known convective wave equation\footnote{However, if we would rewrite the first order equation system of variables $(\sstar p,\bstar u)$ into one equation which involves terms like '$\partial^2 \sstar p/\partial t^2$', '$\Delta \sstar p$', but that still includes linear terms of $\bstar u$ and derivatives of it, this would not meet the present definition of a generalized convective wave equation.}.

Eq.~(\ref{eq:gen_ac_wave_eqn}) represents a strongly coupled generalized convective wave-equation that includes non-linear acoustics  by incorporating the acoustic velocity into the base flow. In the weakly coupled limit, i.e. neglecting the non-linear terms ($\bfm U + {\bstar{u}}/2 \to \bfm U$, $c^2\to \gamma P/\rho$) (and neglecting any sources to keep the  discussion simple), it reduces to the homogeneous equation
\begin{equation}
	\ddb{}{t} \left ( \rho \ddb{\varphi}{t} \right ) - \rho c^2 \Delta \varphi =  0 .
\end{equation}     
Since base flow velocity and density can be unsteady, sound propagation through the unsteady flow field is captured by the wave operator. In the limit of negligible base-flow gradients, it reduces further to the convective wave equation 
\begin{equation}
	\frac{1}{c^2}\frac{\bar D^2\varphi}{Dt^2 } - \Delta \varphi =  0 ,
\end{equation} 
which in the limit of vanishing base-flow $\bfm U \to \bfm 0$, $\bar D/Dt \to \partial/\partial t$, eventually becomes the simple wave equation.

\subsubsection{Base Flow Equations}

The equation system Eqs.~(\ref{eq:inc_rho}-\ref{eq:inc_divU}) are the variable density incompressible Navier-Stokes equations~\cite{guermond2001},

\begin{align}
	\label{eq:inc_rho_orig}
	& \ddb{\rho}{t}  = 0\\
	\label{eq:inc_U_orig}
	&\ddb{\bfm U}{t} + \frac{\nop P}{\rho}- \frac{\nop \cdot \ten{\tau}_{inc}}{\rho} = \bfm 0\\
 \label{eq:inc_divU_orig}
 &\nop \cdot \bfm U = 0,
\end{align}

supplemented by additional flow-acoustic coupling terms $(II)-(IV)$. Furthermore, the divergence of the incompressible shear-stress tensor 
\begin{equation}\label{eq:div_tau_inc}
	\nop \cdot \ten{\tau}_{inc} := \mu \Delta \bfm U
\end{equation}
in Eq.~(\ref{eq:inc_U_orig}) is replaced in Eq.~(\ref{eq:inc_U}) by the compressible expression, e.g. well approximated by Eq.~(\ref{eq:div_tau}). 

The variable density incompressible Navier-Stokes equations describe the dynamics of convective entropy (via Eq.~(\ref{eq:inc_rho_orig})) and vorticity modes (via Eq.~(\ref{eq:inc_U_orig})).

\subsubsection{Alternative Split Systems}\label{ssec:discussion2}

The discussion of the derivation of the split equation system highlights that the final result is not unique; other split systems are possible than the explicitly shown one, that all will lead to an incompressible-like equation system with solenoidal velocity $\nop\cdot\bfm U=0$ enforced by pressure $P$. 

For example, by making the explicit choice $\bfm h = 0$ in Eq.~(\ref{eq:hchoice}) we arrive at Eq.~(\ref{eq:inc_U}) with term $(IV)$ removed. In return, the term '$-(IV)$' will appear in the acoustics related equation~(\ref{eq:APE-u}). Hence, free choice of the vector field $\bfm h$ means that we can freely add an subtract complementary terms to the left-hand sides of Eqs.~(\ref{eq:inc_U},\ref{eq:APE-u}). For example, the incompressible equations could be modified to provide a closer agreement with the variable density equations by replacing the compressible viscous shear stresses with the incompressible ones. To be precise, if we define 
\begin{equation}\label{eq:tau_inc-tau}
	\bstar{\ten{\tau}} := \ten{\tau}-\ten{\tau}_{inc},
\end{equation}   
and add term $\nop \cdot \bstar{\ten{\tau}}/\rho$ to Eq.~(\ref{eq:inc_U}) we reproduce the incompressible viscous stresses of the original variable density equation. Combined with adding '$-(IV)$' we arrive at a form of Eq.~(\ref{eq:inc_U}) even closer to Eq.~(\ref{eq:inc_U_orig}), only differing in terms of the additional coupling term $(III)$,
\begin{equation}\label{eq:U_alternative}
	\ddb{\bfm U}{t} + \frac{\nop P}{\rho}- \frac{\nop \cdot \ten{\tau}_{inc}}{\rho} +
 	\bfm \Omega \times \bstar u + \underbrace{\bstar \omega \times \bfm u}_{(V)} = \bfm 0
\end{equation}
In return, by adding '$(IV)-\nop \cdot \bstar{\ten{\tau}}/\rho$' to the acoustics-related velocity equation, it becomes
\begin{equation}\label{eq:APE-u-alt}
	\pp{\bstar u}{t}  +
  \nop \left ( \bfm U \cdot \bstar u  + \frac{{\bstar u}^2}{2} \right ) + \frac{\nop \sstar p}{\rho}  - \frac{\nop \cdot \bstar{\ten{\tau}}}{\rho} =  \bfm 0,
\end{equation}
where with the help of Eqs.~(\ref{eq:div_tau},\ref{eq:div_tau_inc},\ref{eq:tau_inc-tau}), the viscous term explicitly reads
\begin{equation}
	\nop \cdot \bstar{\ten{\tau}} \simeq  \mu \Delta \bstar u + \frac{1}{3} \mu \nop \left ( \nop \cdot \bstar u \right ).
\end{equation} 
Taking the curl of Eq.~(\ref{eq:APE-u-alt}), we arrive at $\bstar \omega= \nop \times \bstar u\ne 0$, because of the lacking gradient form of the pressure term and presence of the viscous term. Hence, for this set of split equations the velocity decomposition Eq.~(\ref{eq:split_u}) is not any longer based on a Helmholtz decomposition. Therefore, Eq.~(\ref{eq:U_alternative}) also contains the additional coupling term $(V)$.

But the convective term of gradient form excludes growing flow instabilities and thus suppresses turbulence production as part of the acoustic related equations. The compressible velocity based shear stress term effectively includes acoustic boundary layers into the acoustic equation. Altogether, a split equation system with Eq.~(\ref{eq:APE-u-alt}) is equally valuable to a Helmholtz decomposition based approach. The essential key to a valuable split system is that the dominant non-acoustic and acoustic modes are clearly separated\footnote{This resembles the flux vector splitting applied in the solution of the compressible Navier-Stokes equations. There, an approximative split-up of the fluxes into acoustic and non-acoustic contributions is accomplished to tackle each component differently according to their characteristic propagation velocities. Alternative approximative flux formulations are possible as long as the main propagation characteristics are captured. In the split equation approach as presented here, accordingly alternative split formulations are possible as long as each system without coupling terms captures the proper main propagation characteristics.}.

\subsection{Preliminary Conclusions}

To conclude, in form of Eqs.~(\ref{eq:APE-u},\ref{eq:APE-mod_p}) and (\ref{eq:inc_rho}-\ref{eq:inc_divU}) a splitting of the compressible Navier-Stokes equations into two separate systems has been accomplished, each of which is distinguished by acoustic and convective propagation speeds and which are coupled by mutual terms. The coupled system is not limited to very small Mach numbers in use, because it is an exact representation of the compressible Navier-Stokes equations. In addition to the usual formulations, nonlinear effects in the transport terms are taken into account. These deliver e.g. in the APE for higher Mach numbers, non-vanishing time-averaged quantities as a compressible correction to the incompressible flow.

The term $(II)$ in Eq.~(\ref{eq:inc_rho}) is the essential coupling term that describes density variations caused by the volume dilatation of the compressible velocity field. Since density fluctuations are related to acoustic and entropy fluctuations, the term ${\sstar p\nop \rho}/{\rho^2}$ describes an acoustic-entropy coupling or a non-linear acoustic source term. The term $\bfm \Omega \times \bstar u$ can be identified as the essential acoustic-flow coupling term for vortex-dominated flows.


\section{Vortex-Sound in Subsonic Cold Flow}\label{sec:low_subsonic_vortex_sound}

\subsection{Simplified Split System}\label{ssec:spimplified_split_system}

\subsubsection{Mach Number Scaling}

For vortex-sound problems in cold flow of constant ambient temperature $T_0$, pressure $p_0$ and density $\rho_0$, the split equation system can be simplified further. Firstly, the entropy source can be neglected, $\theta \approx 0$. 

Secondly, we simplify the governing equations for the subsonic flow regime by estimating the scaling with powers of Mach number of all split equations terms in the hydrodynamic near field where flow and acoustics are coupled and by appropriately truncating all terms of order $M^n$ for $n\ge N$.       

Let the Mach number defined by appropriate reference velocity $u_\infty=\left | \bfm u_\infty \right |$, $M=u_\infty/c_0$ with the ambient speed of sound defined by $c_0=\sqrt{\gamma p_0/\rho_0}$. For very small Mach numbers $M \to 0$, the flow is entirely represented by the incompressible solution, hence, $\bfm u \to \bfm U \sim  M$ for $M \to 0$. Inevitably, the compressible correction velocity $\bstar{u}$ must scale with a higher Mach number power of at least leading order two, i.e. as a conservative estimate $\bstar u \sim \alpha M^2$. Note, this scaling in general differs from the scaling of the acoustic quantities in the far-field. In the near-field where vortex sound sources are non-zero, the leading order of the compressible correction quantities must comply with those imposed by the sound sources. The factor $\alpha$ is introduced for a later calibration of the magnitude of $\bfm u^*$-terms relative to $\bfm U$-terms.     

The spatial derivatives of incompressible flow quantities scale with $\partial / \partial x_i \equiv \partial_i \sim 1/\delta$, spatial derivatives of compressible quantities generated by incompressible fluctuations $\sim M/\delta$, where $\delta$ indicates a characteristic length scale of the incompressible flow. In consequence, $\partial/\partial t \equiv \partial_t \sim M/\delta$ for time derivatives of both, compressible correction and incompressible flow. 

The incompressible pressure $P$ is defined up to an arbitrary constant by the Poisson equation that results from taking the divergence of Eq.~(\ref{eq:inc_U}). This yields $P\sim M^2$ for the non-constant part of incompressible pressure. Furthermore, the compressible correction can be estimated to scale with $p^*\sim \rho_0 c_0 \left|\bfm u^*\right| \sim \alpha M^2$. Spatial derivatives of pressure $p= P + p^*$ scale with $\partial_i p = \partial_i P + \partial_i p^* \sim M^2/\delta + \alpha M^3/\delta$, i.e. to leading order $\sim M^2/\delta$. Furthermore, $\partial_t p\sim (1+ \alpha) M^3/\delta$. For the decomposition of pressure into the constant ambient part and a rest\footnote{The alternative decompositions $p=p_0+p'$ and $p=P+p^*$ are at the root of some misconceptions persistent in the HAS literature.}, 
\begin{equation}
	p = p_0  + p',
\end{equation} 
it follows that $p'\sim (1+\alpha) M^2$, $\partial_i p' \sim M^2/\delta + \alpha M^3/\delta$, and $\partial_t p' = (1+ \alpha) M^3/\delta$. A similar split up is considered for the density, $\rho = \rho_0 + \rho'$. From $\rho' \approx p'/c_0^2$, it follows $\rho' \sim (1 + \alpha) M^2$, $\partial_i \rho' \sim M^2/\delta + \alpha M^3/\delta$, and $\partial_t \rho' = (1 + \alpha) M^3/\delta$. Based on the Reynolds number of length scale $\delta$, $Re_\delta:=u_\infty \delta/\nu$, the dynamic viscosity can be estimated to scale with $\mu = \rho u_\infty \delta/Re_\delta \sim M / Re_\delta \cdot \delta$.  

Using the previous estimates, the scaling of all terms of the density equation~(\ref{eq:inc_rho}) based on all non-dimensional factors up to Mach number powers $n=5$ reads   
\begin{equation}\label{eq:rho-simple}
	\underbrace{\partial_t \rho'}_{(a)\sim (1+\alpha) M^3} 
	+ \underbrace{U_i \partial_i \rho'}_{(b)\sim M^3+\alpha M^4} 
	+ \underbrace{\rho_0  \partial_i u^*_i}_{(c)\sim \alpha M^3} 
	+ \underbrace{u_i^* \partial_i \rho'}_{(d)\sim M^4 + \alpha M^5} 
	+ \underbrace{\rho'  \partial_i u^*_i}_{(e)\sim (1+\alpha) \alpha M^5 } 
	+  \cdots = 0.
\end{equation}

To analyze the residual pressure equation~(\ref{eq:APE-mod_p}), first we shift the r.h.s. incompressible pressure source term to the l.h.s. and combine it with the residual pressure terms into terms containing only pressure variable $p=P+p^*$. Next, since only gradients of pressure occur, we can replace $p$ by $p'$. The term-wise leading order scaling of the resulting pressure equation reads 
\begin{equation}\label{eq:p-simple}
	\underbrace{\partial_t p'}_{(f)\sim (1+\alpha) M^3}   
	+ \underbrace{U_i \partial_i p'}_{(g)\sim M^3 + \alpha M^4}
	+ \underbrace{\gamma p_0  \partial_i u^*_i}_{(h)\sim \alpha M^3} 
	+ \underbrace{u^*_i \partial_i p'}_{(i)\sim M^4 + \alpha M^5} 
	+ \underbrace{\gamma p' \partial_i u^*_i}_{(j)\sim (1+\alpha) \alpha M^5 } +  \cdots = 0.
\end{equation}

By comparing Eq.~(\ref{eq:rho-simple}) with Eq.~(\ref{eq:p-simple}), it is evident that for $N=4$ (all terms up to leading Mach powers $M^3$ kept) both simplified equations imply the relationship $\rho' = p'/c_0^2$ to strictly hold, i.e., the density equation can be skipped in favor of this isentropic relation. Note, this result does not contradict the initial estimate made for the magnitude of $\rho'$.   

The term-wise study of the incompressible base flow equation~(\ref{eq:inc_U}) yields 
\begin{align}
    \nonumber
	& \underbrace{\partial_t U_i}_{(k)\sim M^2} 
	+ \underbrace{U_j \partial_j U_i}_{(l)\sim M^2 } 
	+ \underbrace{\qfrac{\nop P }{\rho_0}}_{(m)\sim M^2 } 
	- \underbrace{\qfrac{\nop \cdot \ten{\tau}_{inc}}{\rho_0}}_{(n)\sim  M^2 \rec{Re_\delta} } 
	+ \underbrace{\bfm \Omega \times \bfm u^*}_{(o)\sim \alpha M^3 } 
	\\[1.5ex] \label{eq:U-simple} & \quad 
	- \underbrace{\qfrac{\nop P \rho'}{\rho_0^2}}_{(p)\sim (1+\alpha) M^4 } 
	+ \underbrace{\qfrac{\nop \cdot \ten{\tau}_{inc}\rho'}{\rho_0^2}}_{(q)\sim  (1+\alpha) M^4 \rec{Re_\delta} } 	
    + \underbrace{\qfrac{p^*\nop \rho'}{\rho_0^2}}_{(r)\sim M^4 + \alpha M^5}
	- \underbrace{\qfrac{\nop \cdot \ten{\tau}^*}{\rho_0}}_{(s)\sim  \alpha M^5 \rec{Re_\delta} } 
	+ \cdots = 0. 
\end{align} 

Finally, for the resulting compressible velocity correction equation~(\ref{eq:APE-u}) the term-wise scaling analysis in terms of leading order powers gives
 
\begin{equation}\label{eq:u-star-scaling}
	  \underbrace{\partial_t u^*_i}_{(t)\sim \alpha M^3 }
	+ \underbrace{\partial_i \left (  U_j u^*_j \right )}_{(u)\sim \alpha M^3}
	+ \underbrace{\qfrac{\partial_i p^*}{\rho_0}}_{(v)\sim \alpha M^3}
    - \underbrace{\qfrac{\rho'\partial_i p'}{\rho_0^2}}_{(w)\sim (1+\alpha) M^4}	
    - \underbrace{\qfrac{p^*\partial_i \rho'}{\rho_0^2}}_{(x)\sim \alpha M^4 + \alpha^2 M^5}
    + \cdots
	= 
	0.
\end{equation}

\subsubsection{N=3 Subsonic System}

From the term-wise scaling analysis it follows for truncation level $N=3$ that the resulting simplified equations system reduces to the incompressible Navier-Stokes equations, essentially given by the leading-order terms $(k)-(n)$. That is, for small Mach numbers $M<\!\!<$ the compressible Navier-Stokes equations naturally transform to the expected form. For very small Reynolds number $Re<\!\!<$, term $(n)$ will be larger than terms $(k)-(m)$. In this limit, the reduced equation system based on term (n) governs Hele-Shaw flow.

\subsubsection{N=4 Subsonic System}

We consider a coupled simplified equation system up to oder $N=4$ given by terms $(t)-(v)$ of Eq.~(\ref{eq:u-star-scaling}), $(f)-(h)$ of Eq.~(\ref{eq:p-simple}), and $(k)-(o)$ of Eq.~(\ref{eq:U-simple}). Replacing $\sstar p$ in Eq.~(\ref{eq:u-star-scaling}) by $p'-P'$, where $P':=P-p_0$, this yields the coupled subsonic version rewritten together with divergence-free constraint imposed on $U$ in the form 
\begin{align}
  & \pp{\bstar u}{t}  + \nop \left ( \bfm U \cdot \bstar u \right ) + \frac{\nop p'}{\rho_0} = \frac{\nop P'}{\rho_0} \label{eq:ape_momentum}\\
  &\pp{p'}{t} + \bfm U  \cdot \nop p' + \rho_0 c_0^2 \nop \cdot \bstar u = 0  \label{eq:ape_pressure}\\[1.5ex]
  & \pp{\bfm U}{t} + \left ( \bfm U \cdot \nop \right ) \bfm U +\frac{\nop P'}{\rho_0} - \nu_0 \Delta \bfm U 
	 = - \bfm \Omega \times \bstar u \label{eq:incomp_momentum} \\
  & \nop \cdot \bfm U = 0.  \label{eq:incomp_mass} 
\end{align} 
Note, for the ambient density deviations the additional relationship $\rho'=p'/c_0^2$ is in place. Furthermore, $\nu_0 := \mu(T_0)/\rho_0$ based on ambient temperature $T_0$.    

The compressible correction equations (\ref{eq:ape_momentum},\ref{eq:ape_pressure}) correspond to the APE-1 system with incompressible pressure based source term introduced by Ewert \& Schr\"oder~\cite{ewert2003}. Altogether, the simplified system mainly differs from the well validated standard open-loop HAS approach of Kreuzinger \& Schwertfirm~\cite{schwertfirm2012,kreuzinger2013,schwertfirm2014,kreuzinger2016} by the additional acoustic-flow feedback term, shifted here for clarity as an additional source term $-\bfm \Omega \times \bstar u$ to the r.h.s.. Usage of $P'$ implies application of the far-field condition $P'=0$ in the Poisson solver.  

The larger $N$ selected, the higher the Mach number limit up to which the simplified system can be anticipated to be valid. For $N=3$ the incompressible Navier-Stokes equations follow which have an application range up to Mach number $M_{N=3} \approx 0.2 \ldots 0.3$. Due to the higher order of the error term, for $N=4$ the valid range of applicability might be estimated to extend to $M_{N=4} \approx M_{N=3}^{3/4}$, which implies $M_{N=4}\approx 0.3 \ldots 0.4$.  

For truncation levels larger than $N=3$, also variables describing a compressible correction to the incompressible flow variables have to be considered. These terms involve besides the scaling in terms of powers of Mach number also the relative magnitude between compressible correction velocity $\bstar u$ and flow velocity $\bfm U$ as defined by parameter $\alpha$. Values of $\alpha$ larger than unity could imply that higher Mach powers of the terms of compressible correction variables have to be recognized. Large values, e.g., could occur for strong flow-acoustic resonance cases where the magnitude of acoustic particle velocity is limited in the resonant cases only by the amount of damping. For the sake of completeness, therefore the scaling of each term was derived also for the leading order and the next higher $\alpha$ depending term. However, the simulation cases discussed in Section~\ref{sec:validation} reveal that even in flow-acoustic cases with resonances, practically a magnitude of ${O}(1)$ is observed for $\alpha$. Therefore, the same Mach number truncation powers are considered for flow and acoustics.       

It is interesting to note that all alternative formulations of the exact split equations discussed in the previous sections culminate into the same simplified equation set for $N=4$. The simplified set is unique because the Mach-number scaling of the different terms remains invariant if specific terms are shifted between the flow and acoustic equations, refer to the discussion in the previous sections. Furthermore, the additional coupling term $\bstar \omega \times \bfm U$ that would have to be considered for a flow decomposition different from a Helmholtz decomposition, is of order $M^4$, thus would be cancelled for $N=4$. Therefore, the cancellation of terms yields the same simplified set of equations if the same truncation level is applied to both equations sets, i.e., as being used throughout the precedent analysis. 

The equation system (\ref{eq:ape_momentum}-\ref{eq:incomp_mass}) is the main result of this paper. It is used as the governing equations applied for the simulations presented in the validation section.

\subsection{Split Equation Analysis}\label{ssec:seq_analysis}

To preserve the dependence of the incompressible flow time step size on convective flow velocity it is crucial that the coupling terms do not alter the characteristic propagation velocities of the split equation systems so that those of the flow equations become unfavorable dependent on the speed of sound. By recasting the uncoupled APE into pseudo-conservative form, the eigenvalues of the x-derivatives related Jacobian matrix, for example, imply propagation speeds $\lambda_{1,2}=U\pm c_0$  and $\lambda_{3,4}=0$, i.e. two eigenmodes related to acoustic propagation in the flow and two degenerated eigenvalues due to the removal of non-acoustic modes from the APE-system~\cite{ewert2003}. The corresponding system matrix of the flow equations provides eigenvalues defined by flow velocity $U$. Hence, the explicit numerical time step size of the acoustic system is limited by a CFL-number based on a propagation speed of the order of the speed of sound, while that of the flow equations is limited by the maximal flow velocity\footnote{Which is exploited for the simulations in this work by using different time-step sizes for flow and acoustics, where a certain number of acoustic time steps is applied per flow time step.}.      

To prove the conservation of this feature for the coupled system, the simplified coupled equation system of the previous paragraph is recast into primitive form given by  
\begin{equation}\label{eq:Q-system}
  \partial_t \bfm Q + \ten{A} \, \partial_x  \bfm Q + \ten{B}\, \partial_y \bfm Q + \ten{C}\, \partial_z \bfm Q + \ten{H}(\bfm Q)  = \bfm 0.
\end{equation} 
The vector of variables is defined by    
\begin{equation}
   \bfm{Q} =  \left [
   \begin{array}{c}
   \bfm{Q}_1  \\
   \bfm{Q}_2 \\
   \end{array}
   \right ]	, \quad 
   \bfm{Q}_1 =  \left [
   \begin{array}{c}
   \bstar u \\
   p'
   \end{array}
   \right ]	, \quad 
   \bfm{Q}_2 =  \left [
   \begin{array}{c}
   \bfm U \\
   P'  
   \end{array}
   \right ]	.
\end{equation}
Hence, $\bfm Q_1$ is a sub-vector of $\bfm Q$ with all variables of the acoustic system. The sub-vector $\bfm Q_2$ comprises all variables of the flow equations. 
The term $\bfm H(\bfm Q)$ describes the viscous terms for the flow equations with zero components for the acoustic equations. The system matrices are given in the form of block matrices,  
\begin{equation}\label{eq:G2}
\ten{A} =  \left [
   \begin{array}{cc}
   \ten{A}_1 & \ten{K}_1 \\
   \ten{0}  &  \ten{A}_2\\
   \end{array}
   \right ], \;
   \ten{B} =  \left [
   \begin{array}{cc}
   \ten{B}_1 & \ten{L}_1 \\
   \ten{0}  &  \ten{B}_2\\
   \end{array}
   \right ], \;
   \ten{C} =  \left [
   \begin{array}{cc}
   \ten{C}_1 & \ten{M}_1 \\
   \ten{0}  & \ten{C}_2\\
   \end{array}
   \right ].
\end{equation}
Specifically, $\ten{A}_1$, $\ten{B}_1$, and $\ten{C}_1$ are the flux Jacobian related to the acoustic variables for directions $x$, $y$, and $z$, respectively. The matrices $\ten{K}_1$,  $\ten{L}_1$, and $\ten{M}_1$ comprise all terms of the acoustic equations that involve base-flow gradients for the respective directions. The system matrices with superscript $2$ are related to the flow equations. Because of the specific form of the Lamb vector based coupling term, no coupling matrices involving gradients of the acoustic variables are present for the flow equations, $\ten{K}_2 = \bfm 0$. Next, the system matrices of the simplified equation system derived in the previous paragraph are exemplarily discussed for the  $x$-direction. The matrices for the other direction follow accordingly. The $x$-direction related matrices read 
\begin{equation}
   \ten{A}_1 =  \left [
   \begin{array}{cccc}
   U & V & W & \frac{1}{\rho_0} \\
   0 & 0 & 0 & 0\\
   0 & 0 & 0 & 0\\
   \rho_0 c_0^2 & 0 & 0 & U
   \end{array}
   \right ],\quad
   \ten{K}_1 =  \left [
   \begin{array}{cccc}
   u^* & v^* & w^* & -\frac{1}{\rho_0}  \\
   0 & 0 & 0 & 0  \\
   0 & 0 & 0 & 0 \\
   p' & 0 & 0 & 0
   \end{array}
   \right ],\;
\end{equation}
and  
\begin{equation}\label{eq:A2}
   \ten{A}_2 =  \left [
   \begin{array}{cccc}
   U & -\sstar v & -\sstar w & \frac{1}{\rho_0}\\
   0 & U+\sstar u & 0 & 0\\
   0 & 0 &  U+ \sstar u & 0\\
   \rho_0 c_a^2 & 0 &  0 & U
   \end{array}
   \right ].
\end{equation}   
In the system matrix of the flow the components of the Lamb vector coupling terms are showing up as acoustic velocities in $\ten{A}_2$. The pressure related row follows by assuming a weakly compressible formulation based on an artificial propagation speed $c_a>\!\!>c_0$. Hence, the elliptical Poisson problem for incompressible pressure is replaced by the weakly compressible pressure equation $\partial_t P' + \bfm U \cdot \nop P' + \rho_0 c_a^2 \nop \cdot \bfm U=0$. In the limit $c_a \to \infty$ this provides the elliptic solution of the Poisson equation for pressure.  

The specific treatment chosen here for the pressure equation simplifies the further analysis that aims to prove that the eigenvalues of the flow problem are not modified, i.e.  become dependent on the speed of sound $c_0$ due to the two-way coupling with the acoustic problem. This analysis introduces an additional even stiffer eigenvalue, which however does not pose any limitation on the task to be proven. Note, in numerical applications the extra stiffness problem is avoided by using a dedicated efficient Poisson problem to compute pressure, see discussion below.     

The eigenvalues related to system matrix $\ten{A}_2$ derive from the solution of the characteristic polynomial of fourth degree, denoted by $P^4_2(\lambda)=0$, which follows from the determinant of the singular eigenvalue problem, $\mathrm{det} \left ( \ten{A}_2-\lambda \ten{I}_4 \right ) = 0$, where $\ten{I}_4$ is the $4\times 4$ identity matrix. 
  
The uncoupled system matrix of the flow (with acoustic variables $\bstar u$ set to zero) yields the eigenvalues $\lambda_{1,2}=U$ and $\lambda_{3,4}=U \pm c_a$. The first two eigenvalues are related to convective modes. The last two eigenvalues describe left- and right-running waves based on the artificial compressibility wave speed $c_a$. The case with the acoustic velocities included yields $\lambda_{1,2}=U+\sstar u$ and $\lambda_{2,3}=U \pm c_a$.

As part of a fractional step method, the elliptic Poisson problem for pressure is efficiently computed with a fast Poisson solver separately from the flow equations. This way, only the convective eigenvalues related to $\lambda_{1,2}$ are present as part of the first step to update the intermediate flow velocity at the next time level and therefore define the essential velocity scale that restricts the explicit time step size. 

The previous discussion shows that in case of flow resonances with $\alpha = \left | \bstar u \right | / \left | \bfm U \right | = \mathcal{O}(0.1)$, the convection of flow structured is modified by the acoustic particle velocity. However, the general order of magnitude of the time step size needed for the flow velocity update step is not modified by the coupling term.

Note, this discussion is restricted to the eigenvalues of the sub-matrix $\ten{A}_2$ alone. In addition, it has to be demonstrated that the eigenvalues of the complete system matrix $\ten{A}$ with characteristic polynomial $P^8(\lambda)$ do not alter the results separately found for each sub-system matrix. 

In case of $\ten{K}_1\equiv \bfm 0$ in $\ten{A}$, Eq.~(\ref{eq:G2}), the determinant of the complete eigenvalue system related to $\ten{A}$ decomposes into the product of the determinants of the submatrices $\ten{A}_1$ and $\ten{A}_2$, $\mathrm{det} \left ( \ten{A} - \ten{I}_8 \lambda \right)= \mathrm{det} \left ( \ten{A}_1 - \ten{I}_4\lambda \right)\mathrm{det} \left ( \ten{A}_2 - \ten{I}_4\lambda \right)=   0$. Hence the characteristic polynomial of the complete matrix is given by the product of the characteristic sub-polynomials, $P^8(\lambda) = P^4_1(\lambda) P^4_2(\lambda)=0$, which implies $P^4_1(\lambda)=0$ and $P^4_2(\lambda)=0$. Trivially though, the eigenvalues in case of the uncoupled equations are defined by the sub-matrices alone.            

For $\ten{K}_1\ne 0$, we can make use of the box theorem (a.k.a. K\"astchensatz), which states that 
\begin{equation}
   \mathrm{det} \left (
   \begin{array}{cc}
   \ten{P} & \ten{Q} \\
   \ten{0}  &  \ten{R}\\
   \end{array}
   \right ) = \mathrm{det} \left ( \ten{P} \right )  \mathrm{det} \left ( \ten{R} \right ), 
\end{equation}
where $\ten{P}$, $\ten{Q}$, and $\ten{R}$ are matrices of dimension $n \times n$, $n \times m$, and $m \times m$, respectively. The lower left $m\times n$ block matrix is filled with zeros. This result follows e.g. from the Leibniz formula for determinants.

Hence, it infers that the eigenvalues of the complete system matrix $\ten{A}$, Eq.~(\ref{eq:G2}), breaks down into the individual eigenvalue problems related to the main diagonal block matrices $\ten{A}_1$ and $\ten{A}_2$. Thus, the characteristic polynomial still is the product of those of the sub-matrices, $P^8(\lambda) = P^4_1(\lambda) P^4_2(\lambda)=0$. Since the system matrices of the other direction are of similar form, the result also holds for these eigenvalues.

To conclude, in what was to be proven, the present coupling does not alter the eigenvalues present for the decoupled sub-system matrices and as such preserves the order of the  explicit time step size present for the uncoupled systems.

\subsection{Solid Wall Boundary Conditions for Split Velocities}\label{sec:boundaryconditions}

The no-slip boundary condition at a solid wall $\left . \bfm u \right |_w=0$ translates into equivalent conditions for the wall normal and wall tangential split velocity components, respectively, 
\begin{equation}
	\left .  U_n \right |_w= - \left . \sstar{u}_n \right |_w \quad  \mbox{and} \quad \left . U_t \right |_w = - \left . \sstar{u}_t \right |_w. 
\end{equation}
The wall normal component can be modified further taking into account that the Helmholtz decomposition of a velocity field, Eq.~(\ref{eq:helmholtz-2}), is not unique since the gradient of a potential field, which itself is a solution to the Laplace equation $\Delta \sstar \varphi =0$, could be alternately added and subtracted to the split velocity components, so that an alternative set of split velocities results, $\bstar{\check{u}} = \bstar{u} + \nop \sstar \varphi$ and $ \bfm{\check{U}}= \bfm U - \nop \sstar \varphi$, which still complies with the decomposition Eq.~(\ref{eq:split_u}) and also satisfies Eq.~(\ref{eq:helmholtz-1}), i.e. yield a valid Helmholtz decomposition.  Eventually, the Helmholtz decomposition becomes unique if combined with a specific gauge applied to $\sstar \varphi$. For example, utilizing a free space Green's function to compute the velocities related to the vector and tensor potentials of Eq.~(\ref{eq:helmholtz-2})--as e.g. applied in Bio-Savart's induction formula--provides an implicit fix of the gauge. In the present context we can specify the potential field by imposing the solid wall boundary condition
\begin{equation}
	\pp{\sstar \varphi}{n} := \nop \sstar \varphi \cdot \bfm n = - \left .  U_n \right |_w .
\end{equation}              
This yields a non-penetration boundary condition for both alternative split velocities $\bstar{\check{u}}$ and $ \bfm{\check{U}}$.

Note, the alternative split velocities are still governed by the same split equations albeit with altered boundary conditions. Physically, the non-uniqueness problem implies that the hydrodynamic near-field could be associated either to the acoustic or the flow equations, respectively. The current gauge means that the hydrodynamic near-field is simulated as part of the flow equations. 

In what follows, the acoustic equation system is closed by imposing on the irrotational and inviscid compressible correction $\sstar{\check{u}}$ the slip boundary condition
\begin{equation}
\left . \sstar{\check{u}}_n \right |_w=0, \,\,\, \left . \frac{d \sstar{\check{u}}_t}{dn} \right |_w=0 .
\end{equation}

For $\bfm{\check{U}}$ the complete solid boundary condition follows by prescribing the tangential velocity from the acoustic solution, 
\begin{equation}
\check{U}_n=0, \,\,\,\check{U}_t=-\sstar{\check{u}}_t.
\end{equation}
This boundary condition couples the incompressible velocity ${\check{U}_t}$ to the compressible correction ${\check{u}^*_t}$.

A physical interpretation of this coupling can be given by considering a viscous acoustic boundary layer.
For a plane acoustic wave traveling along a wall in $x$-direction an analytical solution is straight forward to derive for a vanishing base-flow, e.g. refer to~\cite{rienstra_2017}. There the (un-split) linearized compressible momentum equation from boundary layer approximation is applied. Fourier transformation in time of perturbation variables yields 
\begin{equation}
  i \omega \hat u = - \frac{1}{\rho_0} \frac{d\hat p}{dx} + \nu \frac{d^2 \hat u}{dy^2},
\end{equation}
where the hat indicates temporal Fourier transforms with convention such that $\partial_t \,\laplace \,i\omega$ is applied.  
With the boundary conditions $\hat u=0$ for $y=0$,  $\hat u=\hat u_\infty$ at $y \rightarrow \infty$ and using  the fact, that the pressure is independent of $y$ in a thin boundary layer the solution is
\begin{equation}
  \label{eq:boundaryu}
  \hat u = \hat u_\infty \left[ 1 - \exp\left( -\frac{(1+i)y}{\sqrt{2\nu/\omega}}\right)\right],
\end{equation}
where $\sqrt{2\nu/\omega}$ is the thickness of the acoustic viscous boundary layer.\\

The same problem reformulated in terms of the subsonic split momentum equations~(\ref{eq:ape_momentum}) and (\ref{eq:incomp_momentum}) reads 
\begin{align}
  i \omega {\hat u}^* = - \frac{1}{\rho_0} \frac{d\hat{p}^\prime}{dx}\\
  i \omega \hat U =  \nu \frac{d^2 \hat U}{dy^2}
\end{align}
with boundary conditions $\hat{U}+\hat{{u}}^*=0$ for $y=0$, $\hat{U} + \hat{u}^*= \hat{u}_\infty$ at $y \rightarrow \infty$. Since the pressure is independent of $y$ it follows that 
\begin{align}
  \label{eq:boundaryustar}
  {\hat u}^* = \hat{u}_\infty 
\end{align}
also independent of $y$. Now the boundary condition for $\hat{U}$ at the wall is
\begin{align}
  \hat{U}(y=0) = - \hat{u}_\infty
\end{align}
and the solution of its momentum equation
\begin{equation}
  \label{eq:boundaryU}
  \hat{ U}  = - \hat{ u}_\infty  \mathrm{exp}\left( -\frac{(1+i)y}{\sqrt{2\nu/\omega}}\right)
\end{equation}
Combining the solutions for $\hat U$ and $\hat{u}^*$ results in the correct solution for the acoustic boundary layer Eq.~(\ref{eq:boundaryu}). 

To conclude, specifying a slip boundary condition for $u^*$ and a no-slip for $u=U+u^*$ results in a simulation result containing the acoustic boundary layers. In this case the numerical grid for $U$ has to be fine enough to resolve them. 

While $u^*$ represents the dilatational part of the acoustic field, its solenoidal part is provided by $U$. The example underlines that although acoustic fluctuations are inevitably connected to flow dilatation, they could be accompanied by vortical fluctuations that must be resolved as well for a complete description.  

The boundary condition can be regarded as a coupling mechanism, which enables the production of vorticity by acoustics in flow field initially at rest. On the contrary the coupling term $\bfm \Omega \times \bfm u^*$ is only different from zero if already vorticity exists.

If the no-slip  boundary condition is applied on $U$ alone, i.e.,
\begin{equation}
\left .  U_n \right |_w= \left . U_t \right |_w = 0,  
\end{equation}
acoustic boundary layers are neglected, which is common practice in acoustic computations. The grid for the solution of $U$ has now only to resolve the flow boundary layer.\footnote{Note, the turbulence driven near-wall mesh resolution of DNC in general is not sufficient to resolve the acoustic boundary, hence, the acoustic boundary layer in general is not resolved either.}


\section{Validation}\label{sec:validation}

The subsonic hydrodynamic/acoustic splitting method defined by the equation system
(\ref{eq:ape_momentum}-\ref{eq:incomp_mass}) is applied to two subsonic vortex sound problems featuring flow-acoustic feedback. A cross-comparison between different settings used for the simulations verifies the two-way coupled approach. The analysis involves to switch between one-way and two-way coupled simulations. From the comparison of these results the presence of the feedback mechanism and its significance for the resulting flow and acoustic fields can be determined. 
A grid convergence study for one of the problems verifies a grid independent solution. On the basis of the time step sizes used to achieve converged solutions for the considered problems, we analyze the impact of the coupled system on numerical stability and if the splitting method does not impose additional stability restrictions compared to an one way coupled hybrid approach, as was theoretically derived in Section~\ref{ssec:seq_analysis}. Furthermore, experimental and numerical results available for both problems from literature are used to validate the results. 

This section is organized as follows. Subsection~\ref{sec:numericalmethod} presents the numerical framework
used for the simulations. As a first problem, in subsection~\ref{sec:caseflute} a generic flute is simulated. The results for a Parker mode around a thick plate in a channel is studied in subsection~\ref{sec:caseparker}. Finally, the relative scaling of flow respectively
acoustic fluctuations in cases of strong resonances are discussed in subsection~\ref{sec:amplitudes}.

\subsection{Numerical Method}
\label{sec:numericalmethod}
The validated run-time coupled open-loop scale-resolving flow and acoustics solver MGLET~\cite{manhart2002} is applied for the simulations. For the flow part, the code solves the incompressible Navier-Stokes equations in the
finite volume formulation on a Cartesian hierarchical grid using a
staggered variable arrangement. The approximation of the derivatives
and the interpolation onto control volume boundaries are accomplished
by second order central schemes.  

A skew symmetric discretization of
the convective term is used to improve the energy conserving properties of the numerical scheme. For time integration, a
third order explicit Runge-Kutta method is
used~\cite{williamson1980}. The incompressibility constraint is
satisfied by solving a Poisson-equation for pressure with an
incomplete lower-upper decomposition and applying a correction step
for velocity and pressure. A geometric multi-grid method
on the hierarchical grid accelerates the convergence of the
Poisson problem.  MGELT is fully parallelized and
known as a reliable tool for accurate and fast direct numerical
simulations (DNS)~\cite{manhart2002,el_khoury2010,el_khoury2010b,schwertfirm2007}
and for large-eddy simulations (LES) with complex geometries
using the immersed boundary method~\cite{peller2006,tremblay2002}. 

In the present application the cut cell variant of the immersed boundary
is used~\cite{kurz2016}, which is a 3D-extension of the concept
presented in~\cite{cheny2010} using cell merging~\cite{hartmann2011}
to avoid time step restrictions due to small cut cell sizes.

Together with the cut cell immersed boundary, different wall
models are available for wall modeled LES (WMLES) which allows for many problems to increase at reasonable cost the range of Reynolds numbers.

The WMLES grid does not exhibit the large aspect ratios and small wall normal mesh spacing, as e.g. indicated in
Figure~\ref{fig:TE_Sketch} for wall resolved LES. Therefore, it becomes feasible to solve the acoustic equation system
(\ref{eq:ape_momentum}, \ref{eq:ape_pressure}) on the same grid with the same
numerical framework as being used for the incompressible Navier-Stokes equations, refer to the discussion in Section~\ref{sec:HADvsDNC}.

Both, the flow and acoustic time step size are chosen based on their
stability limits. Accounting for the shared grid, the stability limit of the acoustic time step is much more restrictive than that for the flow.
Therefore, one flow time step is divided into $n_{sub}$ acoustic sub
time steps. Since one flow time step needs considerably more computational effort than one acoustic time step, the overall computational time increases only moderately relative to a flow simulation without acoustics.

The numerical treatment, involving staggered arrangement of the sound particle velocities, central schemes for spatial operators, and a
fourth order explicit Runge-Kutta method for time integration, ensures low dispersive and dissipative errors. The possibility to simultaneously solve flow and acoustics on the same grid furthermore removes the necessity to interpolate the
source terms between grids and as such avoids numerical artifacts from related interpolation errors. 

To account for the flow-acoustic feedback term, in addition the coupling term $-\bfm \Omega \times \bstar u$ has been implemented as a source term in the momentum balance. The term is formulated as a volume integral over the control cell volume. The computation of vorticity
reuses the already computed components of the velocity gradient tensor, so
that the additional effort to evaluate the term is negligible small.

The numerical boundary conditions do not need to be changed for
neither flow nor acoustics. The feedback term
$-\bfm \Omega \times \bstar u$ is ramped down toward 0 at inflow and open
boundaries, similar to the treatment of the source term $\nabla P$ and the
convection terms in the acoustic equations.

To summarize, the resulting time advancement scheme reads: 

\begin{enumerate}
\item[0.] variables at time level $n$: $(\bfm U,P)^n$, $(\bstar u,p)^n$
\item integration of one flow time step\\
  The acoustic velocity $\bstar u$ in the feedback term $-\bfm \Omega \times \bstar u$ is kept constant during integration. \\
 {\it  result:} variables at new flow time level $(\bfm U,P)^{n+1}$
\item calculation of acoustic source term $\nabla P^{n+1} / \rho_0$
\item integration of $n_{sub}$ acoustic time steps utilizing available source term data\\
      {\it result:} variables at new flow time level $(\bstar u,p)^{n+1}$
\end{enumerate}

For the validation cases at $M= 0.0875$ studied in this work, the compressible velocity in the feedback term is kept constant during the sub steps of the time integration. This procedure formally degrades the numerical order of the time integration applied to the flow equations, but might be acceptable as long as the absolute value of the feedback term remains sufficient small relative to the contribution from all other spatial derivative terms. Based on the Mach number scaling discussed in subsection~\ref{ssec:spimplified_split_system}, the magnitude of term $(o)$ relative to the other spatial terms $(l)-(n)$ in Eq.~(\ref{eq:U-simple}) can be estimated to be one order smaller in the subsonic regime considered in this work. Therefore, the treatment is deemed applicable in the present context. In general, a higher order time integration could be achieved following~\cite{kvaerno1999} or adopting the non-uniform time stepping scheme (NUTS) proposed by Liu et al.~\cite{liu_2014}.

\subsection{Generic flute}
\label{sec:caseflute}

\subsubsection{Reference case description}

\begin{figure}[htp]
  \centering
  \includegraphics[trim=0cm 2cm 0cm 0cm, width=0.6\textwidth]{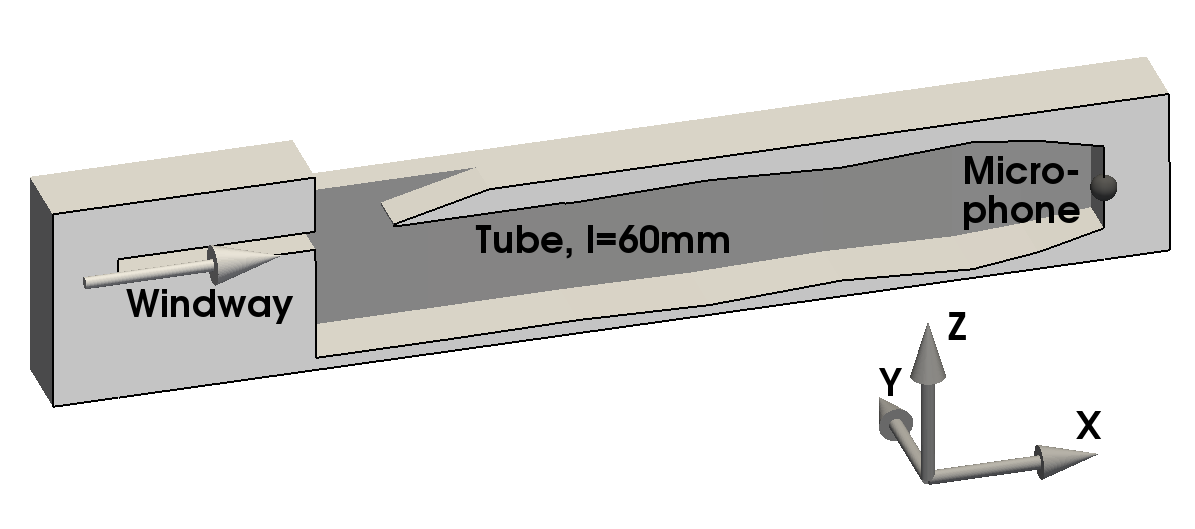}
  \caption{Generic flute, geometry.}
  \label{fig:recorder_sketch}
\end{figure}

The generic flute (see Figure~\ref{fig:recorder_sketch}) represents a model for either the head of a recorder with closed end or a
miniature stopped flute organ pipe. The length of the resonator is
$60\mm$ and the jet hitting the labium exits a windway with $7\mm \times
1.5\mm$ cross section. In a flute, the feedback cycle between flow and
acoustics is the basic sound generation mechanism. The case and
geometry are taken from K\"uhnelt~\cite{kuehnelt2016}. This author studied the
two way energy transfer between flow and acoustics by analyzing
simulations of the full compressible set of Navier-Stokes equations
using a Lattice Boltzmann solver. The velocity in the center of the
windway is $30\mps$ resulting in a Mach number $M=0.0875$. In the
reference simulation the viscosity was increased by a factor of 10
compared to ambient air to reduce the resolution requirements for the
numerical grid. Due to the low Reynolds number ($Re=294$ in the
windway) a uniform grid spacing of $0.175\mm$ and an overall node
number of $14.5\cdot 10^6$ was sufficient to resolve the flow as a DNS  with the Lattice Boltzmann solver without
any additional turbulence model involved~\cite{kuehnelt2016}. The 
sound pressure has been sampled in a probe position located at the stopped end of the
pipe, Figure~\ref{fig:recorder_sketch}.

\subsubsection{Simulation setup}

\begin{figure}[htp]
  \centering
  \includegraphics[width=0.7\textwidth]{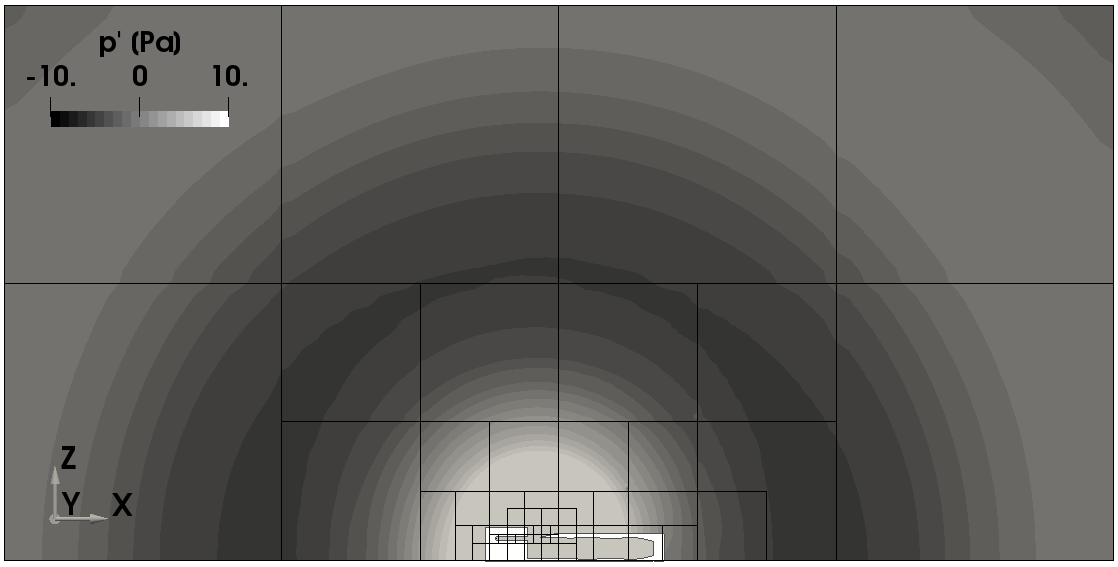}
  \caption{Generic flute. Simulation domain and hierarchical grid structure of grid A1 (coarse grid), Table~\ref{tab:recorder_simulationparameters_modelterms}. Subgrids consisting of $22^3$ cubic cells each are shown. Snapshot of radiated sound field shown as background.}
  \label{fig:recorder_domain}
\end{figure}

For the present simulations a domain of $0.5\m \times 0.5\m \times
0.25\m$ has been chosen, with the generic flute placed at the bottom
center, refer to Figure~\ref{fig:recorder_domain}. The boundary conditions
at the domain edges are set to a hard wall at the bottom and non
reflective openings at all other sides.  The flute itself consists of
no slip walls with no acoustic absorption. A physical time sample duration of $0.05\s$ 
has been simulated for each case, of which only the second half was
used for data evaluation in order to avoid any  transients side effects. Due to the low Reynolds number, the incompressible Navier-Stokes
equations were directly solved without any subgrid-scale model. 

Table~\ref{tab:recorder_simulationparameters_modelterms} summarizes some main simulation parameters. All cases are labeled by a combination of capital letter and single-digit number. The capital letter indicates the specific set of equations used. The single-digit number indicates the resolution of the employed grid, i.e., '1', '2', and '3' for the coarse, medium, and fine grid, respectively. 

In the baseline cases (letter 'A'), the full systems of Eqs.~(\ref{eq:ape_momentum}-\ref{eq:incomp_mass})
with feedback and convection terms in the acoustic equations included is used.
Variants 'B' and 'C' also include the feedback term in the flow equations. Furthermore, in variant 'B' the convective term containing $\bfm U$ is dropped in the compressible Eqs.~(\ref{eq:ape_momentum}, \ref{eq:ape_pressure}). In variant 'C' the full non-linear
acoustic convection term based on $\bfm U + \bstar u/2$  is included according to Eq.~(\ref{eq:APE-u}). 

In case 'D' only a standard open loop approach according to Figure~\ref{fig:open_loop_HAS} was applied by neglecting the feedback term $-\bfm \Omega \times \bstar u$ in the incompressible momentum equation, Eq.~(\ref{eq:incomp_momentum}). Otherwise the same set of equations as in 'A' was solved.

\begin{table}[htbp]
	\centering
	\caption{Generic flute. Simulation parameters.}
	\vspace{2mm}
	\begin{tabular}{@{\arrayrulewidth1.5pt\vline\hspace{3pt}}c@{\hspace{3pt}\arrayrulewidth1.5pt\vline\hspace{3pt}}c|c|c|c|c|c@{\hspace{3pt}\arrayrulewidth1.5pt\vline}}
		\hline
		\noalign{\hrule height1.5pt}
		& $\Delta x_{min}$ & $\Delta x_{max}$ & No. of cells & $\Delta t$ & feedback & convective terms \\
		&                 &                 &              &            & included & in comp. eq.\\
		\noalign{\hrule height1.5pt}   
		A1 & $0.25 \mm$ & $4\mm$ & $2.9\cdot 10^6$ & $2.5\mys$   &  + & + / linear  \\
		\hline
		A2 & $0.1875\mm$ & $6\mm$ & $2.5\cdot 10^6$ & $1.875\mys$&  + & + / linear  \\
		\hline
		A3 & $0.125\mm$ & $4\mm$ & $8.9\cdot 10^6$  & $1.25\mys$ &  + & + / linear  \\
		\noalign{\hrule height1.5pt}
		B1 & $0.25\mm$ & $4\mm$ & $2.9\cdot 10^6$ & $2.5\mys$   &   + & --  \\
		\hline
		C1 & $0.25\mm$ & $4\mm$ & $2.9\cdot 10^6$ & $2.5\mys$    &   + & + / non-linear  \\
		\noalign{\hrule height1.5pt}
		\hline
		D3 & $0.125\mm$ & $4\mm$ & $8.9\cdot 10^6$  & $1.25\mys$ &  -- & + / linear  \\
		\noalign{\hrule height1.5pt}
	\end{tabular}
	\label{tab:recorder_simulationparameters_modelterms}
\end{table}

\begin{figure}[htp]
	\centering
	\includegraphics[trim=0cm 0cm 0cm 0cm, width=0.556\textwidth, angle=180, origin=c]{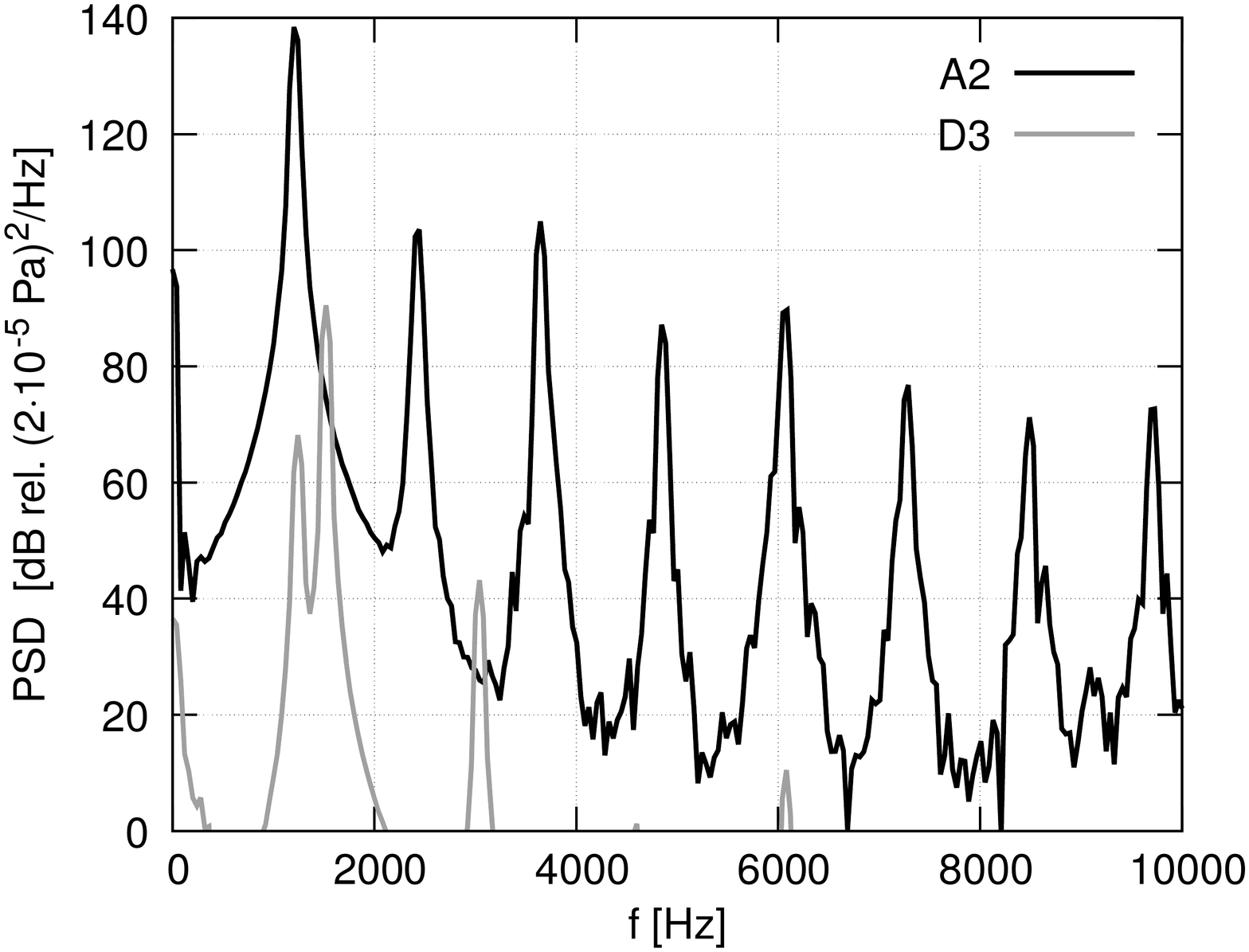}
	\caption{Generic flute. Comparison of sound spectra obtained with and without feedback term $-\bfm \Omega \times \bstar u$. A2, feedback term included on medium grid (black solid line). D3, feedback term not included on fine grid (gray solid line).}
	\label{fig:recorder_feedback_test}
\end{figure}

\subsubsection{Results}

\paragraph{Effect of feedback term}

Figure~\ref{fig:recorder_feedback_test} presents spectra of pressure for two simulations evaluated at the microphone position indicated in Figure~\ref{fig:recorder_sketch}. According to the case nomenclature, D3 (gray solid line in the figure) refers to an open-loop simulation with feedback term $-\bfm \Omega \times \bstar u$ switched off and using the finest mesh resolution. A2 (black solid line) presents the case with feedback term switched on by using the full subsonic system Eqs.~(\ref{eq:ape_momentum}-\ref{eq:incomp_mass}) on the medium grid.  

For A2, the pressure spectrum at the stopped end is dominated by a
fundamental peak at $1212\Hz$ with peak level around $140\dB$. Additionally, a series of higher harmonics shows up with levels at least $30\dB$ below the fundamental peak. Harmonics two and four are located at
$3636\Hz$ and $6060\Hz$. These higher harmonics and the fundamental frequency are close to the eigenfrequencies of the first three passive eigenmodes of the generic
flute, which are at $f_0=1276.5\Hz$, $f_1=3788.6\Hz$ and $f_2=6275.2\Hz$,
as computed using a FEM model in \cite{kuehnelt2016}. The
corresponding spatial structures are approximately the
$\frac{1}{4}\lambda$, $\frac{3}{4} \lambda$ and $\frac{5}{4} \lambda$
modes of the  $60\mm$ pipe. 

In case D3 without feedback, Figure~\ref{fig:recorder_feedback_test} reveals a double-peak structure of significantly reduced peak level at and around the fundamental frequency. The first peak at $1212\Hz$ has a peak level close to $70\dB$, which corresponds to a peak amplitude reduced by more than three orders of magnitude compared to case A2. This peak can be attributed to the fundamental eigenfrequency of the pipe, but exhibiting significantly weaker excitation as in case A2. A secondary small peak is visible slightly above the fundamental frequency at  $1520\Hz$, which can be attributed to coherent flow structure present for case D3 in the jet emerging from the windway.        

Apparently, the feedback term of A2 significantly enhances the energy transfer from the flow to the pipe eigenmode and forces the characteristic frequency of the flow to lock-in to the fundamental frequency so that a high-pitched whistle with fundamental peak at $1212\Hz$  results.  

This results clearly reveals that the simulation with coupling term $-\bfm \Omega \times \bstar u$ included is able to predict flow-acoustic feedback tones that are not properly captured by the open-loop approach.

\begin{figure}[htp]
	\begin{tikzpicture}
	\pgftext{
		\includegraphics[trim=0cm 0cm 0cm 0cm,  width=0.5\textwidth]{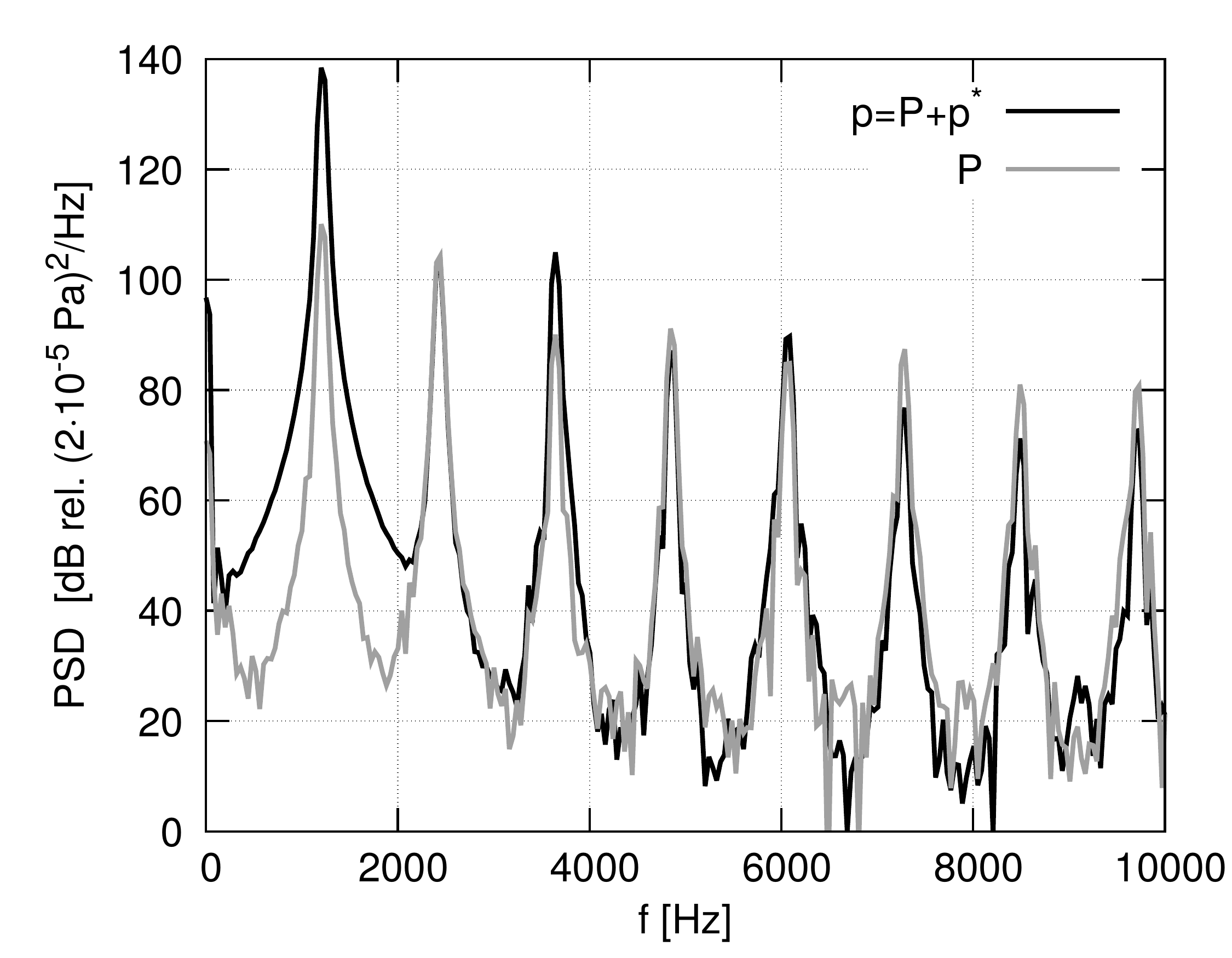}
	}%
	\node[text opacity=1] at (-3.8, 2.5) {\bf(a)};
	\end{tikzpicture}
	\begin{tikzpicture}
	\pgftext{
		\includegraphics[trim=0cm 0cm 0cm 0cm,  width=0.5\textwidth]{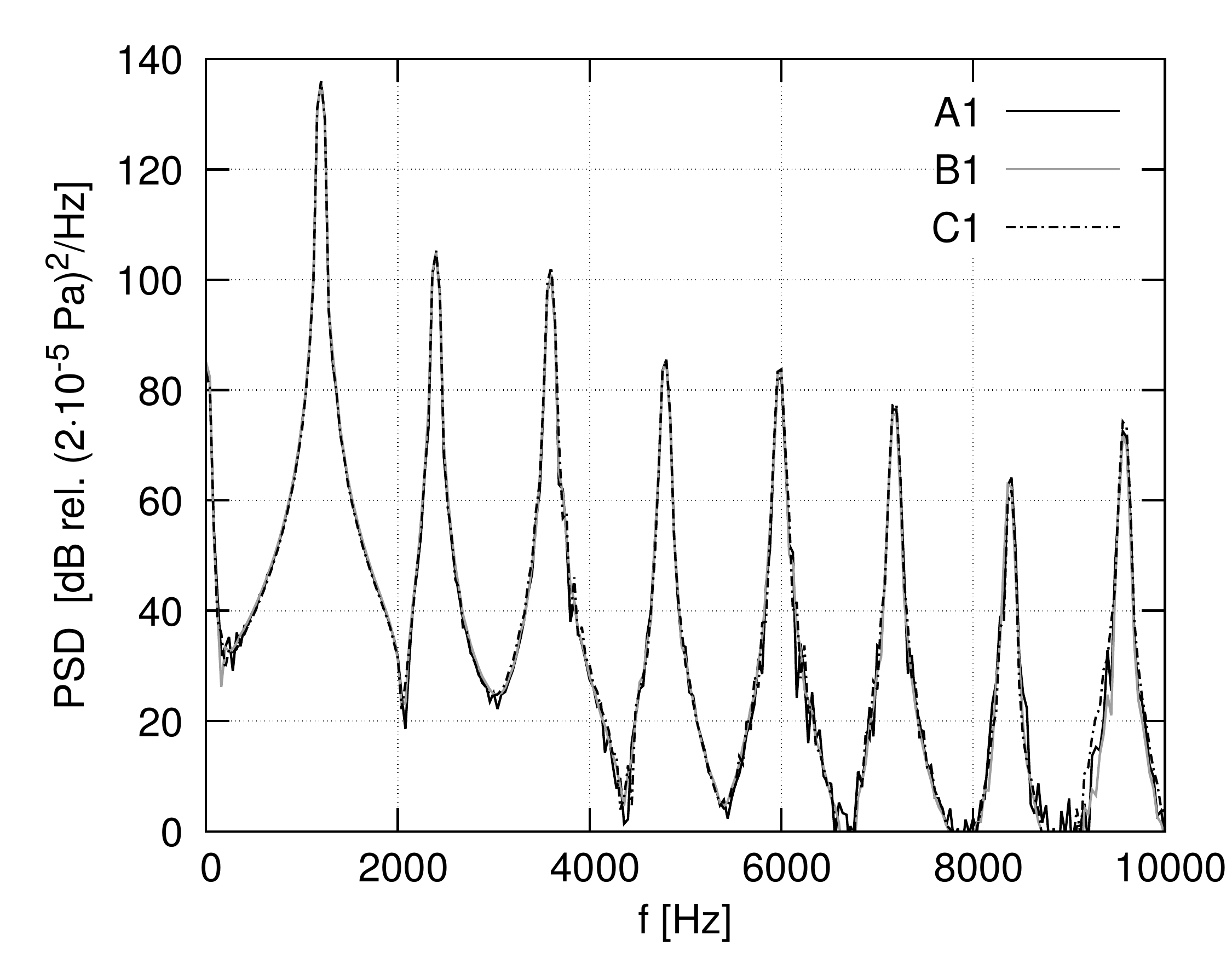}
	}%
	\node[text opacity=1] at (-3.8, 2.5) {\bf(b)};
	\end{tikzpicture}
	\caption{Generic flute. (a) Spectra of compressible ($p$) and incompressible ($P$) pressure at the stopped end for case A2. (b) Different convective terms considered in the acoustic equation system; A1, full subsonic system, Eqs.~(\ref{eq:ape_momentum}-\ref{eq:incomp_mass}); B1, convective terms skipped; C1, non-linear terms according to Eq.~(\ref{eq:APE-u}) included. }
	\label{fig:recorder_gridstudy_convectiveterms}
\end{figure}

Figure~\ref{fig:recorder_gridstudy_convectiveterms}(a) in addition compares the spectra of
compressible (complete) pressure $p$ and incompressible pressure $P$. The figure indicates the pressure peaks of both variables to occur at the same frequencies as both quantities are coupled. However, differences in the
amplitudes are visible. The peaks of $p$ surpass those of $P$
at the fundamental, the 2nd, and the 4th harmonic at $6060\Hz$, i.e. at frequencies corresponding to the pipe eigenfrequencies. At the 5th and
higher harmonics, however, the peaks of complete pressure $p$ are about $10\dB$ lower than those of incompressible pressure $P$. Inevitably, from the pressure decomposition as defined by Eq.~(\ref{eq:p_decomp}) it follows that the compressible pressure correction $\sstar p=p-P$ must be in opposite phase to the incompressible pressure fluctuations at the given microphone position and for these frequencies.

\paragraph{Comparison with compressible simulation and grid convergence study}

\begin{table}[htbp]
	\centering
	\caption{Generic flute. Peak frequency and sound pressure level at the stopped end.}
	\vspace{2mm}
	\begin{tabular}{@{\arrayrulewidth1.5pt\vline\hspace{3pt}}c@{\hspace{3pt}\arrayrulewidth1.5pt\vline\hspace{3pt}}c|c|c@{\hspace{3pt}\arrayrulewidth1.5pt\vline}}
		\noalign{\hrule height1.5pt} & $\Delta x$ & $f$ & SPL \\
		\noalign{\hrule height1.5pt}
		A1 & $0.25\mm$ & $1196\Hz$ & $153.8\dB$ \\
		\hline A2 & $0.1875\mm$ & $1212\Hz$ & $156.7\dB$ \\
		\hline A3 & $0.125\mm$ & $1220\Hz$ & $157.0\dB$ \\
		\noalign{\hrule height1.5pt}
		comp. N.-S.~\cite{kuehnelt2016}&$0.175\mm$  & $1180\Hz$ & $153\dB$ \\
		\noalign{\hrule height1.5pt}
		D3 &$0.125\mm$  & $1520\Hz$ & $108.3\dB$ \\
		\noalign{\hrule height1.5pt}
	\end{tabular}
	\label{tab:recorder_gridstudy}
\end{table}

A grid study was conducted by means of cases A1, A2 and A3. The results have been compared to compressible Navier-Stokes (comp. N.-S.) results  as simulated by K\"uhnelt~\cite{kuehnelt2016} with a Lattice Boltzmann (LB) solver on a uniform Cartesian mesh. Table~\ref{tab:recorder_gridstudy} provides the minimum grid resolution for all cases together with results obtained for the fundamental frequency and overall sound pressure level at the stopped end of the pipe. For completeness, results for case D3 without feedback are also included in the table.  

The table clearly shows that frequency and level match well with the compressible
simulation results from K\"uhnelt~\cite{kuehnelt2016} for all feedback cases A1-A3. The best match between compressible simulation and the present simulation is found for the coarse-grid case A1. The minimum voxel spacing
used for the Lattice Boltzmann method is in between the resolutions of the finer meshes A2 and A3. This would be explicable by the slightly better resolution properties provided by a finite volume approach with staggered arrangement in comparison to a collocation method. However, a grid study was not available from \cite{kuehnelt2016} to provide a conclusive answer yet.

Case D3 indicates that the coherent frequency of the flow does not lock-in to the pipe eigenfrequency and that the sound pressure level is significantly below the reference solution, i.e., the high-pitch whistle is missing.

\paragraph{Relevance of acoustic convective terms}

\begin{figure}[htp]
  \begin{tikzpicture}
  \pgftext{
  \includegraphics[width=0.25\textwidth]{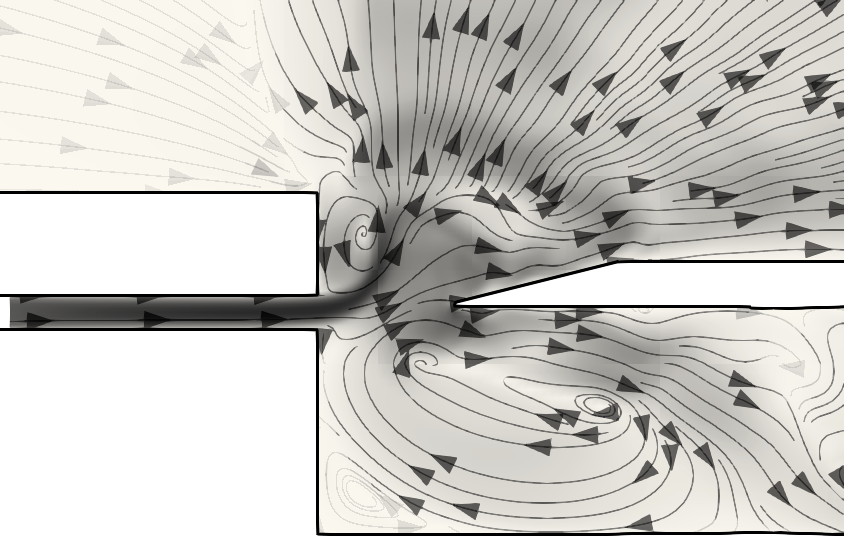}
  \includegraphics[width=0.25\textwidth]{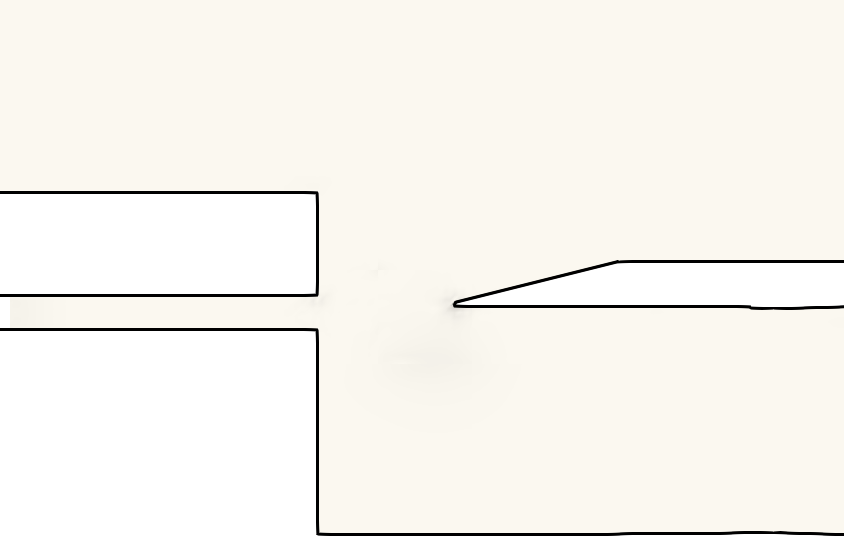}
  \includegraphics[width=0.25\textwidth]{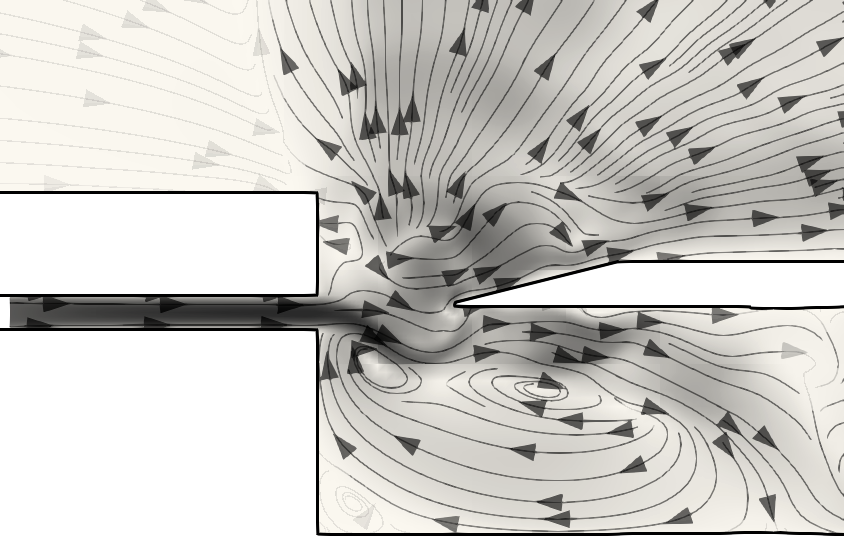}
  \includegraphics[width=0.25\textwidth]{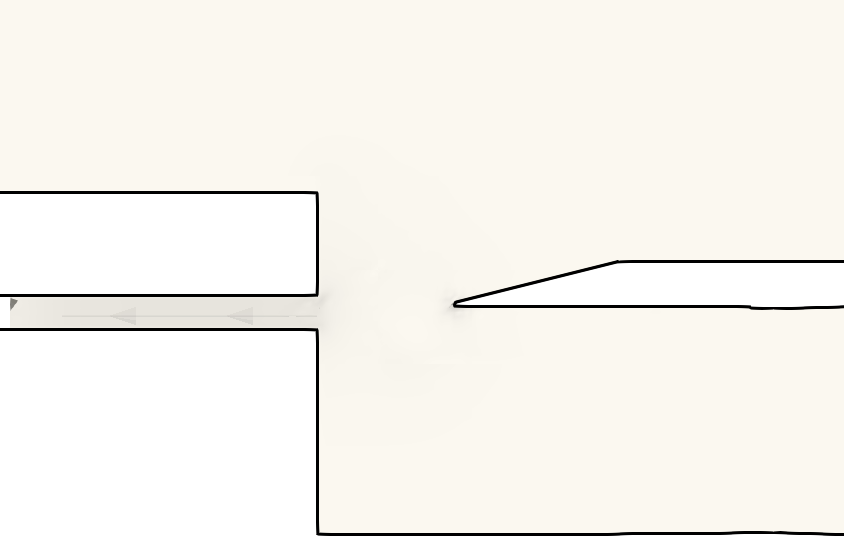}
  }\node[text opacity=1] at (-3.4, -0.8) {$\phi=0$ \hspace{7.5cm}$\phi=\frac{4\pi}{4}$};
  \end{tikzpicture}

  \begin{tikzpicture}
  \pgftext{
  \includegraphics[width=0.25\textwidth]{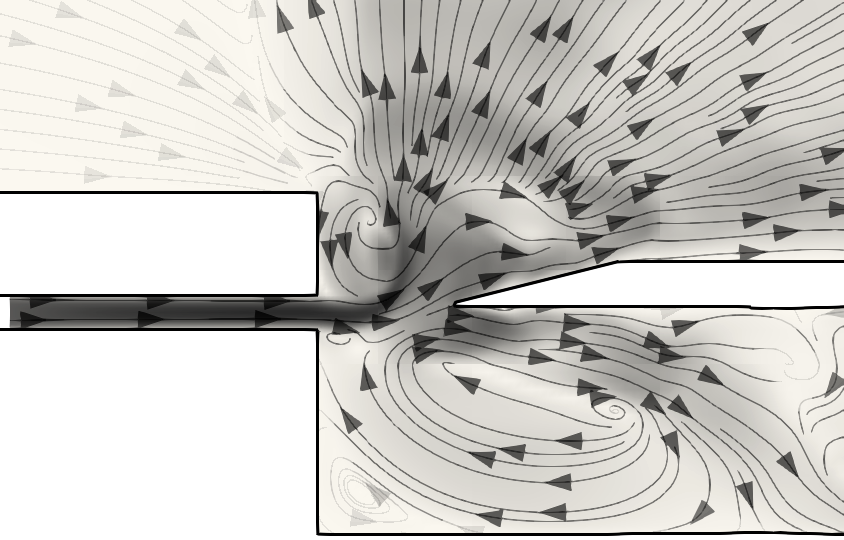}
  \includegraphics[width=0.25\textwidth]{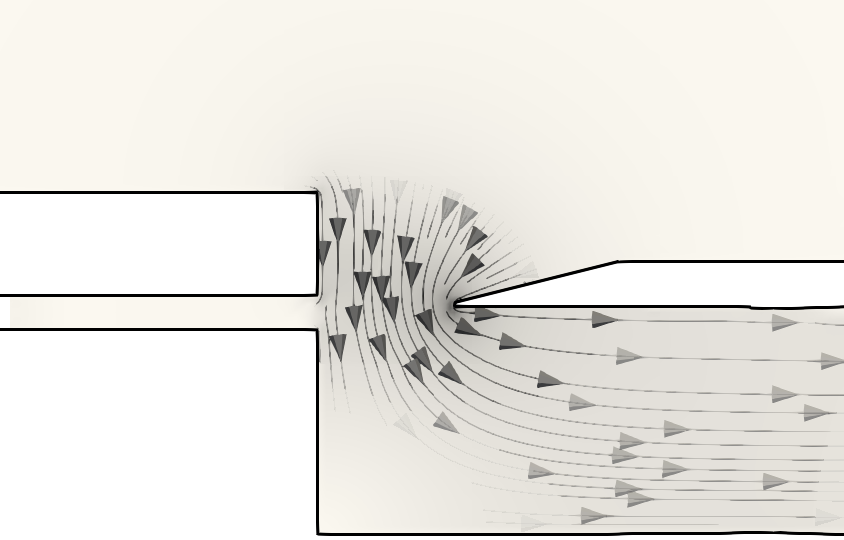}
  \includegraphics[width=0.25\textwidth]{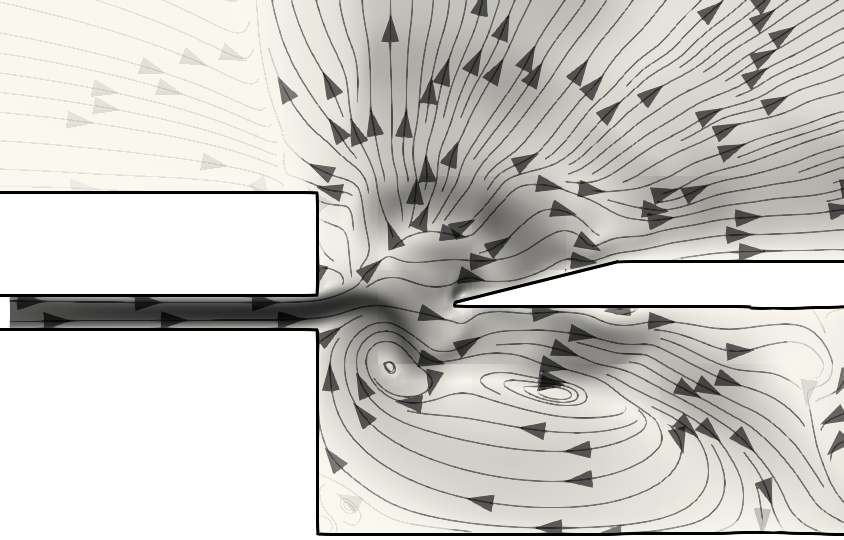}
  \includegraphics[width=0.25\textwidth]{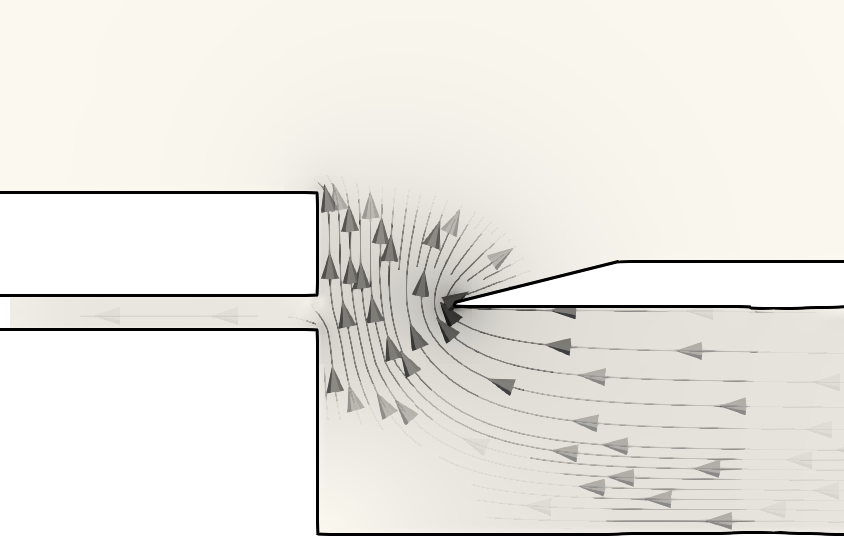}
  }\node[text opacity=1] at (-3.4, -0.8) {$\phi=\frac{\pi}{4}$\hspace{7.5cm}$\phi=\frac{5\pi}{4}$};
  \end{tikzpicture}

  \begin{tikzpicture}
  \pgftext{
  \includegraphics[width=0.25\textwidth]{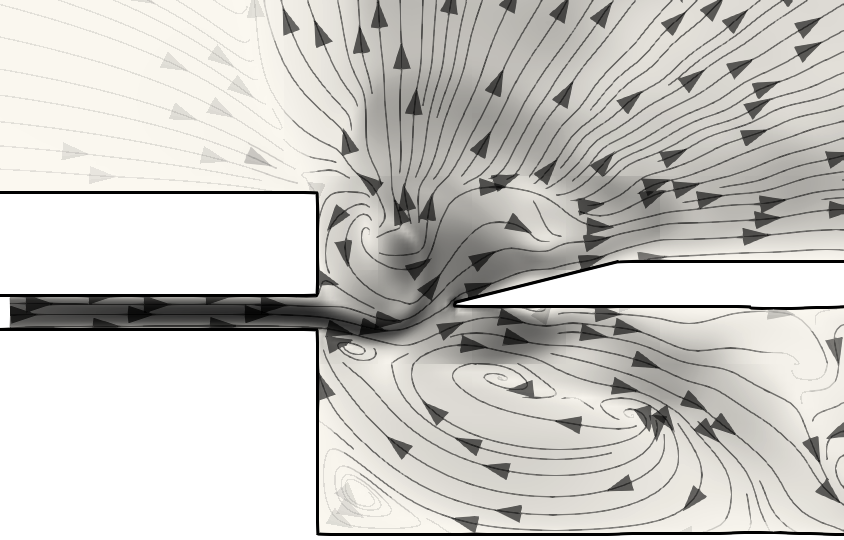}
  \includegraphics[width=0.25\textwidth]{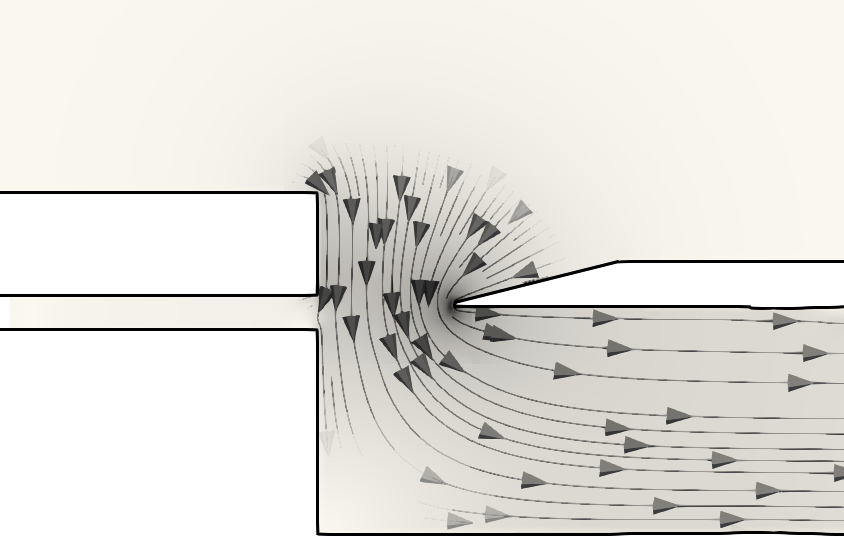}
  \includegraphics[width=0.25\textwidth]{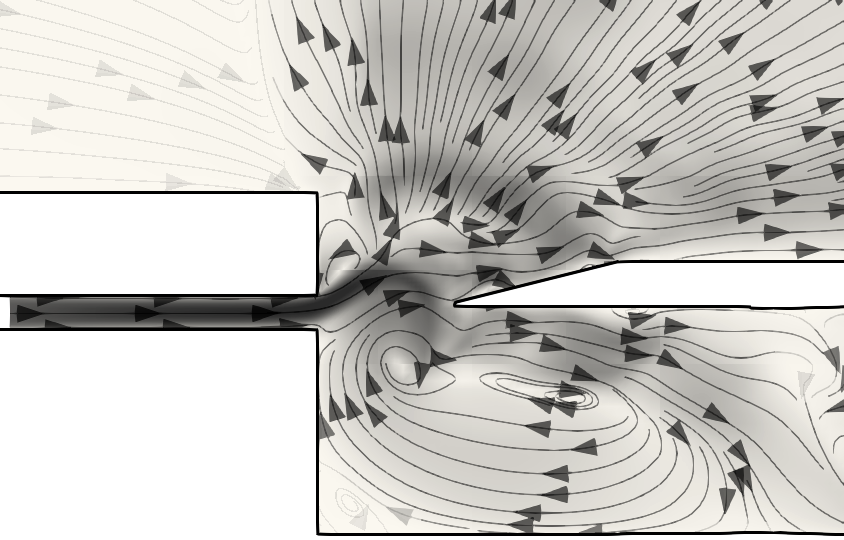}
  \includegraphics[width=0.235\textwidth]{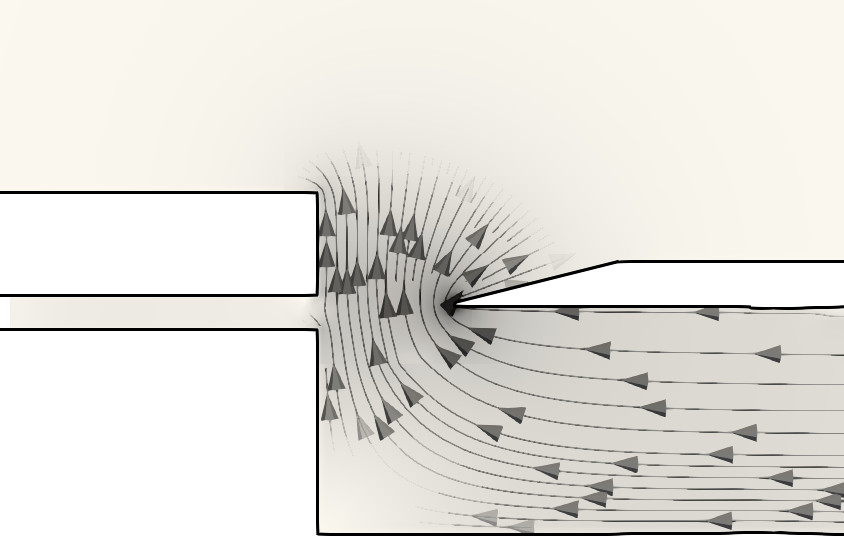}
  }\node[text opacity=1] at (-3.4, -0.8) {$\phi=\frac{2\pi}{4}$\hspace{7.5cm}$\phi=\frac{6\pi}{4}$};
  \end{tikzpicture}

  \begin{tikzpicture}
  \pgftext{
  \includegraphics[width=0.25\textwidth]{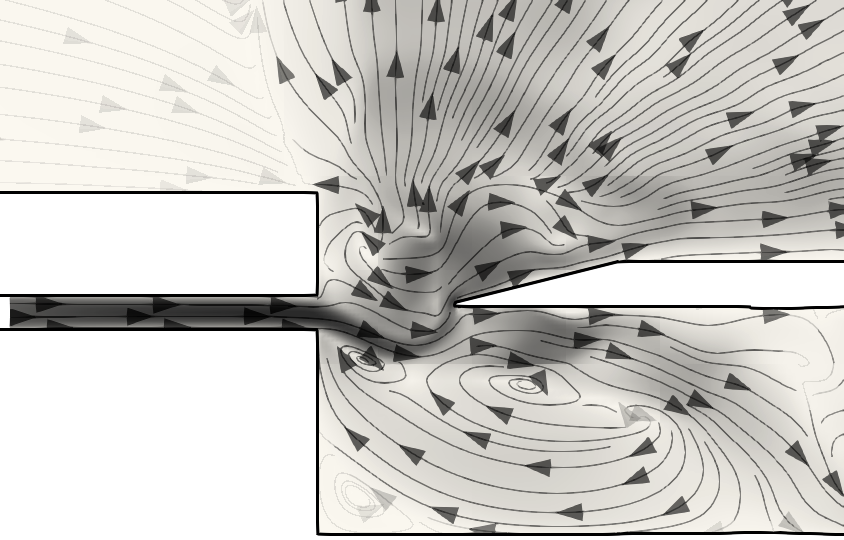}
  \includegraphics[width=0.25\textwidth]{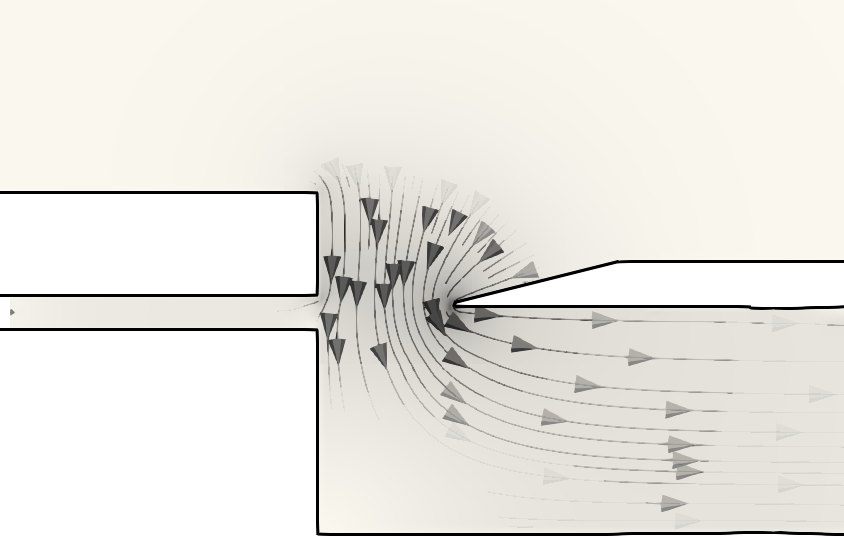}
  \includegraphics[width=0.25\textwidth]{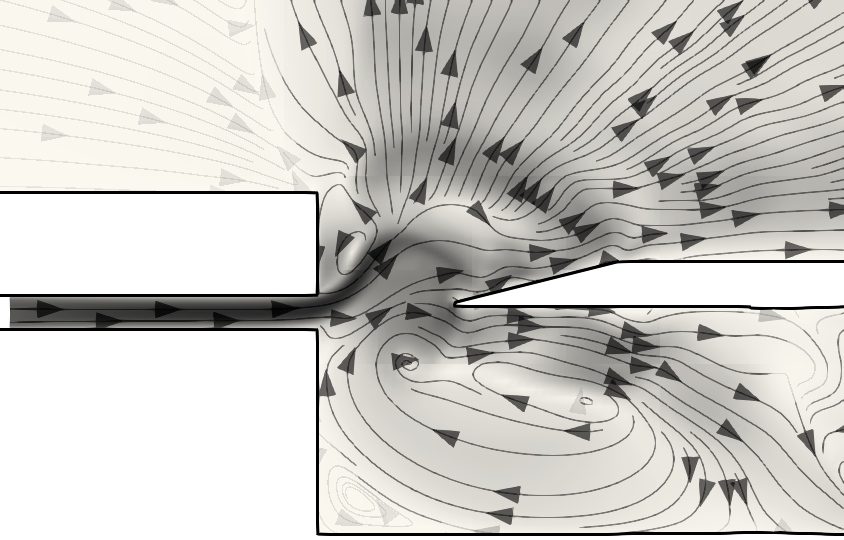}
  \includegraphics[width=0.25\textwidth]{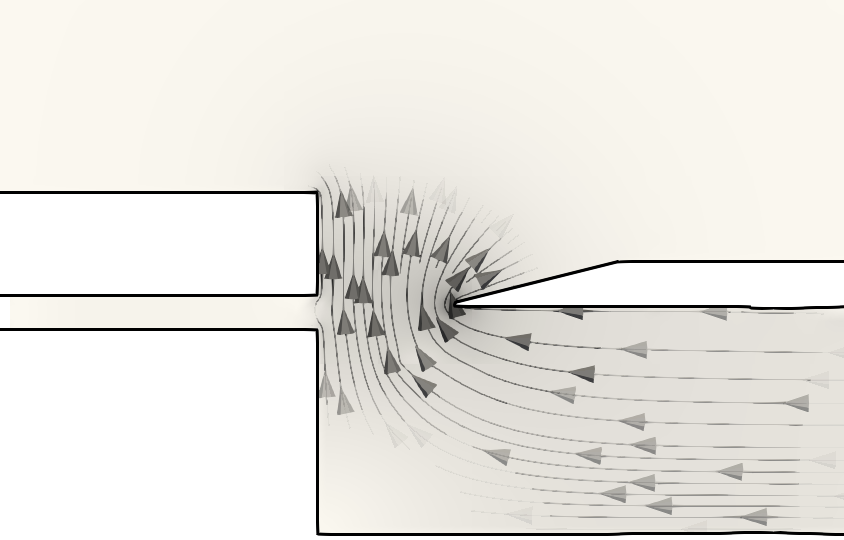}
  }\node[text opacity=1] at (-3.4, -0.8) {$\phi=\frac{3\pi}{4}$\hspace{7.5cm}$\phi=\frac{7\pi}{4}$};
  \end{tikzpicture}

  \caption{Generic flute. Sequence of images showing one cycle at the fundamental frequency with phase angle increment $\Delta\phi=\pi/4$ and phase angle $\phi=0$ related to the maximal upward deflection of the jet. At each phase angle a double image of the velocities in the cut plane $y=0$ (symmetry plane of the flute) is shown with $\bfm U$ on the left and  $\bstar u$ on the right. Arrows indicate the flow direction and the color map ranging from $0$ to $30\mps$ with scale as depicted in Figure~\ref{fig:recorder_u_oneway} defines the magnitude of flow velocity.}
  \label{fig:recorder_cycle}
\end{figure}

The relevance of the convective terms in the acoustics equations is evaluated for the present problem by comparing
cases A1, B1 and C1.  For these cases,  the spectra of $p$ at the stopped
end are effectively identical
(Figure~\ref{fig:recorder_gridstudy_convectiveterms}(b)), showing that the
non-linear convective terms as well as the linear ones in the acoustic
equations are irrelevant for the present problem.  This may be due to the
low Mach number and because deviation of the fluid from rest occur only in a small fraction of the computational domain. Only in the windway and around the mouth of
the flute considerable values of velocity are present that give contributions  to
the convective terms, refer to Figure~\ref{fig:recorder_cycle}. On the contrary, acoustic propagation occurs in the entire domain where the fluid largely is at rest.

\paragraph{Stability of two way coupled equations}

The time step width used for the integration of the flow equations was
based on the stability criteria of the time integration scheme and the
maximum incompressible flow velocity $U$, resulting in a maximal CFL
number of $\approx 1.2$ at a time step width of $\Delta t_F=1.875\mu
s$. The time step width used for the integration of the acoustic
equations was based on the the stability criteria of the time
integration scheme and a maximal transport velocity given by the speed of sound plus convection at the incompressible flow
velocity, $c+U$. The resulting  acoustic time step
width of $\Delta t_A=0.1705 \mys$ is 11 times smaller than
$\Delta t_F$.  The two way coupling between acoustics and flow did not
introduce any additional stability constraints, in agreement with the  theoretical analysis of Section~\ref{ssec:seq_analysis}, so that the same time
stepping constraints could be applied as in the simulations without
feedback.

\paragraph{Discussion of the two way coupling effect}

\begin{figure}[htp]
  \centering
  \includegraphics[width=0.23\textwidth]{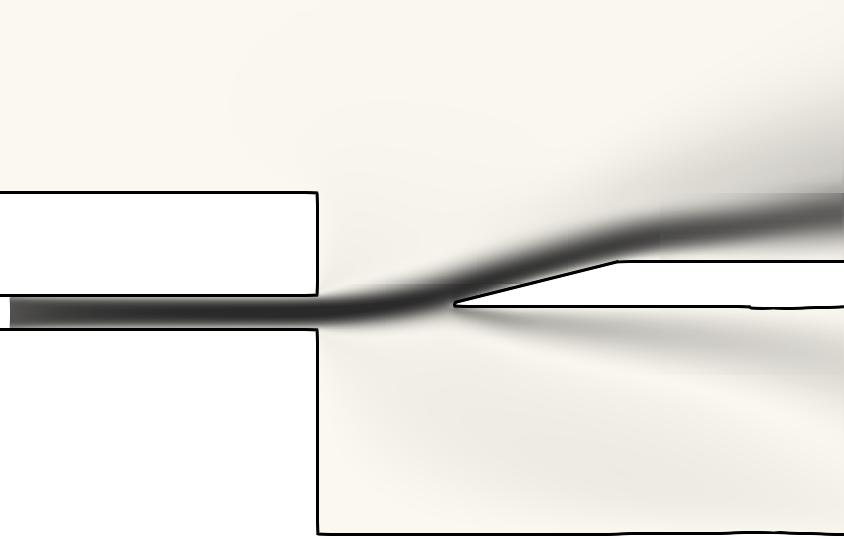}\hspace{1cm}
  \includegraphics[width=0.18\textwidth]{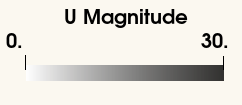}
  \caption{Generic flute, cut plane in the symmetry plane of the flute showing the magnitude of $\bfm U$ with color map ranging from $0$ to $30\mps$. One way coupling, case D3 (no feedback).}
  \label{fig:recorder_u_oneway}
\end{figure}

\begin{figure}[htp]
\centering
  \includegraphics[width=0.3\textwidth]{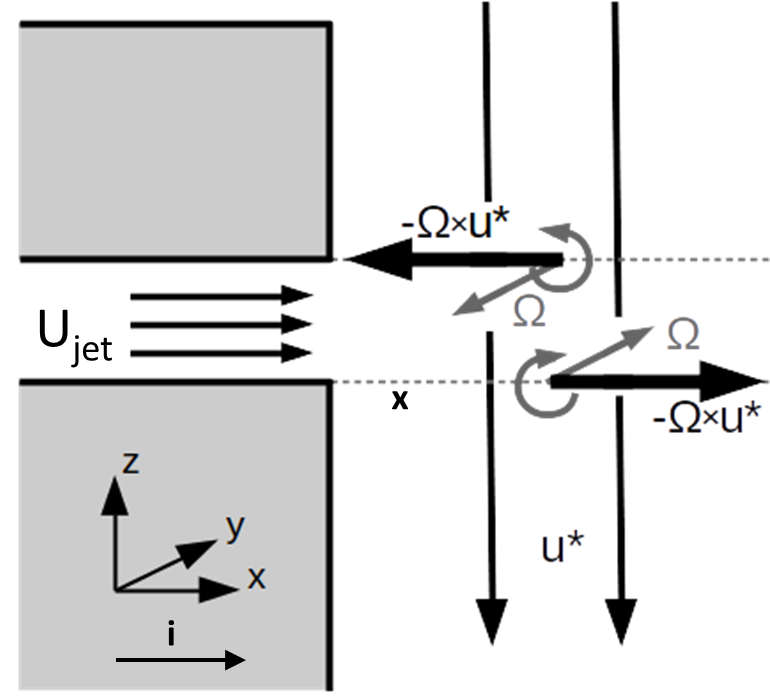}
  \caption{Sketch of the quantities involved in the feedback term $-\bfm \Omega \times \bstar u$. The unity vector in $x$-direction is denoted by $\bfm i$. The cross indicates a point slightly underneath the jet, refer to the discussion in the main body of the text.} 
  \label{fig:recorder_feedback_sketch}
\end{figure}

\begin{figure}[htp]
  \begin{tikzpicture}
    \pgftext{
  \includegraphics[width=0.4\textwidth]{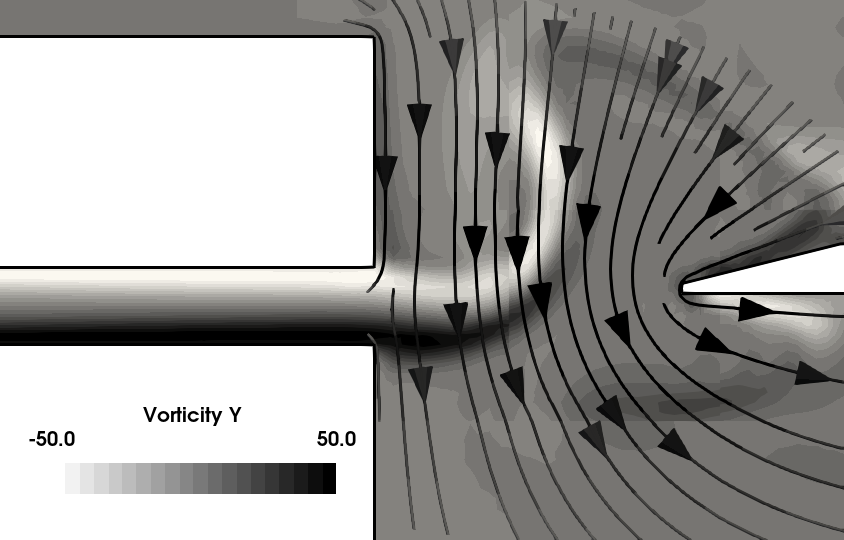}
    }%
    \node[text opacity=1] at (-3.7, 2.0) {\bf(a)};
  \end{tikzpicture}
  \begin{tikzpicture}
    \pgftext{
  \includegraphics[width=0.4\textwidth]{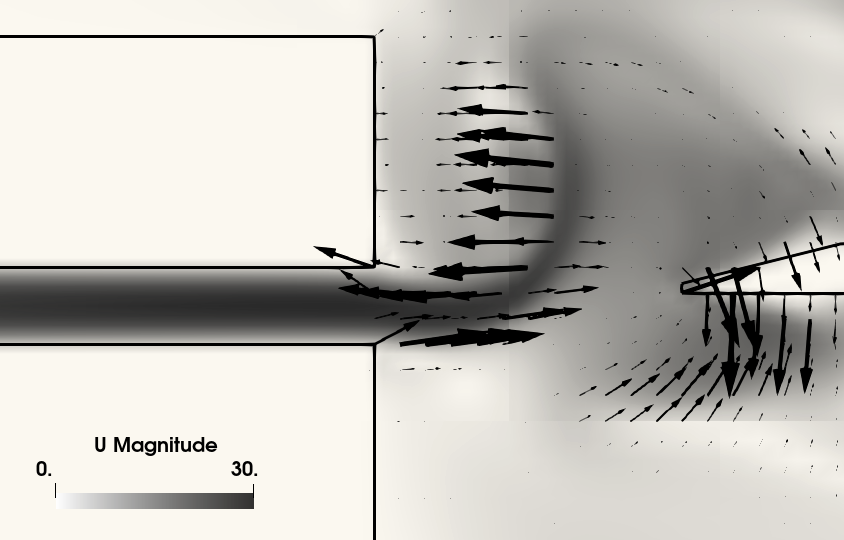}
    }%
    \node[text opacity=1] at (-3.7, 2.0) {\bf(b)};
  \end{tikzpicture}
  \caption{Generic
  flute. Feedback term at phase angle $\phi = \pi/4$. (a) Component $ \Omega_y$ in the symmetry plane $y=0$ with velocity vectors related to $\bstar u$. (b) Magnitude of $\bfm U$ in the symmetry plane and
  arrows indicating $-\bfm \Omega \times \bstar u$.}
  \label{fig:recorder_feedback}
\end{figure}

Figure~\ref{fig:recorder_cycle} shows for case A2 with feedback a sequence of snapshots for one oscillation cycle at the fundamental frequency. The cycle is given by a succession of phase increments $\Delta \phi = \pi/4$, starting at phase angle $\phi=0$ related to the maximally upward deflected jet. For each phase angle, two images are displayed. The left image depicts the incompressible flow velocity field $\bfm U$ in terms of a magnitude map with arrows indicating the flow direction. The left image shows in a similar way the compressible correction velocity $\bstar u$ which mainly defines the sound particle velocity. 

The sequence of images related to the
incompressible flow velocity shows a strong down and upward
flapping of the jet exiting the windway. The right sequence of the compressible correction velocity indicates a magnitude in the immediate vicinity of the wedge that is comparable to the incompressible velocity with values  around $25.5\mps$. It is pointing downwards into the tone hole in the first half period and outward in the second half, indicating that the
hydrodynamic jet is convected by the acoustics.

A detailed look on the acoustics-to-flow feedback mechanism is shown in
Figures~\ref{fig:recorder_feedback_sketch}
and \ref{fig:recorder_feedback}.  The first one sketches the somewhat idealized situation at a time instant when the jet is pointing in horizontal direction at velocity $\bfm{U}_{jet}$ and with downward directed acoustic velocity $\bstar u$. The edges of the plane jet define vortex sheets with vorticity $\bfm \Omega$. The vorticity-vector is pointing in negative $y$-direction at the upper edge. At the lower edge it is oriented in positive direction. The feedback term
$-\bfm \Omega \times \bstar u$ results in a vortex force pointing in negative
$x$-direction at the upper edge of the jet, decelerating the flow. At
the lower edge the direction is reversed, i.e., the flow is
accelerated. 

The feedback term provides an additional local convective acceleration that corresponds to a downward deflection of the jet. Although the feedback term is an incomplete convective term in the momentum equation, it accounts for a convection of vorticity $\bfm \Omega = \nop \times \bfm U$ in the combined velocity field $\bfm u=\bfm U + \bstar u$, as can be seen from the vorticity equation derived by taking the curl of Eq.~(\ref{eq:incomp_momentum}), which yields 
\begin{equation}\label{eq:inc_vorticity}
\pp{\bfm \Omega}{t} + \underbrace{\left ( \bfm U + \bstar u \right )\cdot \nop \bfm U}_{(i)}   = \nu_0 \Delta \bfm \Omega 
+ \underbrace{\bfm \Omega \cdot  \left ( \bfm U + \bstar u \right )}_{(ii)} - \underbrace{\bfm \Omega \nop \cdot  \bstar u}_{(iii)}.   
\end{equation} 
Here, all terms $(i)-(iii)$ contain contributions from the compressible velocity field $\bstar u$. These contributions result from the application of the curl-operator to the feedback term in the derivation of Eq.~(\ref{eq:inc_vorticity}). Term $(i)$ indicates convection with combined velocity $\bfm u$, term $(ii)$ denotes vortex stretching also based on the combined velocity, and term $(iii)$ gives an additional source term from the dilatation of the compressible velocity $\bstar u$. 

For a position slightly underneath but outside the jet, marked by $\bfm{\sf x}$ in Figure~\ref{fig:recorder_feedback_sketch}, the incompressible velocity at time level $t$ is $\bfm U=0$. If the slightly deflected jet has approached the fixed position at time level $t+\delta t$, the incompressible velocity changes to $\bfm U = U_{jet} \bfm i$, where $\bfm i$ shall indicate the unit vector in $x$-direction. Hence, the convective acceleration at the edge of the downward shifting jet is of the order $\delta U_{jet}/\delta t$ with $\delta U_{jet}=U_{jet}-0$ and points in positive $x$-direction. For a point at the upper edge of the jet, slightly inside the jet, similar reasoning yields an equal large convective acceleration pointing in the opposite $x$-direction. Altogether the convective acceleration is as sketched in Figure~\ref{fig:recorder_feedback_sketch} and is a consequence of the convection of vorticity in the combined flow field as defined by the vorticity equation~(\ref{eq:inc_vorticity}).  

Figure~\ref{fig:recorder_feedback} shows the same quantities
as the sketch evaluated from simulation data at phase angle $\phi = \pi/4$ in the symmetry plane $y=0$. Figure~\ref{fig:recorder_feedback}(a) presents the variables present in the feedback term: the vorticity normal to the symmetry plane $\Omega_y$ and the streamlines of acoustic velocity
$\bstar u$ therein. The vorticity is mainly concentrated at the edges of the jet,
the acoustic velocity is pointing downwards into the mouth. Figure~\ref{fig:recorder_feedback}(b) shows the feedback term as arrows and a cut-plane through the
field of incompressible velocity magnitude. In the region where the acoustic
velocity is perpendicular to the jet (close to the exit of the
windway) the feedback term decelerates the flow at the upper edge of
the jet and accelerates it on the lower edge, resulting in a downward
convection of the jet, as described above. Otherwise, in the region where the jet is parallel to the
acoustic velocity (upward pointing end of the jet) the force from the
feedback results in a sidewards bending of the jet.

The flow field for the case without feedback D3 is shown in Figure~\ref{fig:recorder_u_oneway}. In contrast to the strong jet movement in the feedback case, the figure reveals a completely different behavior. In this case the jet
is laying quietly on the labium.  The complete different behavior found for the generic flute simulations
with and without feedback term included underlines the paramount importance of the identified coupling term for a proper prediction of flow-acoustic feedback tones.

The discussion of the feedback term furthermore highlights the general helpfulness of the HAS approach in the analysis of flow-acoustic interaction problems due to the immediately availability of the Helmholtz decomposed flow variables. For example, K\"uhnelt's~\cite{kuehnelt2016} analysis of the sound generation mechanism was based on compressible simulation and the decomposition had to be performed as part of an additional postprocessing step.

\subsection{Parker mode}

\label{sec:caseparker}

\subsubsection{Reference case}

\begin{figure}[htp]
  \centering
  \includegraphics[width=0.85\textwidth]{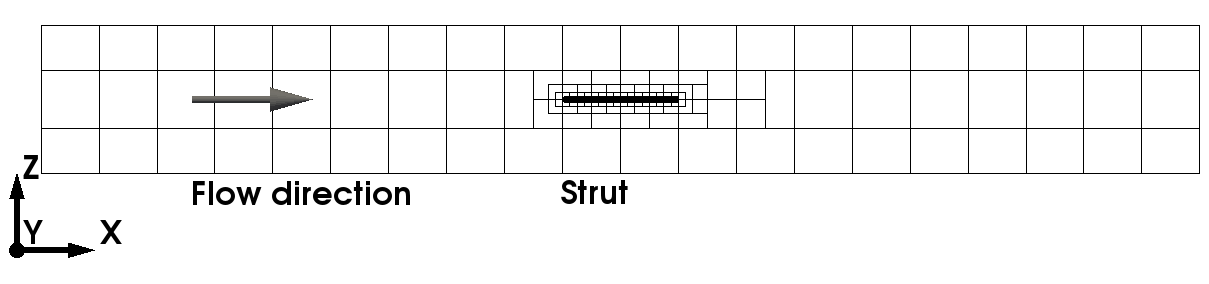}
  \includegraphics[width=0.6\textwidth]{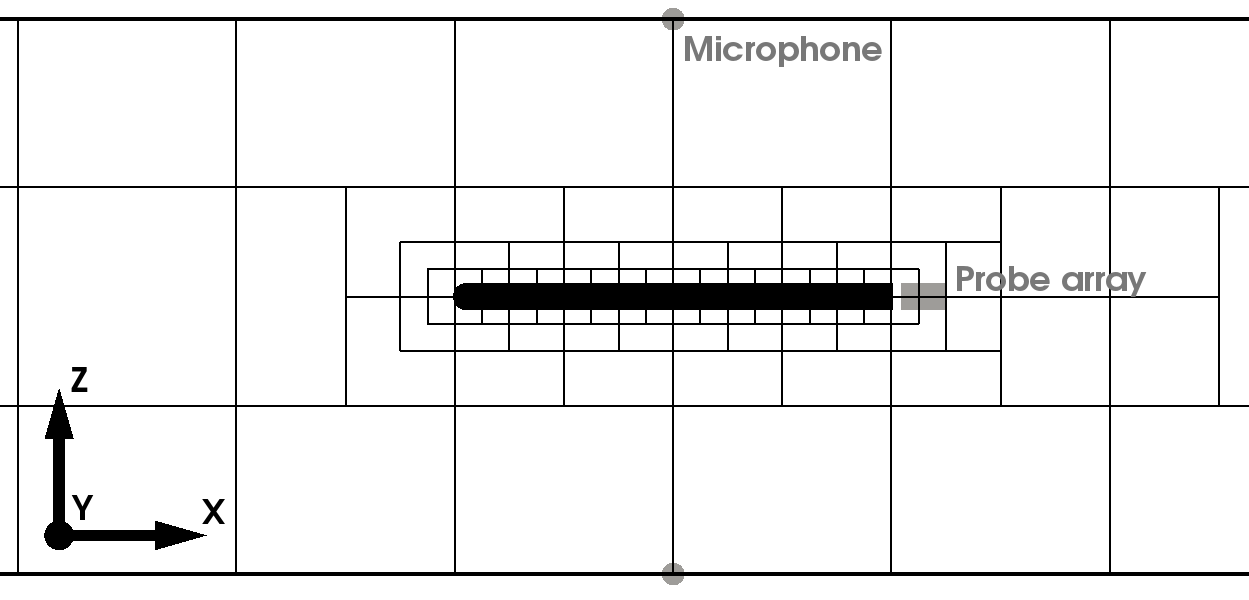}
  \caption{Parker mode. Geometry and grid structure. Top, entire setup. Bottom, enhanced area around the flat plate. Sub grids containing $24^3$ cubic cells are shown.}
  \label{fig:parker_sketch}
\end{figure}

The second simulation problem deals with a flow
exited two dimensional Parker-$\beta$-mode around a thick plate representing a strut in a square channel. The Parker-$\beta$-mode denotes the fundamental acoustic mode around the plate with pressure oscillation in anti-phase above and below the plate. The peak amplitude of pressure is present at the duct walls above and below the center of the plate. The pressure nodes and therefore antinodes of acoustic velocity are located at the leading and trailing edge of the plate. The corresponding eigenfrequency reduces with rising chord length and is always below the cut-on frequency of the duct, which renders it a trapped mode. The spatial structure of the present Parker-$\beta$-mode is shown in Figure~\ref{fig:parker_insitu} (c).

The simulations are compared with reference data taken from the experiment of Welsh et al.~\cite{welsh84}. The setup is sketched in
Figure~\ref{fig:parker_sketch}. The spanwise center plane is shown that comprises the channel geometry with solid walls at the bottom and the top and the plate with rounded leading and square trailing edge located on the center line. By increasing the bulk velocity in the channel, the vortex shedding frequency in the wake of the plate rises. Initially natural vortex shedding with constant Strouhal number takes place, so that the shedding frequency is directly proportional to the velocity.  When the vortex shedding frequency approaches the resonant frequency of the Parker-$\beta$-mode, it suddenly jumps and locks-in to the resonance frequency. After a small plateau, the shedding frequency slowly starts to increase again and eventually resumes a  linear growth at further increased bulk velocity,
refer to the trend shown in Figure~\ref{fig:parker_fshedding}(a).

The channel is $2.56\m$ long with cross section of $0.244\m\times
0.244\m$. The approaching flow is uniform with low turbulence intensity
of 0.04\%. The $t=12.1\mm$ thick plate with semicircular leading and
blunt trailing edge has a chord length of $l=193.6\mm$. The velocities
considered are around $U_b=30\mps$, resulting in a Mach number of
$M=0.0875$ and a Reynolds number based on the plate thickness of
$Re=23700$.

\subsubsection{Simulation setup}

The computational domain of $1.92\m\times 0.096\m\times 0.244\m$ represents a spanwise slice of the channel as depicted in Figure~\ref{fig:parker_sketch}. The plate thickness roughly corresponds to the boundary layer thickness as can be seen in Figure~\ref{fig:parker_welchvisualization}. The width of the computational domain is 8 times the plate thickness. This width is assumed to be sufficiently large to capture the spanwise turbulence length scale for a proper development of the turbulence in the boundary layer and the wake. The grid spacing ranges form $\Delta x =
0.5\mm\dots 4\mm$, resulting in total in 6.5 mio. cells. Refinement is
only present towards the surface of the plate and in its wake (s. Figure~\ref{fig:parker_sketch}).  Based on
the results of the generic flute simulations, convective terms in the
acoustic equations (\ref{eq:ape_momentum}, \ref{eq:ape_pressure})
have been neglected since the Mach number is of equal magnitude for the present problem. 

To measure the wake frequencies an array of sampling probes is placed downstream of the plate trailing edge. It has an extensions of $12.1\mm$ (plate thickness) and $20\mm$ in $z$- and $x$-direction, respectively, and starts $3.25\mm$ downstream of the trailing edge. The acoustic amplitude
of the Parker mode is measured at probes located at the walls of the
channel above and below the center of the plate, refer to Figure~\ref{fig:parker_sketch}.

The flow is computed by wall modeled LES (WMLES), using a standard
Smagorinsky model with a model constant of $C_S=0.1$ and Werner-Wengle wall model \cite{werner1993}. Slip
boundary conditions are applied at the channel walls. Periodicity is considered  
in spanwise direction. A volume flux is prescribed at the inlet plane. The
outlet boundary conditions are Dirichlet for the pressure and von Neumann
for the velocities. For the acoustics part, non-reflecting boundary conditions are used at the inlet and outlet. At the slip walls and the plate surface a small sound absorption coefficient of $\alpha = 0.01$ is prescribed, which defines the ratio of absorbed versus incident sound intensity. To trigger a turbulent boundary layer
at the plate in the WMLES, the wall shear stress
at the leading edge has been artificially increased applying some surface roughness there.

A fixed time step
size for the flow time step was chosen such that the maximal values of
the CFL number fluctuate around 1. Depending on the bulk velocity, this resulted in time step sizes between
$5.38\mys$ and $7.64\mys$. To
fulfill the more strict stability constraint based on the speed of
sound, the acoustic time steps are 8 to 11 times smaller.  Starting
from a uniform flow field, a physical time of $t=0.5\s$ has been
simulated. From this time interval, the first $0.1\s$ have been
neglected in the data evaluation to exclude the initial transient
phase.

Altogether, 18 simulations with different bulk velocities ranging from
$U_b=27\mps$ to $U_b=38\mps$ have been conducted and analyzed. This corresponds to Mach numbers covering a range between $0.079$ and $0.111$.

The simulations have been performed on 80 CPU cores of type Intel(R)
Xeon(R) CPU E5-2670 v2 @ 2.50GHz. To cover one flow time step around 11
$\mathrm{ \mu CPUsec}$ per grid cell had been consumed. This
computational effort includes both, the flow time step and
the integration of the acoustic equations with 8 to 11 acoustic
sub steps, which is causing approximately half of the computational costs. 
The performance number scattered by $\pm 10 \%$ among the
cases, with a clear tendency to lower values at higher $U_b$. This is
reasonable, since at higher Mach number the number of acoustic
sub steps per flow time step reduces. An influence of the
implemented feedback term on the code performance was not observed.

\subsubsection{Results}

\begin{figure}[htb]
  \begin{tikzpicture}
  \pgftext{
  \includegraphics[page=2, trim=0cm 0cm 0cm 0cm, clip, width=0.5\textwidth]{figures/plot_sim_exp_paper.pdf}
  }%
  \node[text opacity=1] at (-3.8, 2.5) {\bf(a)};
  \end{tikzpicture}
  \begin{tikzpicture}
  \pgftext{
  \includegraphics[page=1, trim=0cm 0cm 0cm 0cm, clip, width=0.5\textwidth]{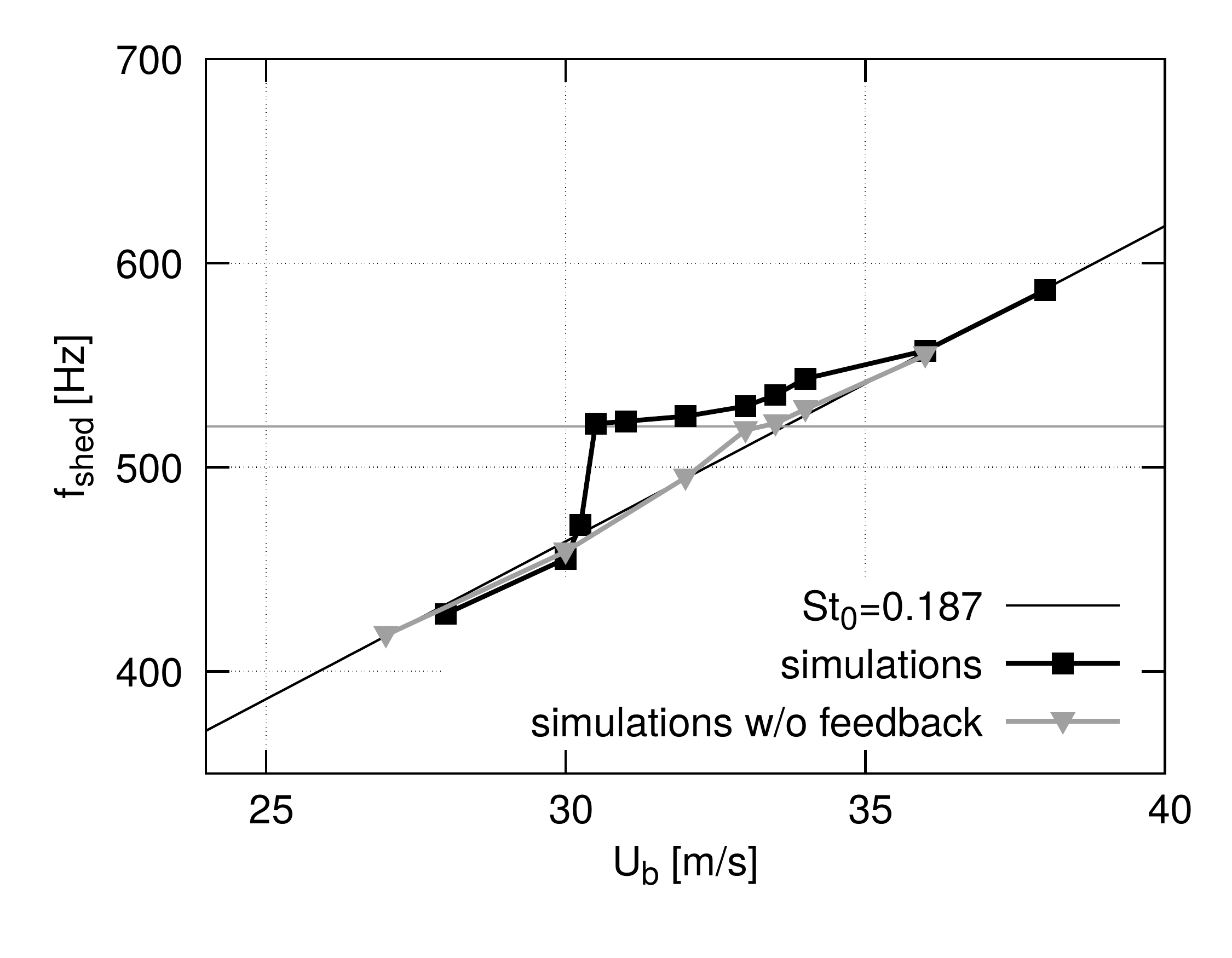}
  }%
  \node[text opacity=1] at (-3.8, 2.5) {\bf(b)};
  \end{tikzpicture}
  \caption{Parker mode. Shedding frequency in the wake. (a) Experimental results, reproduced from Figure~2(b) in \cite{welsh84}; (b) Simulation results. The Strouhal number infers from the slope defined by the data points outside the lock-in plateau (black solid lines), the resonance frequency by the level of the plateau itself (horizontal gray solid lines).}
  \label{fig:parker_fshedding}
\end{figure}

\paragraph{Comparison between experimental and simulated data}

Figure~\ref{fig:parker_fshedding} show the relation between the vortex
shedding frequency and flow velocity. Experimental data are depicted in Figure~\ref{fig:parker_fshedding}(a) and results from the simulations in Figure~\ref{fig:parker_fshedding}(b). Both show the characteristic behavior of the flow case
depending on the increasing bulk velocity: natural vortex shedding, lock in,
hysteresis and again natural vortex shedding.

In addition, Figure~\ref{fig:parker_fshedding}(b) shows results for 
simulations without feedback term. In this case the vortex
shedding frequency is proportional to flow speed over the entire range
of bulk velocities.

Experiment and simulation show the same qualitative behavior. However, the absolute value of the Parker-$\beta$-mode frequency and the natural
vortex shedding frequency (at speeds without feedback) deviate
moderately between experiment and simulation.  

\begin{table}[htbp]
  \centering
  \caption{Parker mode. Scaling parameters.}
  \vspace{2mm}
  \begin{tabular}{@{\arrayrulewidth1.5pt\vline\hspace{3pt}}c@{\hspace{3pt}\arrayrulewidth1.5pt\vline\hspace{3pt}}c|c@{\hspace{3pt}\arrayrulewidth1.5pt\vline}}
    \noalign{\hrule height1.5pt}
    & experiment \cite{welsh84} & simulation \\
    \noalign{\hrule height1.5pt}
     $f_{r\!e\!s}$ & $530\Hz$ & $520\Hz$ \\
    \hline
     $S\!t_0$& $0.213$ & $0.187$ \\
    \hline
     $U_{r\!e\!f}=\frac{f_{r\!e\!s} \cdot t}{S\!t_0}$ & $30.1\mps$ & $33.6\mps$ \\
    \noalign{\hrule height1.5pt}
  \end{tabular}
  \label{tab:parker_scaling}
\end{table}

Table~\ref{tab:parker_scaling} summarizes the resonant frequencies ($f_{res}$) and the Strouhal number ($S\!t_0$) of trailing edge vortex shedding as deduced from experiment and from simulations with feedback. The Strouhal number infers from the slope defined by the data points outside the lock-in plateau in Figure~\ref{fig:parker_fshedding} (black solid lines), the resonance frequency by the level of the plateau itself (horizontal gray solid lines in the figure). Furthermore, as indicated in the table, a reference velocity can be defined at which the lock-in frequency $f_{res}$ coincides with the vortex shedding frequency, $U_{ref}:=f_{res}\cdot t/S\!t_0$, where $t$ denotes the plate thickness.  
In the experiment, the resonant frequency was $f_{res}=530\Hz$ and the natural Strouhal number (if not locked in) of the vortex shedding
was ${S\!t}_0=0.213$. 
With the present numerical setup, the vortex shedding frequency behind the plate was
$S\!t=0.187$ in the WMLES. This is by 11\% lower than the corresponding value in the
experiment, indicating a thicker boundary layer. 

\begin{figure}[htb]
  \centering
  \includegraphics[page=3, trim=0cm 0cm 0cm 0cm, clip, width=0.5\textwidth]{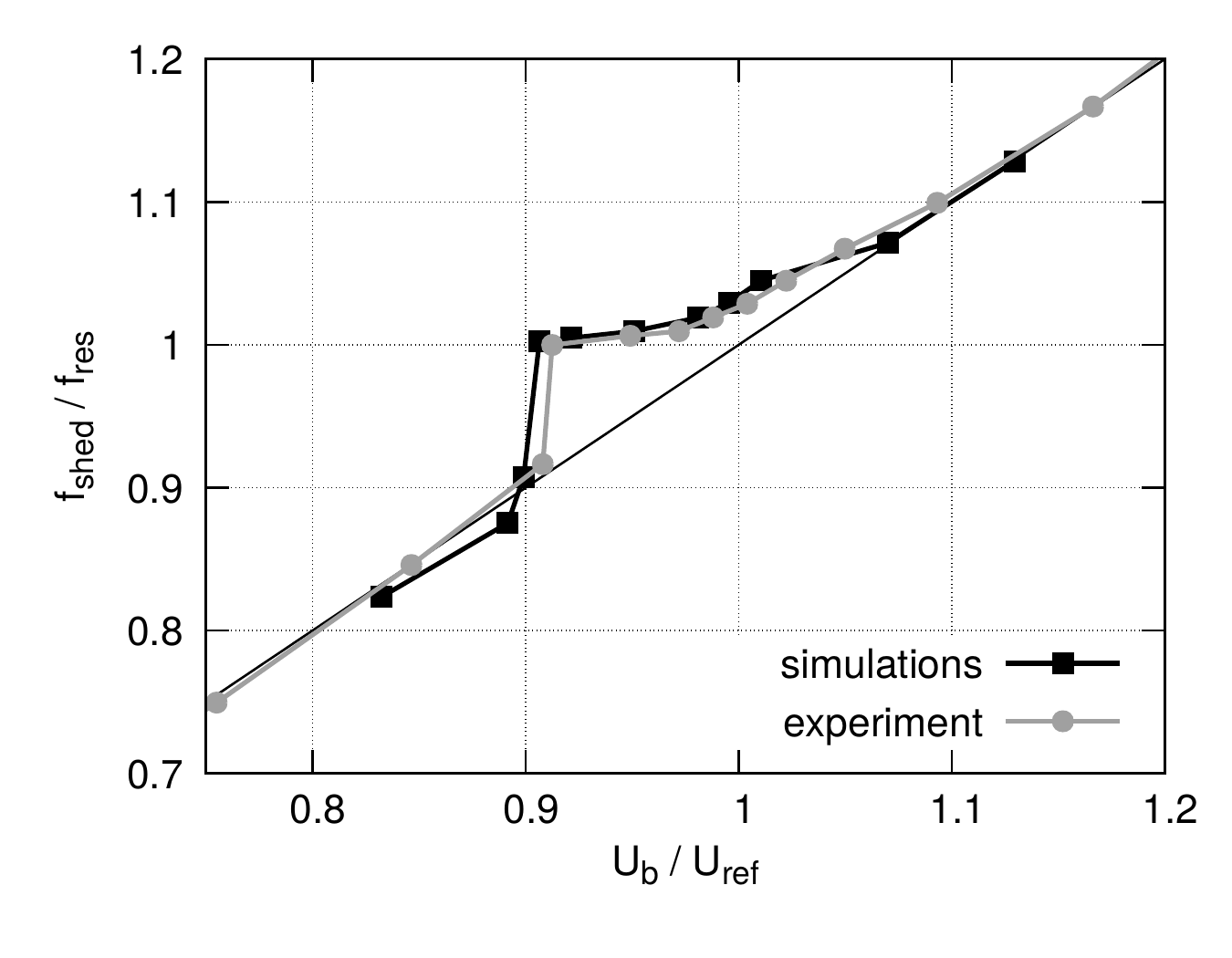}
  \caption{Parker mode. Shedding frequency in the wake, normalized results.}
  \label{fig:parker_fshedding_comparison}
\end{figure}

\begin{figure}[htb]
  \begin{tikzpicture}
  \pgftext{
  \includegraphics[page=5, trim=0cm 0cm 0cm 0cm, clip, width=0.5\textwidth]{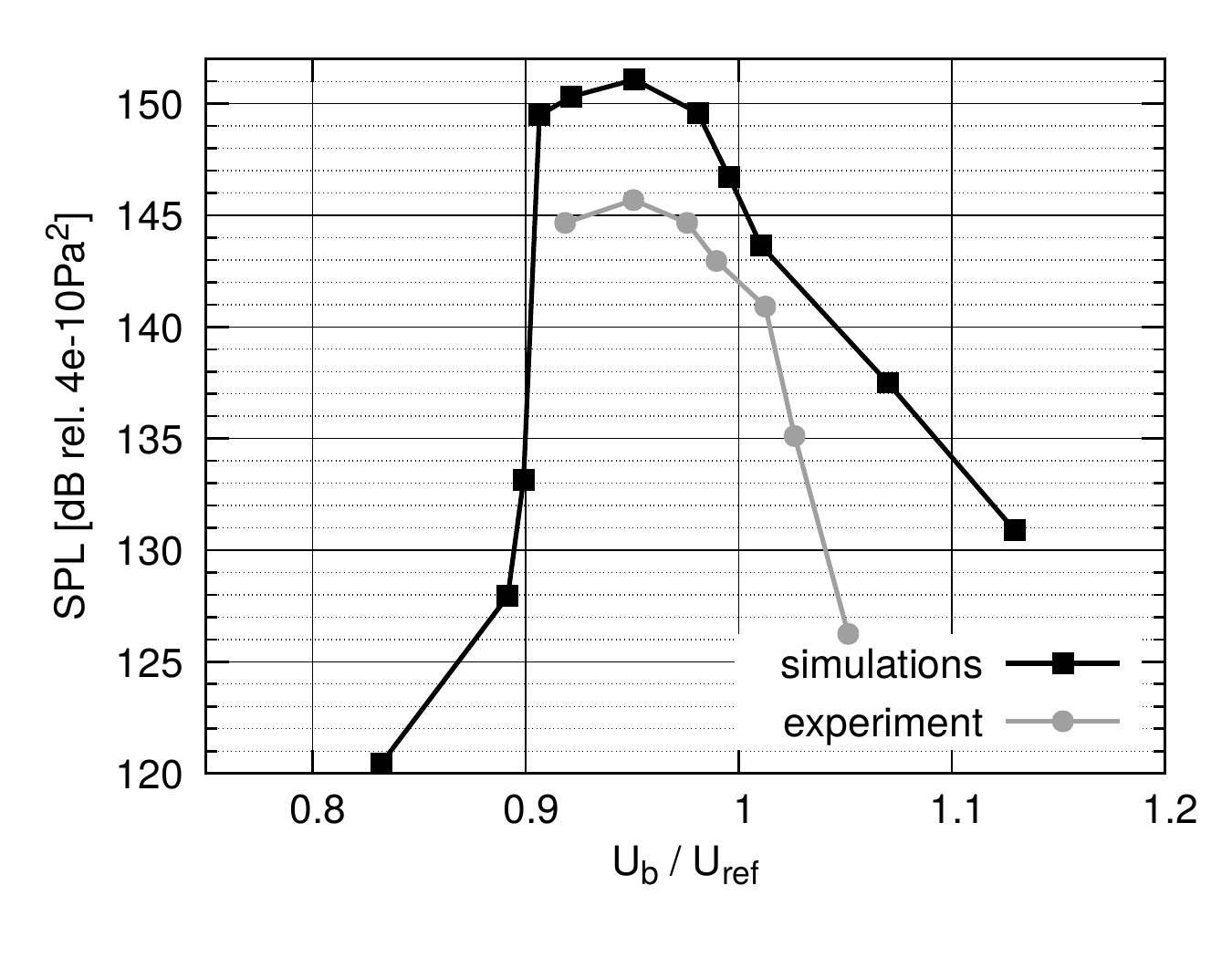}
  }%
  \node[text opacity=1] at (-3.8, 2.5) {\bf(a)};
  \end{tikzpicture}
  \begin{tikzpicture}
  \pgftext{
  \includegraphics[page=4, trim=0cm 0cm 0cm 0cm, clip, width=0.5\textwidth]{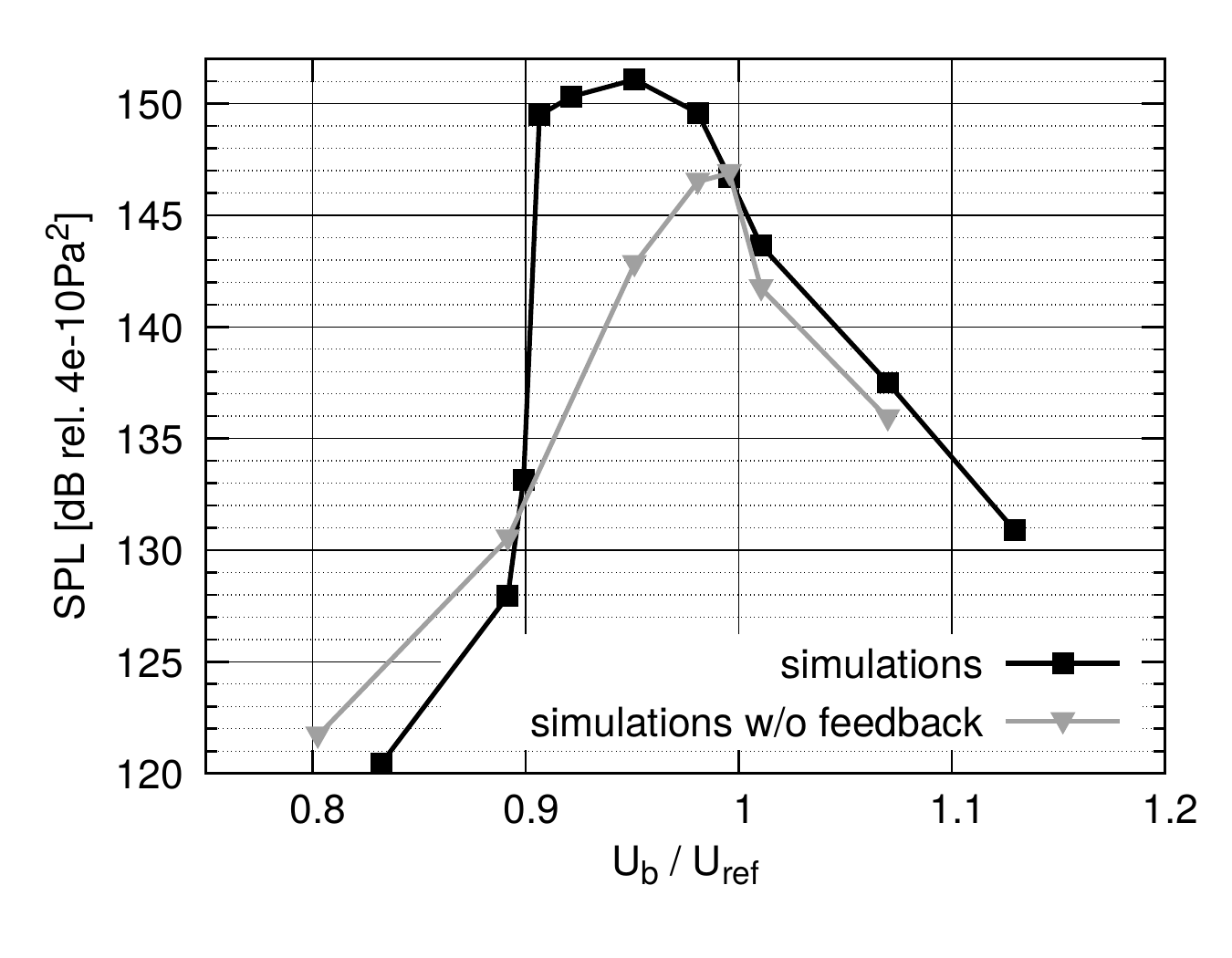}
  }%
  \node[text opacity=1] at (-3.8, 2.5) {\bf(b)};
  \end{tikzpicture}
  \caption{Parker mode. Pressure level at channel wall above the plate. (a) Experimental data reproduced from Figure~2(b) in \cite{welsh84}. (b) Simulation with and without feedback term.}
  \label{fig:parker_spl_comparison}
\end{figure}

To directly compare experiment and simulation, normalized plots using resonance frequency and reference velocity from Table~\ref{tab:parker_scaling} are shown in Figure~\ref{fig:parker_fshedding_comparison}. From the figure it infers that the lock-in effect starts slightly earlier in the simulations compared to the experiment, with a normalized bulk velocity around $U_b/U_{r\!e\!f}=0.903$ for the onset in the simulations and experimental onset at  $U_b/U_{r\!e\!f}=0.910$. However, otherwise the normalized plots reveal an almost identical behavior between experiment and simulation. 

Overall sound pressure level as obtained from experiment and simulation at
the channel wall above the center of the plate are shown in
Figure~\ref{fig:parker_spl_comparison}(a). For the locked-in cases the
qualitative behavior is equal in experiment and simulation, even though the exact amplitude is not
met. The peak value of $151.1\dB$ in the simulations exceeds the
maximal experimental level of $145.5\dB$. The exact peak level depends on the equilibrium between excitation of the acoustic resonance by
the flow and losses. Welsh et al.~\cite{welsh84} mention four loss
mechanisms: non-linear effects in the boundary layers, absorption by
the vortex street, energy absorbed by the channel walls through
vibration and radiation out of the ends of the duct. Wall vibrations are not included in the simulations. The radiation losses are not
modeled correctly, since the exact shape of the channel ends is not
reproduced. Losses by friction and heat transfer at the walls are
approximated by an absorption rate of $\alpha=0.01$. However, because of the ambiguities in the magnitude of the damping effects in the experiment, some deviations in the peak level remain. 

The higher amplitude of the acoustic mode in the simulations could explain
that the lock-in starts already at lower normalized frequency compared to the
experiment as observed in Figure~\ref{fig:parker_fshedding_comparison}.

\paragraph{Differences between natural and locked-in vortex shedding}

\begin{figure}[htb]
	\includegraphics[trim=0cm 0cm 0cm 0cm, width=0.50\textwidth]{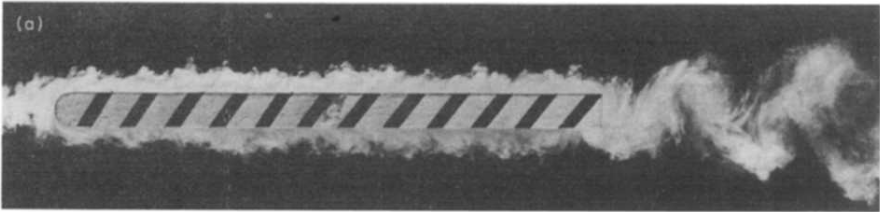}
	\includegraphics[trim=0cm 0.1cm 0cm 0cm, width=0.50\textwidth]{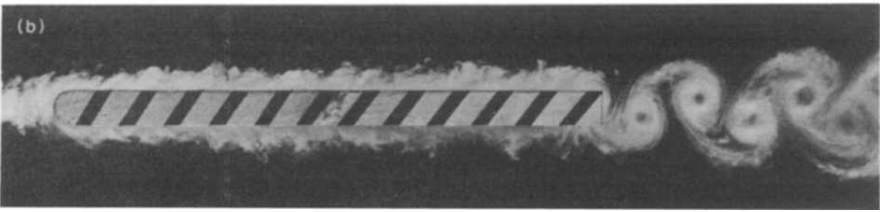}
	\caption{Vortex shedding from a plate with $c/t=16$ and $t=12.1\mm$. (a) Without resonant sound, $St=0.213$, $v_\infty=24.6\mps$; (b) with $\beta$-mode resonant sound, $SPL=145.5\dB$, $St=0.224$, $v_\infty=29.0\mps$. Reprinted from~\cite{welsh84}, Copyright (1984), with permission from Elsevier.}
	\label{fig:parker_welchvisualization}
\end{figure}

\begin{figure}[htb]
	\includegraphics[trim=0cm 3cm 0cm 0cm, clip=true, width=0.5\textwidth]{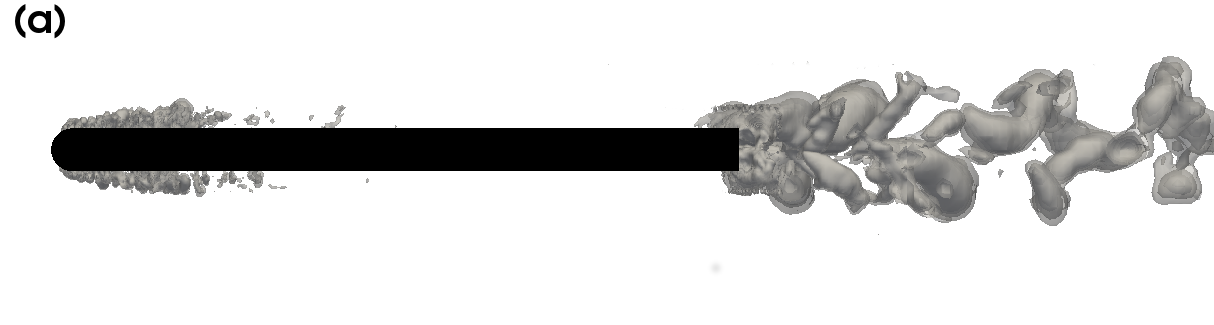}
	\includegraphics[trim=0cm 3cm 0cm 0cm, width=0.5\textwidth]{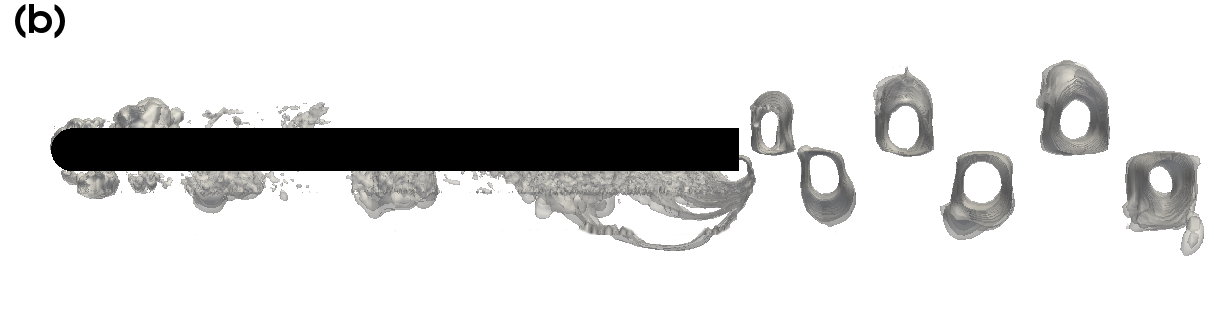}
	\caption{Parker mode. $c_p'=-25\dots -0.075$, iso-surfaces. (a)~$U_b=30\mps$, (b) $U_b=32\mps$.}
	\label{fig:parker_piso}
\end{figure}

To investigate the effect of the feedback on sound pressure level,
simulations without feedback term have been performed as well. Without feedback, the vortex shedding frequency follows a linear trend with increasing flow velocity as shown in Figure~\ref{fig:parker_fshedding}. 
The corresponding scaling of overall sound pressure level with bulk velocity is presented in Figure~\ref{fig:parker_spl_comparison}(b) (gray solid line). The plot shows a peak for the critical bulk velocity $U_b=U_{r\!e\!f}$. The figure includes also the corresponding trend with feedback switched on (black solid line). At critical bulk velocity, the same level is
reached with and without feedback. Since the natural vortex shedding
fits already to the resonance frequency, the feedback has no
additional effect on level. In the velocity ranges of lock-in
between $U_b=30.25\mps$ and $U_b=33.0\mps$ ($U_b/U_{r\!e\!f}=0.90$ and $U_b/U_{r\!e\!f}=0.98$) the amplitudes of fluctuations are with feedback
expectedly higher than without feedback, inevitably yielding higher levels. Above
$U_b=U_{r\!e\!f}$ feedback still causes slightly higher amplitudes. At
velocities below the lock-in the amplitudes with feedback are lower
than the amplitudes without. This indicates that an energy drain from acoustics to the flow is mediated by the feedback mechanism in this frequency range.

Flow visualizations by a smoke probe in the experiment revealed remarkable differences between the flow structures presented for locked-in cases and
cases with natural vortex shedding,
refer to Figure~\ref{fig:parker_welchvisualization}. If locked-in, the vortex
cores are stronger and homogeneous in span-wise direction. The same
difference can be seen in the simulation results, as shown by the
iso-surfaces of fluctuating pressure in Figure~\ref{fig:parker_piso}. At
$U_b=32\mps$ ($U_b/U_{r\!e\!f}=0.95$) flow and acoustics are locked in and the maximal value of
acoustic amplitude is reached (Figure~\ref{fig:parker_piso} b).
$U_b=30\mps$ ($U_b/U_{r\!e\!f}=0.89$) is just below the onset of lock-in
(Figure~\ref{fig:parker_piso} a). The strong acoustic fluctuations not
only change the flow pattern in the wake, but also trigger the separated flow at the leading edge.

\begin{figure}[htb]
  \begin{tikzpicture}
  \pgftext{
  \includegraphics[page=1, trim=0cm 0cm 0cm 0cm, clip, width=0.5\textwidth]{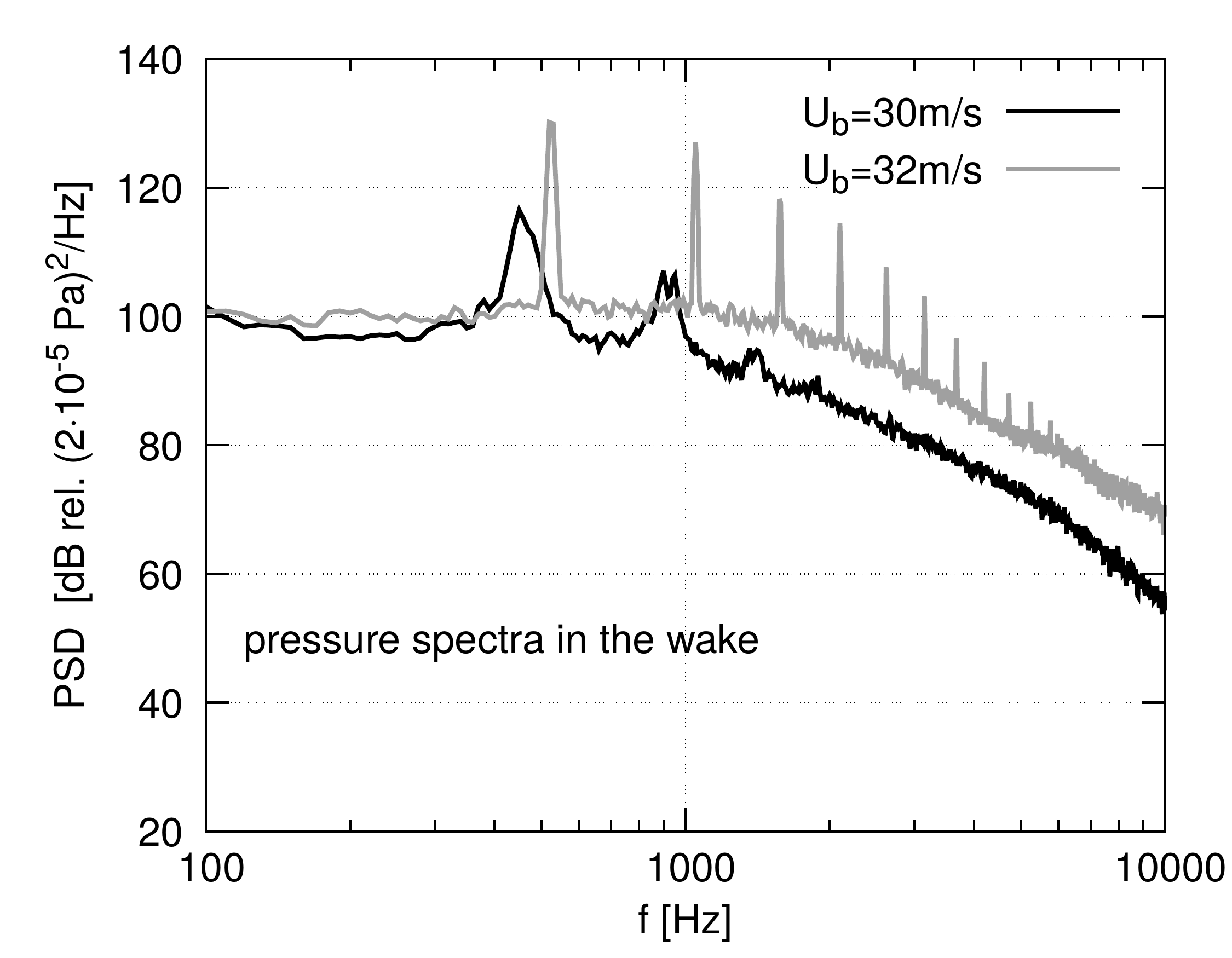}
  }%
  \node[text opacity=1] at (-3.8, 2.5) {\bf(a)};
  \end{tikzpicture}
  \begin{tikzpicture}
  \pgftext{
  \includegraphics[page=2, trim=0cm 0cm 0cm 0cm, clip, width=0.5\textwidth]{figures/plot_spectra_paper.pdf}
  }%
  \node[text opacity=1] at (-3.8, 2.5) {\bf(b)};
  \end{tikzpicture}
  \caption{Parker mode. (a) Pressure spectra in the wake. (b) Pressure spectra above the plate.}
  \label{fig:parker_spec_p}
\end{figure}

\begin{figure}[htb]
  \begin{tikzpicture}
  \pgftext{
  \includegraphics[page=3, trim=0cm 0cm 0cm 0cm, clip, width=0.5\textwidth]{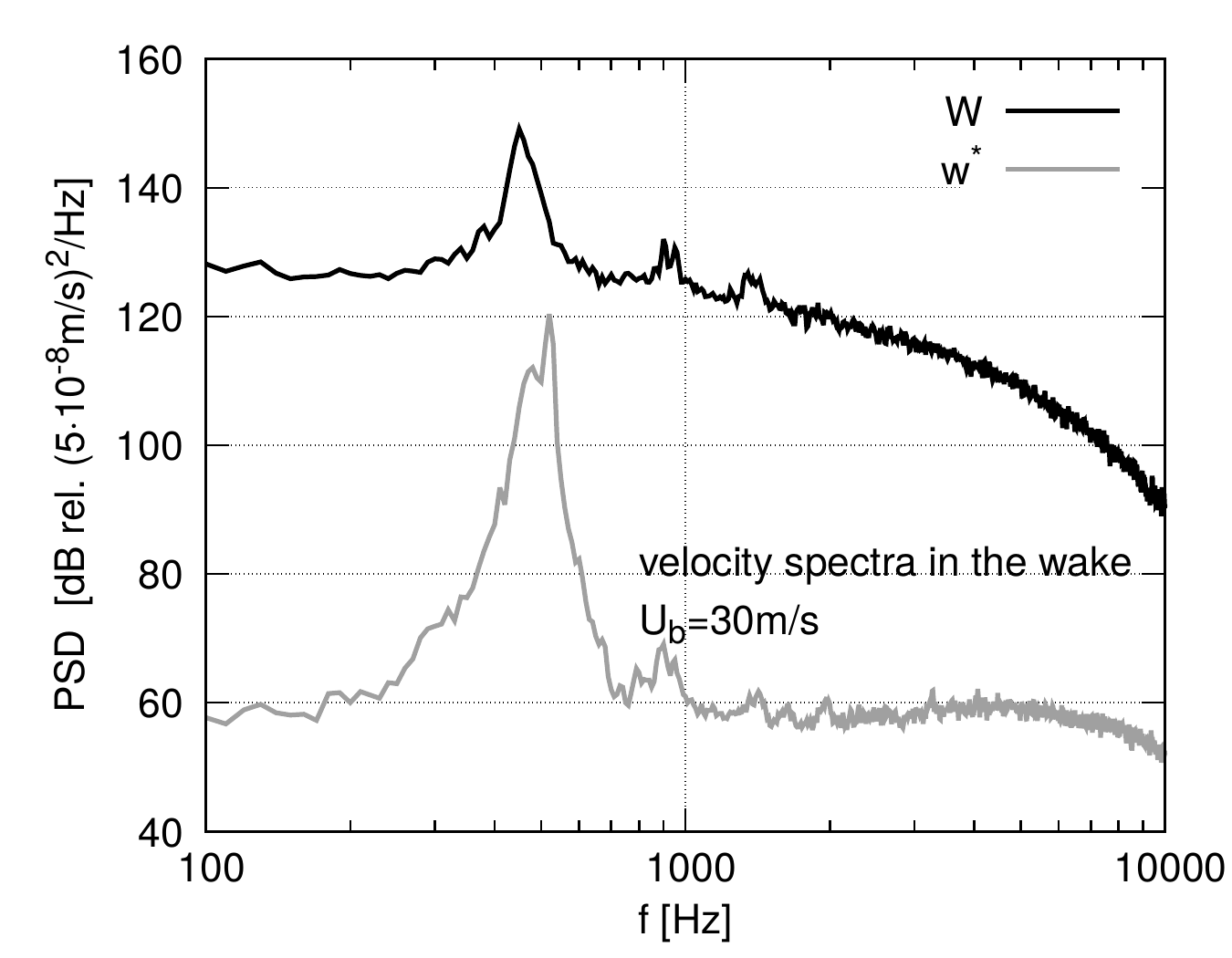}
  }%
  \node[text opacity=1] at (-3.8, 2.5) {\bf(a)};
  \end{tikzpicture}
  \begin{tikzpicture}
  \pgftext{
  \includegraphics[page=4, trim=0cm 0cm 0cm 0cm, clip, width=0.5\textwidth]{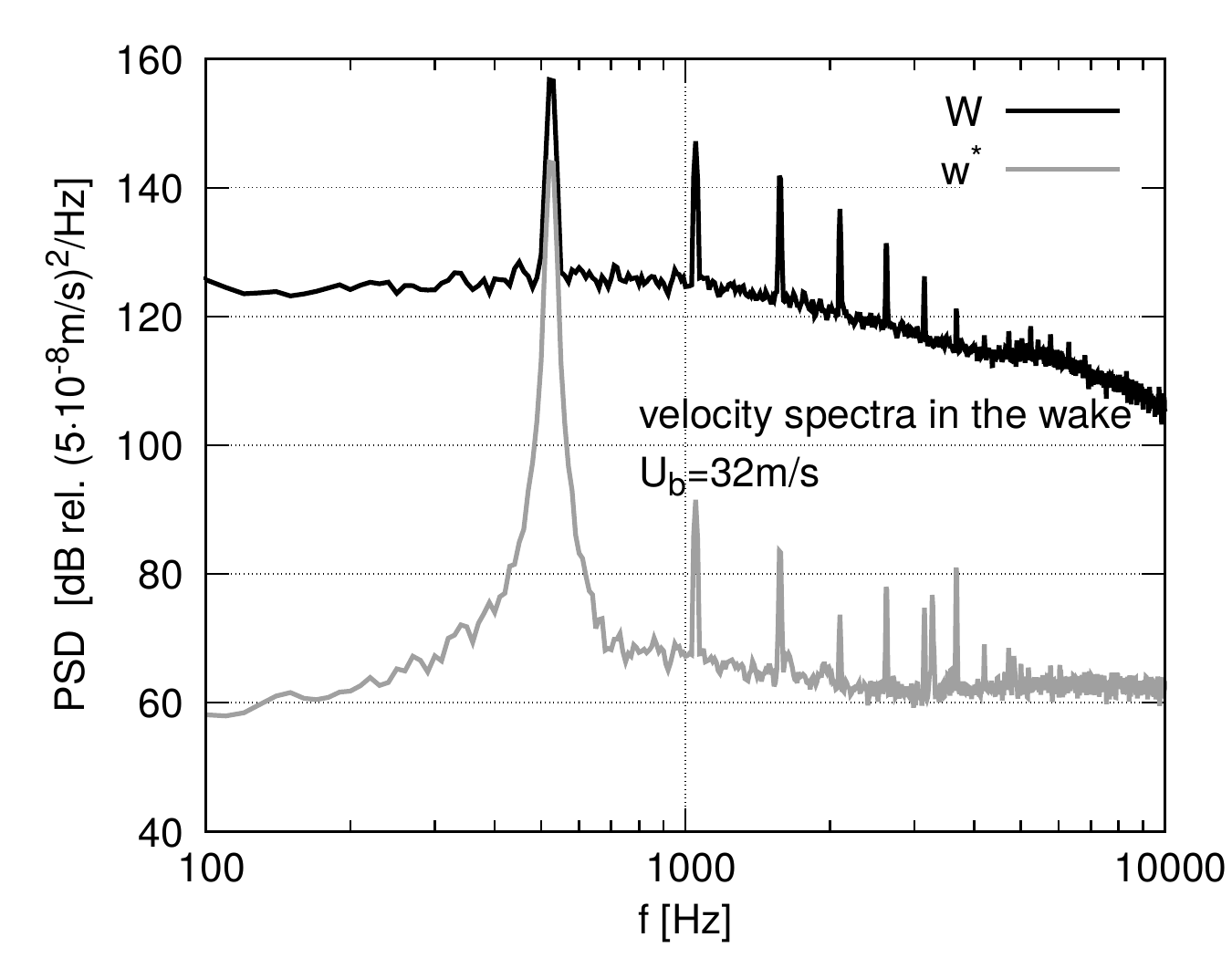}
  }%
  \node[text opacity=1] at (-3.8, 2.5) {\bf(b)};
  \end{tikzpicture}

  \caption{Parker mode. Spectra of normal velocity in the wake. (a) $U_b=30\mps$, (b) $U_b=32\mps$}
  \label{fig:parker_spec_w}
\end{figure}

The pressure spectra in the wake, averaged over all probes of the
array, clearly show the vortex shedding frequency and its higher
harmonics (Figure~\ref{fig:parker_spec_p}(a)). In case of natural vortex
shedding, the peaks are broad and the 3rd harmonic is only hardly
visible. Higher harmonics are not visible. In case of lock-in the peaks
are much sharper. The amplitude of the base peak rises by $13\dB$ and a
series of higher harmonics is visible.  Above and below the plate the
pressure spectra (Figure~\ref{fig:parker_spec_p}(b)) show some peaks
which are at the same frequencies for both velocities, e.g. at $520\Hz$,
$1630\Hz$, $3040\Hz$ and $3600\Hz$. These are the passive acoustic
eigenfrequencies. Additionally, the frequencies of the vortex shedding
are visible: for $U_b=30\mps$ at $455\Hz$ together with a weak first
harmonic; for the locked-in case $U_b=32\mps$ at $520\Hz$ and all higher harmonics.

\begin{figure}
    \centering
  \begin{tikzpicture}
    \pgftext{%
    \includegraphics[width=0.7\textwidth]{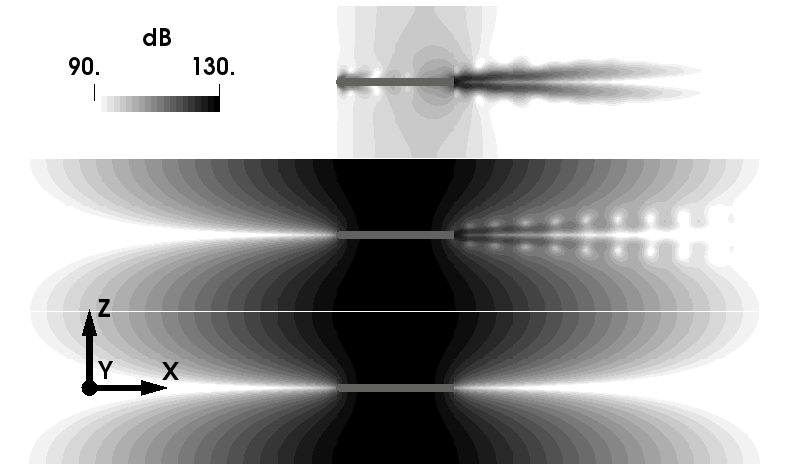}%
    }%
    \node[text opacity=1] at (-5.7, 3.4) {\bf(a)};
    \node[text opacity=1] at (-5.7, 0.9) {\bf(b)};
    \node[text opacity=1] at (-5.7,-1.6) {\bf(c)};
  \end{tikzpicture}
  \caption{Parker mode. Power spectral density at $500\Hz$ with a bandwidth of $50\Hz$, containing the resonance frequency. Different pressure components: (a) $P$, (b) $p$, (c) $p^*$.}
  \label{fig:parker_insitu}
\end{figure}

\subsection{Magnitude of acoustic and flow fluctuations for both simulation problems}
\label{sec:amplitudes}

Finally, the relative magnitude of acoustic and flow fluctuations are discussed for both simulation problems.  Figure~\ref{fig:parker_insitu} shows for the Parker-$\beta$-mode the spatial distribution
of sound pressure level in the channel obtained by separately integrating the power-spectral-density of hydrodynamic pressure $P$, the complete compressible pressure $p$, and the acoustic pressure $p^*$ over a band with center frequency $500\Hz$ and bandwidth $50\Hz$. The contour plots indicate that the acoustic pressure $p^*$ dominates the pressure field outside the wake. 

Figure~\ref{fig:parker_spec_w} presents the vertical components of sound particle velocity $w^*$ and incompressible velocity $W$, respectively, around the edge of the plate, where maximal values are observed. Part (a) in the figure on the left is the non-locked case at $U_b=30\mps$, part (b) indicates a locked-in case at $U_b=32\mps$. Even for the locked-in case the level of sound particle velocity is still $14\dB$ below
the incompressible velocity.

The rms-Values corresponding to the spectra are
$W_{rms}=17.14\mps$ and $w^*_{rms}=3.58\mps$. If the amplitude of acoustic velocities is estimated using the rms-value of pressure $p_{rms}=710\Pa$, which corresponds to a maximal sound pressure level of $151.1\dB$ above the plate as found in the simulations, a value in the same order of magnitude is obtained:  $w^*_{rms}\approx p^*_{rms}/(\rho c)=1.78\mps$.

For the generic flute problem, the pressure spectrum at the stopped end is also
acoustically dominant, refer to Figure~\ref{fig:recorder_gridstudy_convectiveterms}(a). The sound pressure level reaches $156.7\dB$, respectively $p_{rms}=1370\Pa$. Very locally at the sharp edge of the labium, the peak value of
particle velocity $u^*$ is of the same order as the ones of incompressible velocity $U$ with values yielding $25.5\mps$, respectively $30.5\mps$. But this reflects the singular flow behavior in the vicinity of the edge. The maximal value of $u^*$ is not a suitable scale for a general characterization of the flow. Estimating $u^*_{rms}$ using the SPL at the stopped end
results in a particle velocity magnitude in the pressure antinodes of $u^*_{rms}\approx p^*_{rms}/(\rho c)=3.42\mps$, roughly on
order of magnitude lower than the fluctuations in $U$.

To conclude, the simulation of both feedback problems at the common Mach number of $M=0.0875$ exhibit acoustic fluctuations of considerable magnitude relative to the flow. Precisely, the sound particle velocity is about one order of magnitude smaller than the flow velocity. This scaling is in agreement with the Mach number scaling derived in Section \ref{sec:low_subsonic_vortex_sound}, i.e. $U\sim M$ and $u^*\sim \alpha  M^2$, based on a relative scaling magnitude of order one, $\alpha=\mathcal{O}(1)$.

Furthermore, the results from simulation show that the common
assumption of acoustic pressure fluctuations being at low Mach number small compared to
hydromechanical ones, which was already found to be
inappropriate for internal flows in \cite{kreuzinger2013}, does not
hold in the special case of flow involving acoustic feedback with resonances.


\section{Conclusions}\label{sec:conclusions}

A generalized hydrodynamic-acoustic splitting (HAS) method for the
compressible Navier-Stokes equations was introduced that yields an
equivalent set of coupled equations for flow and acoustics. As an
extension to the standard approach based on deriving the governing
equations for the acoustics part by subtracting the incompressible
from the compressible Navier-Stokes equations, the generalized
approach presented in the current work yields also modifications to the governing
flow equations, which deviate from those of incompressible flow by
additional terms that account for the feedback from acoustics to the
flow. A unique simplified version of the split equation system with
feedback is derived that conforms to the compressible Navier-Stokes
equations in the subsonic flow regime, where the feedback reduces to
one additional term in the flow momentum equation. The Mach number
limit of applicability of the simplified split system was estimated to
be given by $M\approx 0.3-0.4$.

The additional feedback term was implemented into an existing scale resolving
run-time coupled code which solves the incompressible Navier-Stokes
equations together with the acoustic perturbation equations. For validation purposes, two
subsonic problems involving flow acoustic
feedback at common Mach number 0.0875 have been simulated. As a first problem, the tone of a generic flute at low Reynolds number is studied. The
results of the hydrodynamic-acoustic splitting based direct numerical
simulation are in good agreement to reference results from Lattice-Boltzmann simulation~\cite{kuehnelt2016}.

The discussion of the feedback term furthermore highlights the general helpfulness of the HAS approach in the analysis of flow-acoustic interaction problems due to the immediately availability of the Helmholtz decomposed flow variables.

The second problem considered the turbulent flow around a
thick plate in a channel at varying Mach numbers which was studied
using wall modeled LES. The simulations showed the lock-in of the vortex
shedding frequency to the Parker-$\beta$-mode resonance in very good
agreement with experimental data from Welsh et al.~\cite{welsh84}. In both cases, turning off the
feedback term in the flow equations resulted in a qualitative different
behavior of the flow. This underlines the flow-acoustic feedback
being of key importance for the studied problems.
The implemented feedback term did not increase the computational effort and also did not negatively affect the stable convective flow based time step size of the flow equations. The findings are in agreement with those also derived from theoretical considerations in the paper.

A combined flow/acoustic simulation effort with about 11 $\mathrm{\mu CPUsec}$ per point and time step was demonstrated. Altogether with the convective time step size and the favorable resolution requirements of the present staggered mesh approach an efficient simulation approach for subsonic noise is accomplished.    

\section*{Acknowledgments}
We would like to express our thanks to Helmut K\"uhnelt for providing us with the geometry of the generic flute.

\bibliographystyle{model1-num-names}
\bibliography{elsarticle_HAS2020}

\end{document}